\documentclass[12pt,letterpaper]{article}
\usepackage{jheppub}

\usepackage{graphicx}
\usepackage{bbm}
\usepackage{amsmath}
\usepackage{amssymb}
\usepackage{mathtools}
\usepackage{amsthm}
\usepackage{float}

\usepackage[labelformat=simple]{subcaption}

\newcommand{\unit}[1]{\hat{\mathrm{\bold{#1}}}}

\DeclareMathOperator{\diag}{diag}

\def\be{\begin{equation}}
\def\ee{\end{equation}}

\newtheorem{Condition}{Condition}

\newtheoremstyle{named}{0.75\baselineskip}{0.75\baselineskip}{\itshape}{}{\bfseries}{.}{.5em}{#3}
\theoremstyle{named}

\begin{document}

\begin{titlepage}

{\flushright ACFI-T25-02 \\ }

\setcounter{page}{1} \baselineskip=15.5pt \thispagestyle{empty}

\bigskip\

\vspace{2cm}
\begin{center}
{\LARGE \bfseries Taxonomy of branes in infinite distance limits}

 \end{center}
\vspace{0.5cm}

\begin{center}
{\fontsize{14}{30}\selectfont Muldrow Etheredge}
\end{center}

\begin{center}
\vspace{0.25 cm}
\textsl{Amherst Center for Fundamental Interactions,\\Department of Physics, University of Massachusetts, Amherst, MA 01003, USA}

\vspace{0.25cm}

\end{center}
\vspace{1cm}
\noindent

I consider flat slices of moduli spaces where the $(-\nabla \log T)$-vectors of particle-towers and branes are constant, and I show that the Emergent String Conjecture constrains these vectors to reside on lattices. In asymptotic limits, this results in exponentially separated, discretized hierarchies of energy scales. I further identify conditions that determine whether a given lattice site must be populated, and I show that only a finite set of configurations satisfies these conditions. I classify all such configurations for 0d, 1d, and 2d moduli spaces in theories with 3 to 11 spacetime dimensions, and I argue that 11d is the maximal spacetime dimension compatible with my assumptions. Remarkably, this classification reproduces the detailed particle and brane content of various string theory examples with 32, 16, and 8 supercharges. It also describes some examples where the assumptions I use are violated, suggesting that my assumptions can be relaxed and the scope of this classification can be expanded. It might also predict new branes. For instance, if heterotic string theory is described by this classification, then it must possess non-BPS branes with D-brane-like tensions. Similarly, if this classification applies to the Dark Dimension Scenario with an extra modulus, then it requires the existence of strings with tensions related to the cosmological constant by $T\lesssim \Lambda^{1/6}$ in 4d Planck units.

\bigskip
\noindent\today

\end{titlepage}

\setcounter{tocdepth}{2}

\hrule
\tableofcontents

\bigskip\medskip
\hrule
\bigskip\bigskip

\section{Introduction\label{s.introduction} }

When parameters in quantum gravity become extreme, the \textbf{\emph{Emergent String Conjecture}} (ESC) \cite{Lee:2019wij} of Lee, Lerche, and Weigand proposes that either extra dimensions decompactify or fundamental strings approach zero-tension. The ESC is a powerful conjecture in the Swampland Program \cite{Vafa:2005ui,Brennan:2017rbf,Palti:2019pca,vanBeest:2021lhn,Grana:2021zvf,Harlow:2022ich,Agmon:2022thq,VanRiet:2023pnx}. It requires strong constraints on the moduli dependence of particle-towers and branes in the string landscape. This paper identifies and studies some of those constraints.

In string theory, the spaces of continuous parameters, or \emph{\textbf{moduli spaces}}, are parametrized by vevs of scalar fields $\phi^a$, and they have metrics $G_{ab}(\phi)$ given by the scalar kinetic matrices that appear in the low-energy EFT Lagrangians,
\begin{align}
	\mathcal L=\frac 1{\kappa^2}\left(R-\frac 12G_{ab}\partial_\mu \phi^a\partial^\mu \phi^b+\dots\right).
\end{align}
The \emph{\textbf{Distance Conjecture}} (DC) \cite{Ooguri:2006in} proposes that these moduli spaces have infinitely long spikes, and that, an asymptotic distance $\phi$ down a spike,\footnote{The literature refers to theories with vevs asymptotically down these spikes as being \emph{\textbf{infinite-distance limits}}, even when the distance down a spike is not actually infinite, but just a large distance in Planck units.} there exist infinite-towers of particles with masses that are exponentially light with $\phi$ in Planck units,
\begin{align}
m_n\sim \exp(-\alpha \phi), \label{e.tower}
\end{align}
where $\alpha$ is an order-1 number. This conjecture has been tested and studied in many situations (e.g., \cite{ Baume:2016psm,Klaewer:2016kiy, Blumenhagen:2017cxt, Grimm:2018ohb,Heidenreich:2018kpg, Blumenhagen:2018nts, Grimm:2018cpv, Buratti:2018xjt, Corvilain:2018lgw, Joshi:2019nzi,  Erkinger:2019umg, Marchesano:2019ifh, Font:2019cxq,  Gendler:2020dfp, Lanza:2020qmt, Klaewer:2020lfg, Etheredge:2022opl, Rudelius:2023mjy,Baume:2023msm, Etheredge:2023odp, Etheredge:2023usk, Aoufia:2024awo, Ooguri:2024ofs, Etheredge:2024amg, Etheredge:2024tok, Baines:2025TBD}). In any infinite-distance limit, the ESC \cite{Lee:2019wij} requires that either KK-modes or string oscillator modes become light with no other towers lighter, and this has been tested and analyzed in various contexts (see, e.g., \cite{Lee:2018urn, Baume:2019sry,Lee:2019xtm, Perlmutter:2020buo, Baume:2020dqd, Xu:2020nlh,Lanza:2021udy, Basile:2023blg, Castellano:2023jjt, Rudelius:2023odg,Bedroya:2024ubj, Calderon-Infante:2024oed, Friedrich:2025gvs}).

The ESC is very sensitive to the spectrum of particle towers and branes within a theory---improper inclusion, deletion, or variation of the moduli-dependence of the mass or tension of a particle tower or a brane in a theory can result in ESC failure. For example, the ESC can be violated under compactification if one does not have branes in the theory that become light fast enough in a given infinite-distance limit \cite{Etheredge:2024amg}. Meanwhile, the ESC can be violated if one adds to a theory a brane that becomes too light too fast, as the brane's oscillator modes could become light faster than KK-modes or string oscillator modes. Thus, the ESC requires a delicate balancing act: in a given asymptotic limit there must exist sufficiently many branes that become light fast enough for the ESC to hold under dimensional reduction, but these branes must not become light too fast. In this paper, I will translate this balancing act into precise rules and classify theories consistent with these rules.

\textbf{\emph{Scalar-charge-to-tension-ratio vectors}}  \cite{Palti:2017elp,Lee:2018spm, Gonzalo:2019gjp, DallAgata:2020ino,Andriot:2020lea,   Benakli:2020pkm, Calderon-Infante:2020dhm, Etheredge:2022rfl, Etheredge:2022opl,   Etheredge:2023odp, Etheredge:2023usk, Etheredge:2023zjk, Castellano:2023jjt, Castellano:2023stg, Etheredge:2024amg, Etheredge:2024tok} of particle towers and branes are useful objects that measure the moduli dependence of their masses and tensions. In a $d$-dimensional theory, these are defined by applying the gradient, with respect to the moduli, to the logarithms of the masses/tensions, via
\begin{align}
	\vec \alpha=-\nabla \log T,
\end{align}
with the $d$-dimensional Planck scale held constant.\footnote{In this paper, when working with a $d$-dimensional theory, I will be setting the $d$-dimensional Planck mass to 1. I will also let the symbol $T$ symbolize both tensions and masses.} These vectors, which I will refer to as $\boldsymbol{\alpha}$\emph{\textbf{-vectors}}, exist in the tangent spaces of moduli spaces, point in directions where tensions decrease fastest, and they promote the $\alpha$'s appearing in the Distance Conjecture \eqref{e.tower} to be vectors. For flat slices of multidimensional moduli spaces where these $\alpha$-vectors are constant, the \emph{\textbf{Convex Hull Distance Conjecture}} of \cite{Calderon-Infante:2020dhm} is the observation that the Distance Conjecture is equivalent to the statement that the convex hull of $\alpha$-vectors of particle towers contains a ball of sufficient radius $\alpha_\text{min}>0$. In \cite{Etheredge:2022opl}, it was argued that $\alpha_\text{min}\geq 1/\sqrt{d-2}$, which is the \emph{\textbf{Sharpened Distance Conjecture}} (Sharpened DC). In more general, non-flat slices of moduli spaces, such as when axions are present, the convex hulls of $\alpha$-vectors are often rotations of polytopes \cite{Etheredge:2022opl,Etheredge:2023odp,Calderon-Infante:2023ler,Etheredge:2023usk}. But, even for flat slices of moduli spaces, sometimes these $\alpha$-vectors are not polytopes and \emph{\textbf{slide}} (i.e., are non-constant) on these flat slices \cite{Etheredge:2023odp, Etheredge:2023usk, Etheredge:2023zjk}.

In this paper, I will restrict my analysis to flat slices of moduli spaces and assume that $\alpha$-vectors are constant, even after compactification or decompactification. These are very strong assumptions, but it is likely that these assumptions can be relaxed with some of my conclusions holding, as I will show that some of the results in this paper apply to theories that do not satisfy these assumptions. Relaxing these assumptions will be the focus of future work.\footnote{For a related discussion on relaxing similar assumptions, see \cite{Etheredge:2024tok}.}

\subsection{Taxonomy review}

In \cite{Etheredge:2024tok}, the ESC was used to derive precise constraints on $\alpha$-vectors of particle towers. Consider an asymptotic limit of a flat $N$-dimensional slice of moduli space. By the ESC and Convex Hull DC \cite{Calderon-Infante:2020dhm}, there must exist at least $N$ towers of either KK-modes or string oscillator modes becoming light, which I will call leading, or \emph{\textbf{principal}}, towers, and I will use the term \emph{\textbf{frame simplex}} to refer to the $(N-1)$-simplex generated by the convex hull of the $\alpha$-vectors of these principal towers. As discussed in \cite{Etheredge:2024tok}, the frame simplex defines a duality frame and vice versa. If all of the principal towers are KK-modes, or if one of the towers is a string oscillator mode, I will call the frame simplex respectively a \emph{\textbf{geometric frame}} or a \emph{\textbf{stringy frame}}. Consider two $\alpha$-vectors of KK-mode principal towers on the same frame simplex, labeled by $i$ and $j$, where each corresponds to decompactification to $D_i$ and $D_j$ dimensions. Then, by \cite{Etheredge:2024tok}, the dot product of the vertices $\vec v_i$ and $\vec v_j$ satisfy the following \emph{\textbf{taxonomy rule}} that governs their dot products and lengths:
\begin{align}
	\vec \alpha_i\cdot \vec \alpha_j=\frac 1{d-2}+\frac 1{D_i-d}\delta_{ij}.\label{e.taxonomy}
\end{align}
The case where vertex $i$ is a string oscillator tower can be obtained from the $D_i\rightarrow \infty$ limit. In this paper, I will rederive and generalize these rules to apply to both subleading towers and branes.

By simultaneously considering multiple directions in moduli space, one can extend the analysis from one duality frame to multiple duality frames by attaching frame simplices. When all of the different frames are considered together, the convex hull of the attached frame simplices generates a \emph{\textbf{tower polytope}}. In the case where the maximum decompactification dimension is 11, these rules were used in \cite{Etheredge:2024tok} to classify all 1d tower polytopes of theories with 10 spacetime dimensions, and all 2d tower polygons of theories with 6 to 9 spacetime dimensions. In this paper, I will extend this classification to theories with 3 to 5 spacetime dimensions. I will also argue that some of the tower polytopes found in \cite{Etheredge:2024tok} are not consistent with the rules I derive.

Principal towers, and thus frame simplices and tower polytopes, are closely related to the species scale $\Lambda$ \cite{vandeHeisteeg:2023ubh, vandeHeisteeg:2023uxj, Calderon-Infante:2023ler, Castellano:2023stg, Castellano:2023jjt}, which is a moduli-dependent UV cutoff scale of the $d$-dimensional EFT. Given a flat slice of moduli space, the \emph{\textbf{species polytope}}  \cite{Calderon-Infante:2023ler} is the convex hull of the set of values that $-\nabla \log \Lambda$ takes in the set of all asymptotic limits of this slice. Using the tower species pattern of \cite{Castellano:2023jjt, Castellano:2023stg}, in \cite{Etheredge:2024tok} the taxonomy rule \eqref{e.taxonomy} for principal towers was used to obtain rules governing and classifying the allowed 2d species polytopes of theories with 6 to 9 spacetime dimensions. In this paper, I will extend this classification to theories with 3 to 5 spacetime dimensions, and I will also argue that some of the species polytopes found in \cite{Etheredge:2024tok} are not consistent with the lattices rules I derive. In \cite{ Rudelius:2023spc,  Etheredge:2024amg}, connections between branes and the species scale were observed which I will further illuminate in this paper.

These taxonomy rules, as well as connections with the species scale, were partially extended to branes in \cite{Etheredge:2024amg}, where it was found that, for a ``suitable" pair of $(p-1)$ and $(q-1)$-branes, where $p\geq q$, their $\alpha$-vectors should satisfy the following \emph{\textbf{brane taxonomy}} rule,
\begin{align}
	\vec \alpha_p\cdot \vec \alpha_q=a_{ij}+1-\frac{q(d-p-2)}{d-2},\label{e.branetaxomony}
\end{align}
where $a_{ij}$ is an integer, and $a_{ij}=1$ when the branes are the same. However, in \cite{Etheredge:2024amg}, it was not fully known when this rule should apply, nor its level of generality.\footnote{In \cite{Etheredge:2024amg}, special cases of this formula were studied, and some bounds on $a_{ij}$ were motivated. When the winding modes of the $(p-1)$ and $(q-1)$-branes from wrapping cycles of a $(p-1)$-torus have $\alpha$-vectors in the lower-dimensional theory on the same frame simplex, special cases of this formula can be derived by applying the taxonomy rule \eqref{e.taxonomy} to these winding modes.} This will be further elaborated in this paper, and these rules \eqref{e.branetaxomony} will be illuminated and generalized.

\subsection{New results}
In this paper, I will generalize the taxonomy rules of \cite{Etheredge:2024tok,Etheredge:2024amg} to extend to subleading particle towers and to branes. I will find that these $\alpha$-vectors reside on lattices, and that these lattices must be sufficiently populated in order for the ESC to hold under dimensional reduction.

In geometric frames, these lattices are obtained by interpreting the light branes and particle towers as string oscillators,\footnote{In this paper, I will primarily focus on string oscillators. Higher dimensional brane oscillators satisfy rules that follow from the rules I will derive for the branes, and they do not significantly affect the results in this paper. Except when explicitly stated otherwise, the only particle towers I will be referring to are towers that are either string oscillator modes, or towers that are dual to KK-modes (such as wrapped branes, or KK-monopoles in 4d, or exotic branes in 3d). Also, except when explicitly stated otherwise, the branes I will be considering will not be instantons. \label{footnote.osc}} wrapped/unwrapped branes, KK-modes, KK-monopoles, or exotic branes \cite{Elitzur:1997zn, Blau:1997du, Hull:1997kb, Obers:1997kk, Obers:1998fb, Eyras:1999at, Lozano-Tellechea:2000mfy, Kleinschmidt:2011vu, Bergshoeff:2011se, Kikuchi:2012za, Hellerman:2002ax, deBoer:2012ma, Etheredge:2024amg}. By consideration of stringy frames under dimensional reduction, I find similar lattices.

\begin{figure}
\centering
\begin{subfigure}{.8\linewidth}
\centering
\includegraphics[scale=1]{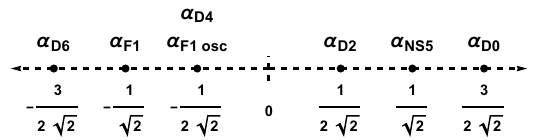}
\caption{Radion lattice of $\alpha$-vectors for  $\frac 12$-BPS branes of M-theory on $S^1$ (i.e., 10d IIA string theory). All of the branes depicted satisfy the radion lattice condition \eqref{e.radion_lattice} with the D0 brane (which is a KK-mode from the M-theory perspective).
}\label{f.1d_radion_lattice}
\end{subfigure}
\begin{subfigure}{.8\linewidth}
\centering
\includegraphics[scale=1]{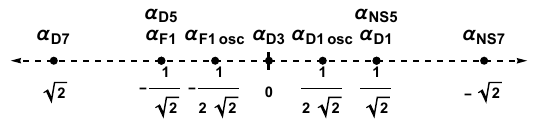}
\caption{Dilaton of $\alpha$-vectors for $\frac 12$-BPS branes in 10d IIB string theory. All of the branes depicted satisfy the dilaton lattice condition \eqref{e.dilaton_lattice} with the oscillators of fundamental strings (and also oscillators of D1 branes).
}\label{f.1d_dilaton_lattice}
\end{subfigure}
\caption{The radion lattice of $\alpha$-vectors of M-theory on $S^1$ (i.e. 10d IIA), and the dilaton lattice of 10d IIB string theory.
}
\label{f.1d_lattices}
\end{figure}

Consider a $d$-dimensional theory with a KK-mode corresponding to decompactification from $d$-dimensions to $D$-dimensions. If $\vec \alpha_{D,d}$ is the $\alpha$-vector of this KK-mode, I argue that the $\alpha$-vector of any non-oscillator $(p-1)$-brane must satisfy the following \emph{\textbf{radion lattice}} condition,
\begin{align}
	\boxed{\vec \alpha_p\cdot \hat \alpha_{D,d}=\frac{p(D-2)-P(d-2)}{\sqrt{(D-d)(D-2)(d-2)}}},\label{e.radion_lattice}
\end{align}
where $P$ is an integer.\footnote{This actually describes multiple lattices simultaneously. For fixed $p$, a lattice is fixed. But, all together, for all branes, there is a superlattice. The structure here is a sum of lattices, or Minkowski sum.} This formula results in $\alpha$-vectors having components in the direction of $\hat \alpha_{D,d}$ that are on a lattice, where $P$ controls the position on the lattice, see Figure \ref{f.1d_radion_lattice}. For the wrapped and unwrapped branes from the $D$-dimensional theory, $P$ can be viewed as the dimension of the brane in the $D$-dimensional theory, but I show that this formula also applies to KK-modes (where $P=0$), KK-monopoles (where $P=D-2$), and exotic branes (where $P>D-2$). Meanwhile, in an emergent string limit with a string oscillator with $\alpha$-vector $\vec \alpha_\text{osc}$, I argue that the $\alpha$-vector of any non-oscillator $(p-1)$-brane must satisfy the following \emph{\textbf{dilaton lattice}} condition,
\begin{align}
	\boxed{\vec \alpha_p(P)\cdot \hat \alpha_\text{osc}=\frac{p}{\sqrt{d-2}}+\frac{\sqrt{d-2}}{2}(p-P)},\label{e.dilaton_lattice}
\end{align}
where $P$ is an integer. I derive this by considering strings under circle reduction, where by T-duality the string winding modes can be interpreted as KK-modes and generate a radion direction with radion lattice conditions \eqref{e.radion_lattice}, and these radion lattice conditions can be uplifted to the higher-dimensional theory to produce the dilaton lattice conditions \eqref{e.dilaton_lattice}. This formula results in $\alpha$-vectors having components in the direction of $\hat \alpha_\text{osc}$ that are on a lattice, see Figure \ref{f.1d_dilaton_lattice}. The dilaton lattice condition \eqref{e.dilaton_lattice} can be reproduced from the radion lattice condition \eqref{e.radion_lattice} by replacing $P$ with $\frac D2(P-p)$ in the radion lattice formula and then taking the $D\rightarrow \infty$ limit. This \emph{\textbf{radion-to-dilaton prescription}} is related to the observation in \cite{Etheredge:2024tok} that emergent string limits behave similarly to decompactification limits where infinitely many dimensions are decompactified.

Since string oscillators have masses that scale with square roots of string tensions, the string oscillators exist in half-integer-multiple superlattices of the radion and dilaton particle lattices,
\begin{align}
	 \hat \alpha_{D,d}\cdot \vec \alpha_\text{osc}=\frac{D-2-\frac P2(d-2)}{\sqrt{(D-d)(D-2)(d-2)}},\qquad \hat \alpha_\text{osc}\cdot	\vec \alpha_\text{osc}'=\frac{1}{\sqrt{d-2}}+\frac{\sqrt{d-2}}4(2-P).
\end{align}
Similarly, oscillator modes of $(p-1)$-branes scale with the $p$-th roots of their respective branes, and thus lie on $1/p$-multiple superlattices of the particle lattices.

\begin{figure}
\centering
\includegraphics{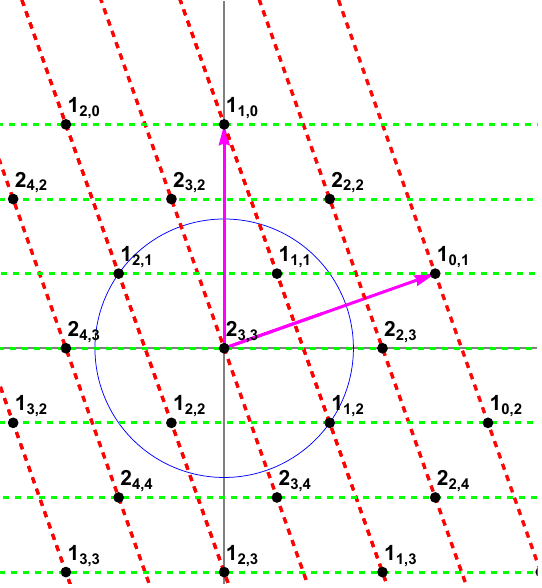}
\caption{Lattice of $\alpha$-vectors for a geometric frame.  Here, $d=4$, and the $1_{0,1}$ and $1_{1,0}$ vertices (highlighted with magenta arrows) form a frame simplex of KK-modes, each corresponding to decompactification to 5d. This frame simplex generates a lattice through the lattice equation \eqref{e.lattice}, where each lattice point is labeled by $p_{P_1,P_2}$, where $p$ is the spacetime dimension of the brane and $P_i$ refers to the position in the lattice. The green and magenta lines are the hyperplanes of points that respectively satisfy equation \eqref{e.radion_lattice} when dot producted with the $1_{0,1}$ and $1_{1,0}$ vertices. The blue circle has radius $1/\sqrt{d-2}$, and must be enclosed in the convex hull of particle towers, as mandated by the Sharpened DC.}\label{f.2d_radion_lattice}
\end{figure}

\begin{figure}
\centering 
\includegraphics{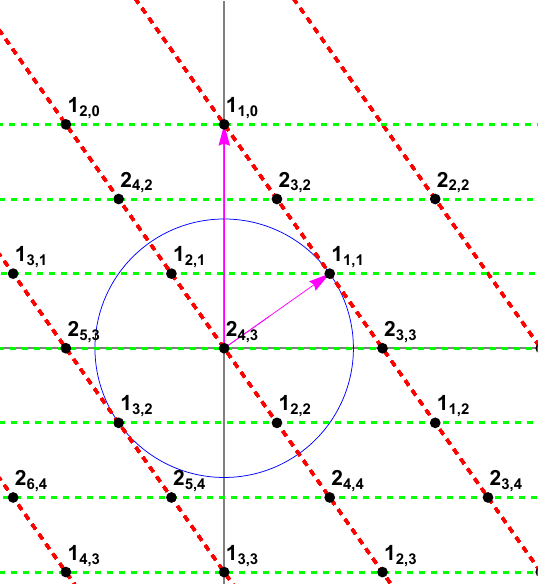}
\caption{Lattice of $\alpha$-vectors for a stringy frame. Here, $d=4$, and the $1_{1,0}$ and $1_{1,1}$ vertices (highlighted with magenta arrows) form a frame simplex of a KK-mode (of decompactification to 5d) and a string oscillator. This frame simplex generates a lattice through the lattice equation \eqref{e.lattice_stringy}, where each lattice point is labeled by $p_{P_1,P_2}$, where $p$ is the spacetime dimension of the brane and $P_i$ refers to the position in the lattice. The green and magenta lines are the hyperplanes of points that respectively satisfy equations \eqref{e.radion_lattice} \eqref{e.dilaton_lattice} when dot producted with the $1_{1,0}$ and $1_{1,1}$ vertices. The blue circle has radius $1/\sqrt{d-2}$, and must be enclosed in the convex hull of particle towers, as mandated by the Sharpened DC.}\label{f.2d_dilaton_lattice}
\end{figure}

The above conditions are for individual principal vertices of KK-modes or string oscillators, but when considering multiple principle vertices at the same time, one obtains multiple radion/dilaton lattice conditions that must be simultaneously satisfied. That is, $\alpha$-vectors need to lie at the intersections of hyperplanes of points satisfying the radion/dilaton lattice conditions. Since the number of vertices generating the frame simplex is equal to the dimension of the flat slice of moduli space, that means that the $\alpha$-vectors need to reside on a lattice of dimension equal to the dimension of the moduli space. See Figures \ref{f.2d_radion_lattice} and \ref{f.2d_dilaton_lattice}.

Precise formulas govern these lattices. Consider a duality frame of an $N$-dimensional flat slice of moduli space of a $d$-dimensional theory. Suppose the frame is a geometric frame, where the $i$th vertex of the frame simplex corresponds to a KK mode decompactifying to $D_i$ dimensions.  With each vertex $\vec v_i$, any $(p-1)$-brane must then satisfy the lattice conditions \eqref{e.radion_lattice} with an integer $P_i$. As derived in Section \ref{s.lattice}, this, together with the generators of the frame simplex satisfying the taxonomy rule \eqref{e.taxonomy}, implies that any $\alpha$-vector is governed by the \emph{\textbf{lattice formula}},
\begin{align}
	\boxed{\vec\alpha_p(\vec P)=\sum_i\left(\frac{D_i-d}{D -2}P +p-P_i\right)\vec v_i},
	\label{e.lattice}
\end{align}
where
\begin{align}
	P=\sum_i\left(P_i-p\right)+p,\qquad D=\sum_i\left(D_i-d\right)+d.
\end{align}
For an example of a frame simplex and the resulting lattice, see Figure \ref{f.2d_radion_lattice}.

Suppose that the $i$-th vertex of a frame simplex is a string oscillator vertex, with $\vec v_i=\vec v_\text{osc}$. We can obtain a stringy lattice equation from \eqref{e.lattice} by following the radion-to-dilaton prescription of replacing $P_{i}$ with $\frac {D_i}2(P_i-p)$ and taking the $D_{i}\rightarrow \infty$ limit. This produces the stringy lattice formula,
\begin{align}
	\vec\alpha_p(\vec P)=\sum_{j\neq i}\left(\frac{D_j-d}{2}(P_i-p)+p-P_j\right)\vec v_j+\left(\tilde P-\frac{\tilde D-2}{2}(P_i-p)\right)\vec v_i,\label{e.lattice_stringy}
\end{align}
where
\begin{align}
	\tilde P=p+\sum_{j\neq i}(P_j-p),\qquad \tilde D=d+\sum_{j\neq i}(D_j-d).
\end{align}
For an example of a 2d-lattice with a stringy frame, see Figure \ref{f.2d_dilaton_lattice}.

These lattices satisfy various identities. For example, the dot product rules between various lattice points from a geometric frame satisfy
\begin{align}
	\boxed{\vec \alpha_p(\vec P) \cdot \vec \alpha_{p'}(\vec P')=\frac{p p' }{d-2}-\frac{P P' }{D -2} +\sum_i  \frac {(P_i-p)(P_i'-p)}{D_i-d}}.\label{e.lattice_product}
\end{align}
The $i$-th and $j$-th principal vertices $\vec v_i$ and $\vec v_j$ can be obtained by setting $p=1$ and respectively setting $P_k=1-\delta_{ki}$ or $P_k=1-\delta_{kj}$. Inserting these two vertices into the lattice product equation \eqref{e.lattice_product} returns the taxonomy dot product relation \eqref{e.taxonomy} of \cite{Etheredge:2024tok}, and thus \eqref{e.lattice_product} generalizes the taxonomy rule \eqref{e.taxonomy}. Since the lattice formula \eqref{e.lattice_product} also applies to branes, it generalizes the brane taxonomy rules of \cite{Etheredge:2024amg}.

Many lattice sites need to be empty in order for the ESC to be satisfied, since having $\alpha$-vectors be too long results in particle towers or branes becoming light too fast. Meanwhile, sufficiently many lattice sites need to be populated for the ESC or Sharpened DC (and also the Brane Distance Conjecture (Brane DC) of \cite{Etheredge:2024amg}) to be satisfied under dimensional reduction.

In this paper, I introduce the notion of a \textbf{\emph{principal}} brane, which is a $(p-1)$-brane at a particular lattice site that is required to be present in order for the Sharpened DC and ESC\footnote{I will actually use a slightly stronger version of the ESC, proposed by \cite{Klaewer:2020lfg}, that supposes that, for flat slices of moduli spaces of dimension greater than $1$, emergent string limits are accompanied by KK-modes at the string scale. } to hold upon toroidal reduction on a $(p-1)$-torus. Under this assumption, I argue that a $(p-1)$-brane is principal if it satisfies the following condition:\footnote{As a reminder, I am assuming that there is no sliding throughout this paper, even after compactification or decompactification.}
\begin{Condition}[Principality]
	Given a lattice site $\vec \alpha_p$ for a $(p-1)$-brane in a $d$-dimensional theory, this site is principal if it satisfies the following two conditions:
	\begin{enumerate}
		\item $\vec \alpha_p$ satisfies the \textbf{principal norm condition,}
		\begin{align}
			\boxed{\alpha_p^2=\frac 1n+1-p+\frac{p^2}{d-2}},\label{e.principal_norm}
		\end{align}
		where $n$ is either a positive integer or $\infty$. Branes satisfying \eqref{e.principal_norm} are called \textbf{principal norm candidates}.
		\item For every other principal norm candidate $(q-1)$-brane present in the theory with $q\leq p$, $\vec \alpha_p$ satisfies the \textbf{principal product condition} with $\vec \alpha_q$,
		\begin{align}
			\boxed{\vec \alpha_p\cdot \vec \alpha_q\leq 1-q+\frac{p q}{d-2}}.\label{e.principal_product}
		\end{align}
	\end{enumerate}
\end{Condition}
In particular, if a $(p-1)$-brane is principal, I argue that the lattice product equation \eqref{e.lattice_product} simplifies, and that every other $(q-1)$-brane satisfies
\begin{align}
	\vec \alpha_{p}\cdot \vec \alpha_{q}=\frac{Q-q }{d-D_p-p+1}+\frac{p q}{d-2},
\end{align}
for some integers $Q$ and $D_p$.

Given a frame simplex and resulting lattice, the principal norm condition is a Diophantine equation with only a finite set of lattice points allowed as principal norm candidates. However, the principal product condition does not always require all of these principal norm candidates to be present in the spectrum. Often only subsets of the principal norm candidates mutually satisfy the principal product conditions, and thus different theories are distinguished by different choices. Also, as I point out in this paper, the known examples of sliding in \cite{Etheredge:2023odp} involve $\alpha$-vectors sliding between these different choices.

\begin{figure}
\centering 
\includegraphics{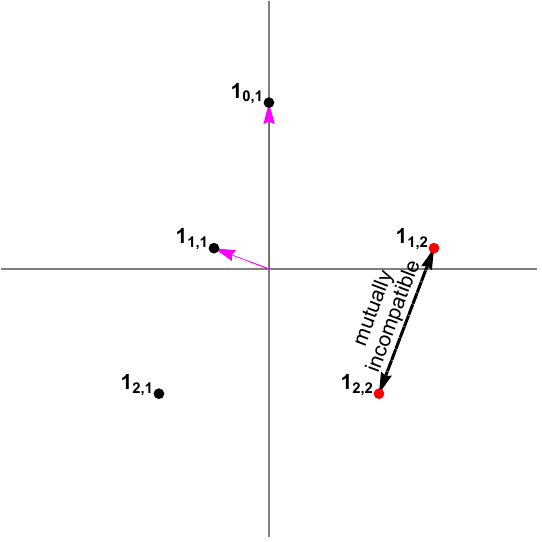}
\caption{Example of principal norm candidates that do not all mutually satisfy the principal product condition. Here $d=9$, and the magenta arrows point at the frame simplex generated by the $1_{1,0}$ KK-mode (corresponding to decompactification to 10d) and the string oscillator mode $1_{1,1}$. This example matches the dilaton-radion profile of any of the 10d string theories compactified on a circle. The points $1_{1,2}$, $1_{2,1}$, and $1_{2,2}$ satisfy the principal norm conditions, but the points $1_{2,1}$ and $1_{2,2}$ do not satisfy the principal product condition with each other. In the landscape, IIA and IIB string theory are obtained by respectively either keeping $1_{2,1}$ and discarding $1_{2,2}$, or keeping $1_{2,2}$ and discarding $1_{2,1}$, and the heterotic theories involve sliding between these choices.}\label{f.dilatonchoice}
\end{figure}

To see an example of different ways the principal product conditions can be satisfied given a set of principal norm candidates, consider the theory in Figure \ref{f.dilatonchoice}. In this example, there is a 2d lattice of a 9d theory where one of the frame simplex vertices (the $1_{1,0}$ vertex) is a KK-mode of decompactification to 10d, and the other frame simplex vertex is a string oscillator vertex (the $1_{1,1}$ vertex). This example describes the dilaton-radion profile of all of the five superstring theories compactified on a circle. In this example, the points $1_{1,2}$, $1_{2,1}$, and $1_{2,2}$ are principal norm candidates, but the points $1_{2,1}$ and $1_{2,2}$ do not satisfy the principal product condition with each other, and thus the principal product condition is obtained by choosing one of two options. In the string landscape, each of these two options are realized, or there is sliding between the two choices:
\begin{enumerate}
	\item There could be a particle tower at $1_{2,1}$ but no particle tower at $1_{2,2}$, just like in IIA string theory on a circle (where the $1_{2,1}$ tower corresponds to D0 branes). As discussed in Section \ref{s.classification}, this uniquely fixes the higher-dimensional principal branes, and they exactly match the branes of IIA string theory on a circle.
	\item There could be a particle tower at $1_{2,2}$ but not at $1_{2,1}$, just like IIB string theory on a circle, where the $1_{2,1}$ particle tower corresponds to wrapped D1-branes. As discussed in Section \ref{s.classification}, this uniquely fixes the higher-dimensional principal branes, and they exactly match the  branes of IIB string theory on a circle.
	\item There could be sliding between the $1_{2,2}$ and $1_{2,2}$ points. This happens in heterotic string theory on a circle with no Wilson lines.
\end{enumerate}

Given these assumptions, in Section \ref{s.classification}, I classify all duality frames, principal branes, and tower/species polytopes for all 0d, 1d, and 2d moduli spaces of theories between 3 and 11 spacetime dimensions, and I argue that 11d is the maximum spacetime dimension where the ESC can be preserved under dimensional reduction. This generalizes the classification of tower/species polytopes in \cite{Etheredge:2024tok}, and also provides a classification of the behavior of higher-dimensional branes in these theories. I also compare this classification with string landscape examples with 32, 16, and 8 supercharges in various spacetime dimensions. If it turns out that sliding occurs between different choices of principal norm candidates, then my classification also classifies theories with this type of sliding.

As an application of my classification, I examine the Dark Dimension Scenario \cite{Montero:2022prj} with an extra noncompact modulus, which I motivate with an argument involving exotic branes.\footnote{I motivate the existence of an extra modulus by requiring the ESC to hold upon compactifying our universe to 3d and having a sufficient class of exotic branes to satisfy the ESC. This requires some assumptions about exotic branes, see Sections \ref{s.lattice.exotic} and \ref{s.dark_dimension}.} My classification of principal branes in 2d moduli spaces, or equivalently the ESC, requires the existence of strings whose mass scale $m_s$ is bounded above by the Dark Dimension KK mass $m_\text{KK}$ by $m_s\lesssim m_\text{KK}^{1/3}$, in 4d Planck units. In the Dark Dimension Scenario, the KK-mode scale is related to the cosmological constant $\Lambda$ by $m_\text{KK}\sim \Lambda^{1/4}$, and thus the strings I find have mass scales bounded above by $m_s\lesssim \Lambda^{1/12}$. This is the GUT scale explored in \cite{Heckman:2024trz}.

An outline of this paper is as follows. In Section \ref{s.lattice}, I use the ESC to argue that $\alpha$-vectors of particle towers and branes reside on lattices, and I derive formulas governing these lattices and generalize the taxonomy rules of \cite{Etheredge:2024amg, Etheredge:2024tok}. In Section \ref{s.principal}, I consider the ESC under dimensional reduction and derive principality conditions that require certain lattice points to be populated. In Section \ref{s.classification}, I classify all duality frames, lattices, principal branes, and tower/species polytopes for 0d, 1d, and 2d moduli spaces of theories with spacetime dimensions between 3 and 11, and I argue that 11d is the maximum spacetime dimension consistent with my assumptions. I compare this classification with examples of theories with 32, 16, and 8 supercharges in the string landscape. In Section \ref{s.dark_dimension}, I argue that, if these assumptions and this analysis applies to the Dark Dimension Scenario with an extra modulus, then there must exist a string with a tension related to the cosmological by $T\lesssim \Lambda^{1/6}$.

\emph{Shortly before the completion of this paper, a similar paper \cite{Grieco:2025bjy} was released. There, it was found that $\alpha$-vectors of particle towers and strings were related to each other through lattice constraints. It would be interesting to explore further whether the observations in \cite{Grieco:2025bjy} are examples of the lattice taxonomy rules proposed in this paper, and whether these lattice taxonomy rules explain the bottom-up observations of 4d $\mathcal N=1$ theories in \cite{Lanza:2021udy}.}

\section{Lattice rules\label{s.lattice}}

Given a flat slice of moduli space where $\alpha$-vectors are constant, I argue in this section that the ESC requires that the $\alpha$-vectors of particle towers and branes are fixed to reside on a lattice.

I begin by analyzing one-dimensional non-compact moduli spaces. By the ESC, in an asymptotic limit of a 1d moduli space, the modulus is either a radion or a dilaton. I first consider the case of a radion. In a decompactification limit where the radion becomes large, non-oscillator particle towers and branes can be interpreted as KK-modes, wrapped/unwrapped branes, KK-monopoles, or exotic branes. I show that this forces the $\alpha$-vectors to have discrete, lattice-valued radion components given by \eqref{e.radion_lattice}. I next consider the case of a dilaton. In a weak-coupling limit of a dilaton, where a fundamental string approaches zero tension, I invoke an argument involving dimensional reduction to show that $\alpha$-vectors of particle towers and branes have discrete dilaton components given by \eqref{e.dilaton_lattice}. Upon circle reduction, the winding modes of strings are T-dual to KK-modes, and thus define radion directions in the lower-dimensional theory. The radion lattice formula \eqref{e.radion_lattice} applies to the branes in the lower dimensional theory, and this can be uplifted to dilaton lattice rules \eqref{e.dilaton_lattice} for branes in the higher-dimensional theory.

After analyzing 1d moduli spaces, I analyze multi-dimensional flat moduli spaces that have several radions and possibly a dilaton. For each radion and dilaton, $\alpha$-vectors of particle towers and branes need to have lattice-valued radion and dilaton components, and this requires the $\alpha$-vectors to reside on higher-dimensional lattices. Given a frame simplex, I derive a formula \eqref{e.lattice} that explicitly expresses the $\alpha$-vectors in terms of the $\alpha$-vectors of the towers that generate the frame simplex. With this, I show that $\alpha$-vectors of particle towers and branes satisfy various identities, including generalizations of the taxonomy rules of particle towers and branes observed in \cite{Etheredge:2024tok, Etheredge:2024amg}.

I conclude this section with a discussion on how exotic branes populate sites within these lattices.

\subsection{Radion lattice conditions\label{s.lattice.rules}}

Consider compactification of a $D$-dimensional theory on a rigid torus to a $d$-dimensional theory, where $\rho$ is the canonically normalized radion controlling the volume of the torus. A $(P-1)$-brane in the $D$-dimensional theory that wraps cycles of the torus to become a $(p-1)$-brane in the $d$-dimensional theory has a tension in $d$-dimensional Planck units given by \cite{Etheredge:2022opl}
\begin{align}
	T_p^{(d)}\sim \exp\left(-\frac{p(D-2)-P(d-2)}{\sqrt{(D-d)(D-2)(d-2)}}\rho \right) T^{(D)}_P. \label{e.tension_reduction}
\end{align}
This tension scales exponentially with the radion, and the prefactor in front of the radion is a function of $D$, $d$, $P$, and $p$. 

It is useful to express this radion dependence using $\alpha$-vectors. In the $d$-dimensional theory, consider the $\alpha$-vector $\vec \alpha_{D,d}$ of the KK-modes related to decompactification back to $D$-dimensions. In the limit that the KK-mode becomes light, wrapped and unwrapped branes have tensions governed by \eqref{e.tension_reduction}, and this means that their $\alpha$-vectors will have projections on the direction of $\vec \alpha_{D,d}$ satisfying 
\begin{align}
	\boxed{\vec \alpha_p(P)\cdot \hat \alpha_{D,d}=\frac{p(D-2)-P(d-2)}{\sqrt{(D-d)(D-2)(d-2)}}}.
\end{align}
This is the \emph{\textbf{radion lattice}} condition \eqref{e.radion_lattice} from the Introduction.

The radion lattice condition \eqref{e.radion_lattice} extends beyond wrapped/unwrapped branes, and applies also to KK-modes, KK-monopoles, and exotic branes. For example, the $\alpha$-vectors of KK-modes satisfy \eqref{e.radion_lattice} with the $P=0$ and $p=1$, as if they are instantons in the $D$-dimensional theory that become particles under dimensional reduction. KK monopoles are codimension-3 branes and have $\alpha$-vectors in the opposite directions of KK-modes \cite{Etheredge:2024amg}, and this is produced by setting $p=d-3$ and $P=D-2$ in the radion lattice formula \eqref{e.radion_lattice}, as if a codimension-2 brane in the $D$-dimensional theory wraps $D-d+1$ cycles to become codimension-3 brane in the $d$-dimensional theory. In subsection \ref{s.lattice.exotic}, I will argue that there also can exist a class of codimension-2 exotic branes with $p=d-2$ but with $P$ can be greater than $D-2$. It is strange that KK-modes, KK-monopoles, and exotic branes have the formulas work out like this, and it would be very interesting to have a deeper or physical explanation for these numerical properties.

The radion lattice formula \eqref{e.radion_lattice} implies that, in the $\hat \alpha_{D,d}$ direction, the $\alpha$-vectors branes have lattice-valued components. More technically, the lattice in \eqref{e.radion_lattice} is a sum of lattices (i.e. a Minkowski sum), where one lattice involves $p$ multiples of $\sqrt{\frac{D-2}{(D-d)(d-2)}}$, and the other lattice involves $-P$ multiples of $\sqrt{\frac{d-2}{(D-d)(D-2)}}$. The first lattice is related to the spacetime dimension $p$ of the brane in the $d$-dimensional theory, and the other lattice is related to the spacetime dimension $P$ of the brane in the $D$-dimensional theory.

Let us first consider the radion lattice for particle towers, where $p=1$. The $P=0$ site has radion component given by\footnote{This is formally the formula for an ``instanton wrapping the circle to produce a tower", but, as mentioned earlier, a physical interpretation of this statement is lacking.}
\begin{align}
	\hat \alpha_{D,d}\cdot \vec \alpha_{p=1}(P=0)=\sqrt{\frac{D-2}{(D-d)(d-2)}}.
\end{align}
Next, the $P=1$ site is obtained from particles in the $D$-dimensional theory that do not wrap any of the cycles of the torus, and their radion components are given by
\begin{align}
	\hat \alpha_{D,d}\cdot \vec \alpha_{p=1}(P=1)=\sqrt{\frac{D-d}{(D-2)(d-2)}}.
\end{align}
Next, the $P=2$ site is obtained from a tower of winding modes from strings wrapping one-cycles of the torus. Higher values of $P$ for particles can be obtained by having $(P-1)$-branes wrap $(P-1)$-cycles. For a fully wrapped brane, $P=D-d+1$ and $p=1$, providing
\begin{align}
	\hat \alpha_{D,d}\cdot \vec \alpha_{p=1}(P=D-d+1)&=-\sqrt{\frac{D-2}{(D-d)(d-2)}}.
\end{align}

While the radion lattice \eqref{e.radion_lattice} describes string tensions, it does not necessarily describe their oscillator masses. In $d$-dimensional Planck units, string oscillator masses scale with the square-roots of string tensions. Thus $\alpha$-vectors for oscillators can be obtained by setting $p=2$ in \eqref{e.radion_lattice} and then dividing by 2, giving
\begin{align}
	\hat \alpha_{D,d}\cdot \vec \alpha_\text{osc}=\frac{D-2-\frac P2(d-2)}{\sqrt{(D-d)(D-2)(d-2)}}.
\end{align}
String oscillators can thus be viewed as residing on the half-integer-multiple superlattice of the radion particle lattice.\footnote{As mentioned in the Introduction, I will not focus on oscillator modes of higher-dimensional branes. However, since these branes are governed by the lattice condition \eqref{e.radion_lattice}, the oscillators of $(p-1)$-branes reside within $1/p$-integer-multiple superlattices of the radion particle lattice, since in Planck units the oscillators of $(p-1)$-branes have masses that scale with the $p$-th roots of the brane tensions.}

\begin{figure}
\centering
\begin{subfigure}{.49\linewidth}
\centering
\includegraphics[scale=.8]{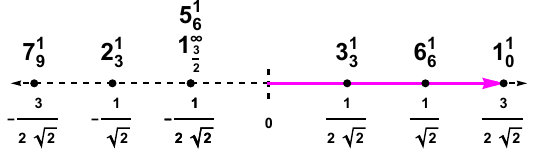}
\caption{$(1)$-frame, generated by $1_{0}^1$. }
\label{f.1d.radion_example.1i}
\end{subfigure}
\begin{subfigure}{.49\linewidth}
\centering
\includegraphics[scale=.8]{figs/1d/10d_IIA.pdf}
\caption{Lattice of $\alpha$-vectors for $\frac 12$-BPS branes in 10d IIA string theory.}\label{f.1d.radion_example.IIA}
\end{subfigure}
\begin{subfigure}{.49\linewidth}
\centering
\includegraphics[scale=.8]{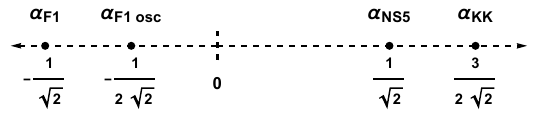}
\caption{Radion and dilaton lattices of known branes of 10d heterotic $\mathrm{E_8\times E_8}$ string theory.}\label{f.1d.radion_example.HE}
\end{subfigure}
\caption{ Radion lattice of $\alpha$-vectors in 10d IIA and heterotic $\mathrm{E_8\times E_8}$ string theory. Figure \ref{f.1d.radion_example.1i} lists the $\frac 12$-BPS $\alpha$-vectors of IIA string theory, labeled by $p_{P}^n$, for 1d moduli spaces of 10d theories with a decompactification limit to 11d. $p$ refers to the spacetime dimension of the brane, $P$ refers to the coordinates of the brane with respect to the frame, and $n$ labels the number of dimensions that decompactify from the KK-modes that come from the winding modes of the brane. The magenta arrows point to the frame simplex vector $1_0^1$. The names of these IIA string theory branes are depicted in Figure \ref{f.1d.radion_example.IIA}. The known branes in $\mathrm{E_8\times E_8}$ heterotic string theory also satisfy these lattice conditions and are depicted in Figure \ref{f.1d.radion_example.HE}. }
\label{f.1d.radion_example}
\end{figure}

For an example of a radion lattice, see Figure \ref{f.1d.radion_example}, which depicts the $\alpha$-vectors of IIA string theory in 10d, where the D0 brane from an M-theory perspective is the KK-mode framing the lattice.

\subsection{Dilaton lattice conditions}

This discussion so far has dealt with KK-modes and their resulting radion lattices, but one can argue that there also exist lattice conditions for dilatons in perturbative string limits. Roughly speaking, such dilaton lattices count the number of string couplings that appear in the tension formula of a brane in string units. For example, given a fundamental string associated with a dilaton, that string has a tension that scales as 1 with the dilaton in string units, D-branes scale with $1/g_s$ in string units, and NS5-branes scale with $1/g_s^2$ in string units, and the conversion from string units to Planck units involves the introduction of $g_s$ factors that depend on the spacetime dimensions of the branes.

Before I present a rigorous argument for dilatonic lattices, I review a heuristic way of guessing their form. As was discussed in \cite{Etheredge:2024tok}, perturbative string limits in some ways behave as decompactification limits of infinitely many dimensions. One can use this to guess the form of the dilaton lattice. Given a $(p-1)$-brane, consider the radion lattice \eqref{e.radion_lattice} in the $D\rightarrow \infty$ limit in the case where $P$ scales linearly with $D$ by $P\sim \frac {\Delta }2D$, for some $\Delta \in \mathbb R$. For now, $\Delta $ could be any real number, but later on I will present rigorous reasoning to fix $\Delta $ to be an integer. In that case, the KK-lattice equation in the $D\rightarrow\infty$ limit with $P/D$ fixed becomes
\begin{align}
	\vec\alpha_p\cdot \hat \alpha_{\infty,d}=\frac{p}{\sqrt{d-2}} -\frac {\Delta \sqrt{d-2}}2,\qquad\text{(na\"ive)}.\label{e.dilaton_lattice_naive}
\end{align}
The above formula is a guess, based on (na\"ively) viewing emergent string limits as decompactification limits of infinitely many dimensions. I will now show that indeed it is the correct formula and that $\Delta \in \mathbb Z$. 
 
Consider an emergent string limit in a $d$-dimensional theory where the light string has an $\alpha$-vector of $\vec\alpha_2^{(d)}$. The direction of $\vec \alpha_2^{(d)}$ defines a dilaton direction in the $d$-dimensional theory. Consider compactifying this theory on a circle. Then this string produces a winding mode, which in a T-dual frame is a KK-mode, and thus branes in the $(d-1)$-dimensional should satisfy the radion lattice conditions \eqref{e.radion_lattice} with this winding mode. This winding mode has an $\alpha$-vector of
\begin{align}
	\vec \alpha_{2\rightarrow 1}^{(d-1)}=\frac{-d+4}{\sqrt{(d-2)(d-3)}} \hat \rho  +\vec\alpha_2^{(d)},
\end{align}
where $\rho$ is the radion controlling the radius of the circle. Consider next a $(p-1)$-brane in the $d$-dimensional theory that wraps the circle and produces a $(p-2)$-brane in the $(d-1)$-dimensional theory. If $\vec \alpha^{(d)}_p$ is the $\alpha$-vector of this brane in the $d$-dimensional theory, then in the $(d-1)$-dimensional theory this wrapped brane has an $\alpha$-vector of
\begin{align}
	\vec \alpha_{p\rightarrow p-1}^{(d-1)}=\frac{p-d+2}{\sqrt{(d-2)(d-3)}}\hat \rho  +\vec \alpha_p^{(d)}.
\end{align}
Since $\vec \alpha_{2\rightarrow 1}^{(d-1)}$ can be viewed as a KK-mode decompactifying one dimension, that means that the radion lattice conditions apply to this vertex. Thus, the radion lattice conditions \eqref{e.radion_lattice} constrict the dot product between the winding mode and $\vec \alpha_{p\rightarrow p-1}^{(d-1)}$ to be
\begin{align}
	\vec \alpha_{2\rightarrow 1}^{(d-1)}\cdot\vec \alpha_{p\rightarrow p-1}^{(d-1)}=\frac{(p-1)(d-2)-(P-2)(d-3)}{\sqrt{(d-2)(d-3)}}\sqrt{\frac{d-2}{d-3}}=\frac{(p-1)(d-2)}{d-3}-P+2,
\end{align}
for some integer $P$. This dot product is also equal to
\begin{align}
	\vec \alpha_{2\rightarrow 1}^{(d-1)}\cdot\vec \alpha_{p\rightarrow p-1}^{(d-1)}=\frac{(d-p-2)(d-4)}{(d-2)(d-3)}+\vec \alpha_{2}^{(d)}\cdot\vec \alpha_{p}^{(d)},
\end{align}
where $\vec \alpha_{2}^{(d)}\cdot\vec \alpha_{p}^{(d)}$ is the dot product between the $\alpha$-vectors of the string and the $(p-1)$-brane in the $d$-dimensional theory. Thus, the dot products between the $\alpha$-vectors of the string and the $(p-1)$-brane in the $d$-dimensional theory satisfy
\begin{align}
	\vec \alpha_2^{(d)}\cdot \vec \alpha_p^{(d)}&=\frac{(p-1)(d-2)}{d-3}-P+2-\frac{(d-p-2)(d-4)}{(d-2)(d-3)}=\frac {2p}{d-2}+p-P.
\end{align}
The $\alpha$-vectors of the string oscillators are given by
\begin{align}
	\vec \alpha_\text{osc}^{(D)}=\frac 12\vec \alpha^{(D)}_2,
\end{align}
since the masses of the oscillators scale with the square root of the tension, $m_\text{osc}\sim \sqrt T_\text{str}$. Thus, the $\alpha$-vectors of emergent string limits satisfy the following lattice conditions,
\begin{align}
	\vec \alpha_\text{osc}\cdot \vec \alpha_p=\frac{p}{d-2}+\frac{p-P}2.
\end{align}
Or, using $|\vec \alpha_\text{osc}|=1/\sqrt{d-2}$,
\begin{align}
	\boxed{\hat  \alpha_\text{osc}\cdot \vec \alpha_p=\frac{p}{\sqrt{d-2}}+\frac {(p-P) \sqrt{d-2}}2 }. 
\end{align}
This is the \emph{\textbf{dilaton lattice}} condition \eqref{e.dilaton_lattice} from the Introduction. This confirms the na\"ive dilaton lattice guess in \eqref{e.dilaton_lattice_naive}, where $\Delta=P-p$. For example, in 10d string theory, fundamental strings have $p=2$ and $P=2$, D$(p-1)$-branes have $P=p+1$, and NS5-branes have $P=p+2=8$.

One can also convert the radion lattice condition \eqref{e.radion_lattice} into the dilaton lattice condition \eqref{e.dilaton_lattice} using what I will call the  \emph{\textbf{radion-to-dilaton prescription}}. This involves replacing $P$ with $\frac 12(P-p)D$ in the usual KK-lattice formula, which produces
\begin{align}
	\hat \alpha_{D,d}\cdot \vec \alpha_{p,\frac 12(P-p)D}=\frac{p(D-2)-\frac 12(P-p)D(d-2)}{\sqrt{(D-d)(D-2)(d-2)}},
\end{align}
and then taking the $D\rightarrow \infty$ limit. This results in the dilaton lattice condition \eqref{e.dilaton_lattice},
\begin{align}
	\lim_{D\rightarrow \infty}\left(\hat \alpha_{D,d}\cdot \vec \alpha_{p,\frac 12(P-p)D}\right)=\frac p{\sqrt{d-2}}+\frac{(p-P)\sqrt{d-2}}{2}=\hat  \alpha_\text{osc}\cdot \vec \alpha_p.
\end{align}
As I will later discuss, this prescription also applies in converting higher-dimensional lattices of geometric frame simplices into lattices of stringy frame simplices.

The dilaton lattice equation \eqref{e.dilaton_lattice} describes the positions of strings, but it does not describe the positions of their oscillators. To find the positions of their oscillators, one can set $p=2$ and divide \eqref{e.dilaton_lattice} by 2, since string oscillator masses scale with the square roots of string tensions. For a string labeled by $p=2$ and some number $P$, one can obtain its oscillator by dividing $p$ and $P$ by 2 in the dilaton lattice formula \eqref{e.dilaton_lattice},
\begin{align}
	\hat  \alpha_\text{osc}\cdot \vec \alpha_\text{osc}'=\frac{1}{\sqrt{d-2}}+\frac {(1-P)\sqrt{d-2}}4 .
\end{align}

\begin{figure}
\centering
\begin{subfigure}{.49\linewidth}
\centering
\includegraphics[scale=.8]{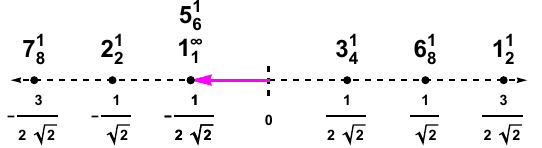}
\caption{$(\infty)$-frame, generated by $1_{1}^\infty$, and a decompactification limit $1_{3}^1$.} \label{f.1d.dilaton_example.i1}
\end{subfigure}
\begin{subfigure}{.49\linewidth}
\centering
\includegraphics[scale=.8]{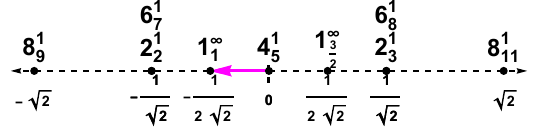}
\caption{$(\infty)$-frame, generated by $1_{1}^\infty$, and another emergent string limit $1_{3/2}^\infty$. }
\label{f.1d.dilaton_example.i}
\end{subfigure}
\begin{subfigure}{.49\linewidth}
\centering
\includegraphics[scale=.8]{figs/1d/10d_IIB.pdf}
\caption{Lattice of $\alpha$-vectors for $\frac 12$-BPS branes in 10d IIB string theory.}\label{f.1d.dilaton_example.IIB}
\end{subfigure}
\begin{subfigure}{.49\linewidth}
\centering
\includegraphics[scale=.8]{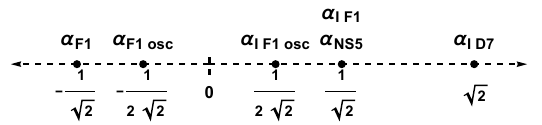}
\caption{Radion and dilaton lattices of known branes of 10d heterotic SO(32) string theory. }\label{f.1d.dilaton_example.HO}
\end{subfigure}
\caption{Dilaton lattice of $\alpha$-vectors in IIA, IIB, and SO(32)/Type I heterotic string theory, labeled by $p_P$. Figures \ref{f.1d.dilaton_example.i1} and \ref{f.1d.dilaton_example.i} respectively list the $\frac 12$-BPS $\alpha$-vectors of IIB and IIA string theory, labeled by $p_{P}^n$, for 1d moduli spaces of 10d theories with a stringy frame simplex $1_1^\infty$. $p$ refers to the spacetime dimension of the brane, $P$ refers to the coordinates of the brane with respect to the frame, and $n$ labels the number of dimensions that decompactify from the KK-modes that come from the winding modes of the brane. The magenta arrows point to the frame simplex vector $1_1^\infty$. Figure \ref{f.1d.dilaton_example.i1} has also a geometric frame, and is consistent with the geometric frame analysis and 10d IIA string theory (and the known branes of $\mathrm{E_8\times E_8}$ heterotic string theory), depicted in Figure \ref{f.1d.radion_example}. Figure \ref{f.1d.dilaton_example.i} has two stringy frames and depicts the brane content of IIB string theory, with explicit names of branes depicted in Figure \ref{f.1d.dilaton_example.IIB}. Figure \ref{f.1d.dilaton_example.i} also governs the known branes in SO(32) heterotic string theory, which are depicted in Figure \ref{f.1d.dilaton_example.HO}.}
\label{f.1d.dilaton_example}
\end{figure}

Both IIA and IIB provide simple examples of dilaton lattices, and these are depicted in Figure \ref{f.1d.dilaton_example}.

\subsection{Higher-dimensional lattices}

For higher-dimensional moduli spaces, the radion and dilaton lattice conditions are enriched. For a duality frame of a flat $N$-dimensional moduli space, the associated $(N-1)$-dimensional frame simplex is generated by $N$ principal tower vectors. Each vertex of the frame simplex is either a KK-mode or string oscillator, and each vertex thus points in the direction of a radion or a dilaton. The theory's particle towers and branes then have $\alpha$-vectors that satisfy the radion or dilaton lattice conditions with each KK or string oscillator vertex, and this requires the $\alpha$-vectors to reside on an $N$-dimensional lattice.

More explicitly, consider an $N$-dimensional moduli space of a $d$-dimensional theory, and suppose that $\vec v_i$ is the $i$-th vertex generating the frame simplex, and suppose that this vertex is a KK-mode related to decompactification to $D_i$ dimensions. Then, all of the $\alpha$-vectors of other particle towers and branes satisfy the lattice conditions with respect to that vertex,
\begin{align}
	\vec \alpha\cdot \hat v_i=\frac{p(D_i-2)-P(d-2)}{\sqrt{(D_i-d)(D_i-2)(d-2)}}.\label{e.vertex_radion_lattice}
\end{align}
For a single vertex, this condition implements the condition that the $\alpha$-vectors reside on a lattice of hyperplanes of fixed values of the $i$-th radion. In order for this lattice condition to be satisfied for all of the vertices $\{\vec v_i\}$ of the frame simplex simultaneously, the condition is that the $\alpha$-vectors lie on intersections of multiple hyperplanes. In general, for an $N$-dimensional moduli space, the lattice conditions for the $N$ vertices generating the frame simplex is that the $\alpha$-vectors of particle towers and branes lie on points where the $N$ hyperplanes intersect. For an example of such a lattice, see Figure \ref{f.2d_radion_lattice}. A similar argument holds if one of the frame simplex vertices is a string oscillator mode, where $D_i\rightarrow \infty$ from the radion-to-dilaton prescription, and similar lattices can be obtained, see Figure \ref{f.2d_dilaton_lattice}. 
I now derive explicit formulas and identities governing these lattices.

Let $\{\vec v_i\}$ be the $N$ vectors that generate a frame simplex. Defining the matrix
\begin{align}
	\mathbf V=\begin{pmatrix}
		\vec v_1\\\vdots\\\vec v_N
	\end{pmatrix},
\end{align}
the lattice conditions for a $(p-1)$-brane's $\alpha$-vector are\footnote{I here consider the case where the frame is geometric. A stringy frame simplex can be derived using the radion-to-dilaton prescription.}
\begin{align}
	(\mathbf V \vec \alpha_p)_i=\vec v_i\cdot \vec \alpha_p=\frac{p(D_i-2)-P_i(d-2)}{(D_i-d)(d-2)}.\label{e.lattice_matrix}
\end{align}
To obtain the solutions $\vec \alpha_p$ to \eqref{e.lattice_matrix}, we need the inverse of $\mathbf V$. To find it, consider the Gram matrix $\mathbf G$ of the vectors $\vec v_i$. Since these vectors $\vec v_i$ are generators of the frame simplex, they satisfy the taxonomy rule \eqref{e.taxonomy} with each other, and thus the Gram matrix is given by
\begin{align}
	\mathbf G=\mathbf V\mathbf V^T,\qquad (\mathbf G)_{ij}=\vec v_i\cdot \vec v_j=\frac 1{d-2}+\frac 1{D_i-d}\delta_{ij}.
\end{align}
It is useful to write the Gram matrix $\mathbf G$ as
\begin{align}
	\mathbf{G}=\mathbf D+\frac {\mathbf {1}}{d-2},\qquad \mathbf D\equiv \diag\left(\frac 1{D_1-d},\dots,\frac 1{D_N-d}\right),\qquad (\mathbf 1)_{ij}\equiv 1,
\end{align}
since then the Sherman-Morrison formula provides the inverse of $\mathbf G$,
\begin{align}
	\mathbf G^{-1}=\mathbf D^{-1}-\frac{\mathbf D^{-1}\mathbf 1 \mathbf D^{-1}}{D-2},\qquad D\equiv d+\sum_i (D_i-d).
\end{align}

Since $\mathbf G$ is symmetric, $(\mathbf G^{-1})^{T}=\mathbf G^{-1}$, and thus
\begin{align}
	\mathbf V^{-1}=\mathbf V^T \mathbf G^{-1}.
\end{align}
Explicitly,
\begin{align}
	(\mathbf V^{-1})_{ij}=\mathbf V^T_{ij}(D_j-d)-\mathbf V^T_{ik}\frac{ (D_k-d)(D_j-d)}{D-2},
\end{align}
where in this formula there is no sum on $j$'s. Applying $\mathbf V^{-1}$ to both sides of \eqref{e.lattice_matrix} provides an explicit formula for the $\alpha$-vectors of $(p-1)$-branes,
\begin{align}
	(\alpha_p)_i=\mathbf V^T_{ij}\left(\frac{D_j-d}{D-2}P+p-P_j\right),\qquad 
	P\equiv p+\sum_i(P_i-p).
\end{align}
This produces the geometric \emph{\textbf{lattice formula}} \eqref{e.lattice} from the Introduction:
\begin{align}
	\boxed{\vec\alpha_{p}(\vec P)=\sum_i\left(\frac{D_i-d}{D -2}P +p-P_i\right)\vec v_i}.
\end{align}
For example, setting $p=1$ and $P_j=1-\delta_{ij}$ recovers the $i$-th principal vertex $\vec v_i$.

This analysis has been for a geometric frame simplex, but a similar formula can be obtained from a stringy frame simplex. Suppose that the $i$-th vertex is a string oscillator vertex, with $\vec v_i=\vec v_\text{osc}$. We can obtain the lattice formula by using the radion-to-dilaton prescription of replacing $P_{i}$ with $\frac {D_i}2(P_i-p)$ and taking the $D_{i}\rightarrow \infty$ limit, which produces
\begin{align}
	\vec\alpha_{p_{\vec P}}=\sum_{j\neq i}\left(\frac{D_j-d}{2}(P_i-p)+p-P_j\right)\vec v_j+\left(\tilde P-\frac{\tilde D-2}{2}(P_i-p)\right)\vec v_i,
\end{align}
where
\begin{align}
	\tilde P=p+\sum_{j\neq i}(P_j-p),\qquad \tilde D=d+\sum_{j\neq i}(D_j-d).
\end{align}

\subsubsection{Pericenter}
The lattice formula \eqref{e.lattice} can be conveniently expressed in terms of the pericenter of the frame simplex.

We first compute the frame simplex's pericenter. The facet generated by the vectors $\vec v_i$ is given by
\begin{align}
	\vec \alpha(x)=\sum_i x_i \vec v_i,\qquad \sum_i x_i=1,\qquad x_i\geq 0.\label{e.facet}
\end{align}
Using the taxonomy rule \eqref{e.taxonomy} and \eqref{e.facet}, the length of $\vec \alpha(x)$ is given by
\begin{align}
	\alpha^2(x)=\frac 1{d-2}+\sum_i\frac{x_i^2}{D_i-d}.
\end{align}
The pericenter is where this is minimized. Let $I,J=1,\dots,N-1$, and $i,j=1,\dots,N$, and let $x_{N}=1-\sum_I x_I$. Then
\begin{align}
	\alpha^2(x)=\frac{1}{d-2}+\sum_I \frac{x_I^2}{D_I-d}+\left(1-\sum_I x_I\right)^2\frac 1{D_N-d}.
\end{align}
This is minimized at the pericenter, or where
\begin{align}
	0=\frac 12\nabla_I \alpha^2(x)= \frac{x_I}{D_I-d}+\frac {\sum_J x_J-1}{D_N-d}.
\end{align}
We can write this as the matrix equation
\begin{align}
	\mathbf M\vec x=\left(\mathbf N+\frac {\mathbf 1}{D_N-d}\right)\vec x=\frac {\vec 1}{D_N-d},\qquad \vec 1_i\equiv 1,\qquad \mathbf N\equiv \diag\left(\frac 1{D_I-d}\right).
\end{align}
To find $x_I$, the inverse of $\mathbf M$ can be found using Sherman-Morrison formula,
\begin{align}
	\mathbf M^{-1}&=\mathbf N^{-1}-\frac{\mathbf N^{-1}\mathbf 1 \mathbf N^{-1}}{D -d}.
\end{align}

So, $\alpha^2(x)$ is minimized at
\begin{align}
	(\vec x)_I=\left (\frac 1{D_{n+1}-d}\mathbf M^{-1}\vec 1\right )_I=\frac{D_I-d}{D -d}. \label{e.minimizexI}
\end{align}
This argument involved treating $x_N$ as a dependent coordinate and finding the minima using the other coordinates, but one could have solved for another coordinate $x_i$ and have $x_N$ governed by an analog of \eqref{e.minimizexI}. Thus, more generally, the pericenter is determined by evaluating \eqref{e.facet} at the coordinates
\begin{align}
	x_i=\frac{D_i-d}{D -d},
\end{align}
resulting in the location of the pericenter $\vec \alpha_\text{pc}$ being at
\begin{align}
	\vec \alpha_\text{pc}=\sum_i \frac{D_i-d}{D -d}\vec v_i.
\end{align}
With this, the lattice formula \eqref{e.lattice} takes the form
\begin{align}
	\vec\alpha_p(\vec P)=P\frac{D -d}{D -2} \vec \alpha_\text{pc}+\sum_i (p-P_i)\vec v_i.\label{e.lattice_pericenter}
\end{align}

For a stringy facet where $\vec v_i$ is a string oscillator, we follow the radion-to-dilaton prescription by replacing $P_i$ with $\frac 12 \Delta D_i$ in \eqref{e.lattice_pericenter} and then taking the $D_{i}\rightarrow \infty$ limit. This produces
\begin{align}
	\vec\alpha_p(\vec P)&=\frac {\tilde D-d}2\Delta \tilde{\vec \alpha}_\text{pc}+\sum_{j\neq i}\left(p-P_j\right)\vec v_j+\left(\tilde P-(\tilde D-2)\frac 12\Delta \right)\vec v_\text{osc},
\end{align}
where $\tilde{\vec{\alpha}}_\text{pc}$ is the pericenter of the non oscillator vertices, $\Delta$ is an integer, and
\begin{align}
	\tilde D\equiv d+\sum_{j\neq i}(D_j-d),\qquad \tilde P\equiv p+\sum_{j\neq i}(P_i-p).
\end{align}

\subsubsection{Lattice form}
One can write the lattice formula \eqref{e.lattice} in a more explicit form of a lattice. An $N$-dimensional lattice can be defined as the set of vectors of the form
\begin{align}
	\vec \alpha=\vec \alpha_0+\sum_i\Delta_i \vec b_i,\label{e.lattice_generic}
\end{align}
where $\vec \Delta\in \mathbb Z^N$.

We can reorganize the $\alpha$-vectors in the lattice formula \eqref{e.lattice} into the following form,
\begin{align}
	\vec\alpha_p(\vec P)=p\frac{D -d}{D -2}\vec \alpha_\text{pc}+\sum_i (P_i-p)\left(\frac{D -d}{D -2}\vec \alpha_\text{pc} -\vec v_i\right).\label{e.lattice_form}
\end{align}
The form of the $ \alpha$-vectors in \eqref{e.lattice_form} matches exactly the form of a lattice in \eqref{e.lattice_generic}, with
\begin{align}
	\Delta_i=P_i-p,\qquad \vec \alpha_0=p\frac{D -d}{D -2}\vec \alpha_\text{pc},\qquad \vec b_i=1_i\frac{D -d}{D -2}\vec \alpha_\text{pc}- \vec v_i.
\end{align}

\subsection{Dot products and identities}
Given the lattice formula \eqref{e.lattice} and the taxonomy rules \eqref{e.taxonomy}, one can compute dot product rules between $\alpha$-vectors on the lattice. Since this lattice describes both branes and all particle towers (including subleading towers), the dot products I derive generalize the tower taxonomy rule \eqref{e.taxonomy} of  \cite{Etheredge:2024tok}, since \eqref{e.taxonomy} applies only to principal towers that generate the frame simplex.

Using the following identities,
\begin{align}
	\vec \alpha_\text{pc}^2&=\frac 1{d-2}+\frac{1}{D -d}=\frac{D -2}{(D -d)(d-2)},\\
	\vec \alpha_\text{pc}\cdot \vec v_i&=\sum_j \frac{D_j-d}{D -d}\left (\frac 1{d-2}+\frac1{D_j-d}\delta_{ij}\right )=\vec \alpha_\text{pc}^2,\\
	\sum_{ij} (p-P_i)  (p-P_j')\vec v_i\cdot \vec v_j&= \frac {(P -p)(P '-p)}{d-2}+\sum_{i}\frac{(P_i-p)(P_i'-p)} {D_i-d},
\end{align}
one can derive the  \emph{\textbf{lattice dot product formula}} \eqref{e.lattice_product} from the Introduction,
\begin{align}
	\boxed{\vec \alpha_p(\vec P) \cdot \vec \alpha_{p'}(\vec P')=\frac{p p' }{d-2}-\frac{P P' }{D -2} +\sum_{i}  \frac {(P_i-p)(P_i'-p)}{D_i-d}}.
\end{align}
This generalizes the taxonomy rules of \cite{Etheredge:2024tok, Etheredge:2024amg}. 

We can recover the tower taxonomy rule \eqref{e.taxonomy} from the lattice dot product formula \eqref{e.lattice_product}. To do this, note the dot product between the principal vectors $\vec v_i$ and $\vec v_j$  of the frame simplex can be obtained by setting $p=p'=1$, and $P_k=\delta_{ik}$ and $P_k'=\delta_{jk}$,  and this produces
\begin{align}
	\vec \alpha_p(\vec P) \cdot \vec \alpha_{p'}(\vec P')=\frac 1{d-2} + \frac {\delta_{ij}}{D_i-d},
\end{align}
which is precisely the taxonomy rule \eqref{e.taxonomy} of \cite{Etheredge:2024tok} for principal towers of the frame simplex!

The dot product formula \eqref{e.lattice_product} produces a length-squared formula of lattice points satisfying
\begin{align}
	 \alpha_p^2(\vec P)&=\frac{p ^2}{d-2}-\frac{P ^2}{D -2} +\sum_i  \frac {(P_i-p)^2}{D_i-d}.
\end{align}
The differences between vectors satisfy a simple formulat,
\begin{align}
	\vec\alpha_p(\vec P)-\vec\alpha_{p'}(\vec P')=\vec\alpha_{p-p'}(\vec P-\vec P').
\end{align}
Additionally, there is a scaling identity,
\begin{align}
	\vec\alpha_{\lambda p}(\lambda \vec P)=\lambda \vec\alpha_p(\vec P).
\end{align}
This means that $\lambda=1/2$ can be used to relate strings and their oscillator modes, where $p=2$ and $1$.

\subsection{Exotic branes \label{s.lattice.exotic}}

There are several kinds of codimension-2 branes one can obtain when compactifying a $D$-dimensional theory on a circle to obtain a $(d=D-1)$-dimensional theory. For example, codimension-2 and codimension-3 branes in the $D$-dimensional theory that respectively wrap or do not wrap the circle produce codimension-2 branes in the $d$-dimensional theory. Exotic branes \cite{Elitzur:1997zn, Blau:1997du, Hull:1997kb, Obers:1997kk, Obers:1998fb, Eyras:1999at, Lozano-Tellechea:2000mfy, Kleinschmidt:2011vu, Bergshoeff:2011se, Kikuchi:2012za, Hellerman:2002ax, deBoer:2012ma, Etheredge:2024amg} are an additional type of codimension-2 brane one can obtain in the $d$-dimensional theory, and these can be subtle from the $D$-dimensional theory's perspective, as they do not come from wrapped or unwrapped branes from the $D$-dimensional theory.

Suppose that the $D$-dimensional theory has a toroidal decompactification limit to a $(D+n)$-dimensional theory. Then, the $d$-dimensional theory can be described by compactification of a $(D+n)$-dimensional theory on a $(D+n-d)$-dimensional torus. For each 1 cycle of this $(D+n-d)$-torus, there can exist KK-monopoles, and they can wrap or not wrap the other cycles of the torus in ways that produce codimension-2 branes in the $d$-dimensional theory that are not described as wrapped/unwrapped branes from the original $D$-dimensional theory. Thus, some of these monopoles can produce codimension-2 branes in the $d$-dimensional theory that are exotic from the perspective of the $D$-dimensional theory. This extends the accessible values of $P$ for codimension-2 branes in the lattice formulas.

Consider a $d$-dimensional theory with two KK vertices $\vec v_1$ and $\vec v_2$ forming a frame simplex, corresponding to toroidal decompactifications to respectively $D_1$ and $D_2$ spacetime dimensions. This means that there exists a decompactification limit to a $(D_1+D_2-d)$-dimensional theory. From the perspective of the $(D_1+D_2-d)$-dimensional theory, one could compactify $D_1$-dimensions on a $(D_2-d)$-torus, obtain a $D_1$-dimensional theory with $(D_1-3)$-dimensional KK-monopoles, and then further compactify on a $(D_1-d)$-torus to the $d$-dimensional theory. The $(D_1-3)$-dimensional KK-monopoles from the $D_1$-dimensional theory can reduce to, in the $d$-dimensional theory, exotic branes with $(P_1,P_2)=(D_1-3,D_1+D_2-d-2)$, and such branes are exotic from the perspective of the $D_2$-dimensional theory. Alternatively, one could instead compactify the $(D_1+D_2-d)$-dimensional theory to a $D_2$-dimensional theory on a $(D_1-d)$-torus, obtain $(D_2-3)$-dimensional KK-monopoles, and then further compactify on a $(D_2-d)$-torus to the $d$-dimensional theory, and obtain $(D_2-3)$-dimensional branes in the $d$-dimensional theory with coordinates $(P_1,P_2)=(D_1+D_2-d-2,D_2-3)$, which are exotic from the perspective of the $D_1$-dimensional theory.

\begin{figure}
\centering
	\includegraphics{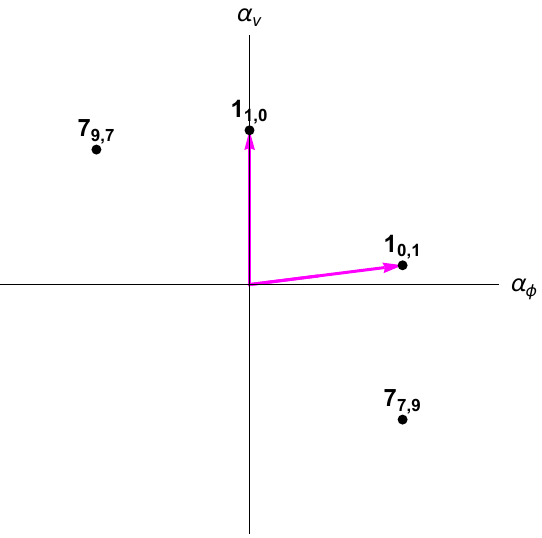}
	\caption{Dilaton-radion components of codimension-2 $\alpha$-vectors of IIA string theory on a circle. The magenta arrows point to the frame simplex generators $1_{1,0}$ and $1_{0,1}$, respectively corresponding to the KK-modes and the D0-brane. From the perspective of 10d IIA string theory, the $7_{9,7}$ brane is the D6-brane, but the $7_{7,9}$ brane is exotic. } \label{f.exotic}
\end{figure}

As an example, consider Type IIA string theory on a circle. With IIA on a circle, the unwrapped D6-brane produces a codimension-2 brane in the 9d theory, but there is also another type of exotic codimension-2 brane. Consider the M-theory perspective. With this perspective, the 9d theory can be viewed as coming from an 11d theory on a 2-torus, which allows for two types of codimension-2 branes in the 9d theory. This is because there are two different orders one could circle reduce to obtain the 9d theory. One could first compactify the $x^{10}$ circle, obtain the D6-brane as the KK-monopole in the 10d theory, and then compactify the $x^9$ circle, where the D6-brane produces a codimension-2 brane in the 9d theory. Alternatively, one could first compactify the $x^9$ circle, obtain a KK-monopole in the resulting 10d theory, and then compactify the $x^{10}$ circle, and obtain a codimension-2 brane that is distinct from the D6-brane. The codimension-2 brane obtained in this second way is exotic from the perspective of IIA string theory \cite{Etheredge:2024amg}, see Figure \ref{f.exotic}.

The type of exotic branes I have here described depend on the existence of decompactification limits in the higher-dimensional theories. Thus, for the existence of this class of exotic branes, multi-dimensional moduli spaces are needed. This will be important in motivating the existence of an extra modulus in the Dark Dimension scenario in Section \ref{s.dark_dimension}. However, if there exist other types of exotic branes not described by the above mechanism, the conclusions involving exotic branes in this paper should be revisited.

\section{Principal branes and the Sharpened Distance Conjecture \label{s.principal}}

Some lattice sites need to be populated in order for the ESC to be preserved under dimensional reduction. That is, there are some lattice sites such that, if there is no brane with an $\alpha$-vector at that site, then the ESC will fail under dimensional reduction. These sites, which I call \textbf{\emph{principal}}, are the focus of this section.

In this section, I identify two conditions that, if both satisfied, imply that a lattice site is principal.\footnote{As mentioned in the Introduction, I am considering only flat slices of moduli space, and I am assuming that there is no sliding, even under dimensional reduction or decompactification.} The first condition is the principal norm condition \eqref{e.principal_norm} (which is a Diophantine equation related to lengths of $\alpha$-vectors), and the second condition is the principal product condition \eqref{e.principal_product} (which is a constraint related to the dot products of $\alpha$-vectors). Later, in Section \ref{s.classification}, I will use these principality conditions to classify all of the allowed lattices, duality frames, tower/species polytopes, and principal branes of 0d, 1d, and 2d moduli spaces in diverse dimensions that are consistent with my assumptions.

These principal lattice sites satisfy various identities. In particular, I will show that, given a principal lattice site $\vec\alpha_\text{princ}$ and any other lattice site $\vec \alpha$, the dot product expression \eqref{e.lattice_product} between these sites simplifies. I will also discuss connections between principal branes and the species polytope, and I will also discuss how principal branes can sometimes require the existence of instantons and possibly motivate a distance conjecture for instantons in future work.

\subsection{Principal towers}
Let us first revisit some properties related to principal towers, and study when a particular tower is necessary for the Sharpened DC or the Emergent String Conjecture.

Consider a $d$-dimensional theory that satisfies the Sharpened DC, and consider a particle tower with an $\alpha$-vector of $\vec \alpha_1$. Does the theory still satisfy the Sharpened DC if this tower is deleted?

If $\vec \alpha_1$ has a length satisfying
	\begin{align}
		\vec \alpha_1^2>\frac 1{d-2}, \label{e.SDCtowergtr}
	\end{align}
and if, for every other $\alpha$-vector $\vec \alpha_1'$ of any other tower,
\begin{align}
		\vec \alpha_1\cdot \vec \alpha_1'\leq \frac{1}{d-2},\label{e.SDCtowerprod}
\end{align}
then deletion of the $\vec \alpha_1$ tower results in the failure of the Sharpened DC. To see this, consider the direction $\hat \alpha_1$ of this tower. Then, the two above conditions imply that every other tower $\vec \alpha_1'$ has a component in the direction $\hat \alpha_1$ satisfying
\begin{align}
	\vec \alpha_1'\cdot \hat \alpha_1< \frac 1{\sqrt{d-2}}.
\end{align}
Thus, deletion of the $\vec \alpha_1$ tower results in failure of the Convex Hull Distance Conjecture of \cite{Calderon-Infante:2020dhm} with the Sharpened DC bound \cite{Etheredge:2022opl}. Thus, $\vec \alpha_1$ must be the $\alpha$-vector of a principal tower!

If one requires the theory to satisfy not just the Sharpened DC, but also the ESC, then one can diagnose whether a tower is a principal tower required by the ESC by having the tower satisfy a particular length condition and satisfying dot product conditions with a subset of the particle towers. First, consider a tower with an $\alpha$-vector satisfying \eqref{e.SDCtowergtr}. If it has a chance of being a principal tower, then by the taxonomy rule \eqref{e.taxonomy} it must have a length of the form
\begin{align}
	\vec \alpha_1^2=\frac 1n+\frac 1{d-2},\label{e.princtowernorm}
\end{align}
for some positive integer $n$. By the above analysis, it is required by the Sharpened DC if, for every other particle tower $\alpha$-vector $\vec \alpha_1'$, equation \eqref{e.SDCtowerprod} is satisfied. However, due to the ESC, one needs to only check equation \eqref{e.SDCtowerprod} with towers that themselves could be principle, i.e., that also satisfy \eqref{e.princtowernorm} (with potentially different integers $n$). This is because, if the tower $\vec \alpha_1$ is deleted and there is any hope of the ESC being satisfied, it must be satisfied by other towers satisfying \eqref{e.princtowernorm}.

Thus, the tower $\vec \alpha_1$ is required by the ESC if its length satisfies \eqref{e.princtowernorm} and, for every other particle tower with $\alpha$-vector $\vec \alpha_1'$ satisfying \eqref{e.princtowernorm}, equation \eqref{e.SDCtowerprod} is satisfied. I will refer to such towers as \emph{\textbf{principal}}.

\subsection{Principal branes}
Given a duality frame and the resulting lattice governing the positions of $\alpha$-vectors, which sites in the lattice must be populated? I now identify conditions on the lattice that can require certain lattice points to be populated by branes.\footnote{Some of the results here might apply to, or at least affect, the situation where there is sliding and may present an avenue towards the classification of sliding. See future work.}

Define a \textbf{\emph{principal brane}} as a brane with spacetime dimension $p$ in a $d$-dimensional theory such that, upon reduction to a $(d-p+1)$-dimensional theory on a $(p-1)$-torus, the deletion of its winding modes from the theory results in failure of the Sharpened DC and ESC. If one has a principal brane, then the resulting winding modes are principal towers. I now identify some sufficient conditions for a brane to be principal that can be checked in the $d$-dimensional theory.

\subsubsection{Principal norm condition}
Consider wrapping a $(p-1)$-brane on a rigid $(p-1)$-torus to produce winding modes. If $\rho$ is the radion controlling the volume of that torus, then the resulting $\alpha$-vector $\vec \alpha_{p\rightarrow 1}^{(d-p+1)}$ in the $(d-p+1)$-dimensional theory is related to the $\alpha$-vector $\vec \alpha_p^{(d)}$ in the $d$-dimensional theory by
\begin{align}
	\vec \alpha_{p\rightarrow1}^{(d-p+1)}&=\vec \alpha_p^{(d)}+\frac{d-2-p(d-p-1)}{\sqrt{(p-1)(d-2)(d-p-1)}} \hat \rho.\label{e.ap1}
\end{align}
If this tower of winding modes is principal in the $(d-p+1)$-dimensional theory, then it must satisfy the tower taxonomy rule \eqref{e.taxonomy} of \cite{Etheredge:2024tok},
\begin{align}
	(\vec \alpha_{p\rightarrow 1}^{(d-p+1)})^2=\frac 1n+\frac 1{d-p-1}.\label{e.ap1norm}
\end{align}
Equations \eqref{e.ap1} and \eqref{e.ap1norm} require that the $(p-1)$-brane in the $d$-dimensional theory needs to satisfy the following \emph{\textbf{principal norm}} condition 
\begin{align}
	\boxed{(\vec \alpha_p^{(d)})^2=\frac 1n+1-p+\frac{p^2}{d-2}.}
\end{align}
This is \eqref{e.principal_norm} from the Introduction, and this is a Diophantine equation. I will call lattice sites satisfying the principal norm condition \eqref{e.principal_norm} \emph{\textbf{principal norm candidates}}. All principal branes must satisfy this condition, and tower polytopes are constructed out of them, and in later sections of this paper, I will classify these branes in various contexts.

\subsubsection{Principal product condition}

In the previous subsubsection, I identified the length that a $(p-1)$-brane's $\alpha$-vector must have in order for it to have the necessary length to be a principal tower upon reduction. But, the brane might not be necessary for the ESC after dimensional reduction, because in the lower-dimensional theory there could be other particle towers that become light and are the principal towers that result in the ESC being satisfied. I now identify a product condition that, if satisfied by a principal norm candidate brane, results in that brane being required in the higher-dimensional theory in order for the ESC to hold in the lower-dimensional theory.

Consider the $\alpha$-vector of a particle tower $\vec \alpha_{p\rightarrow1}^{(d)}$ in a $d$-dimensional theory that comes from a fully wrapped $(p-1)$-brane in a $(D=d+p-1)$-dimensional theory. The $\alpha$-vector of this tower is
\begin{align}
	\vec \alpha_{p\rightarrow 1}^{(d)}=\vec \alpha_p^{(D)}-(d-3)\sqrt{\frac{D-d}{(D-2)(d-2)}}\hat \rho,
\end{align}
where $\rho$ the radion controlling the volume of the torus and $\vec \alpha_p^{(D)}$ is the $\alpha$-vector of the $(p-1)$-brane in the $D$-dimensional theory.

If
\begin{align}
	(\vec \alpha_{p\rightarrow1}^{(d)})^2>   \frac 1{d-2},\label{e.ap1length}
\end{align}
and also, for every other tower vector $\vec \alpha_1'^{(d)}$ of particle towers satisfying \eqref{e.princtowernorm},
\begin{align}
	\vec \alpha_{p\rightarrow1}^{(d)}\cdot \vec \alpha_1'^{(d)}\leq \frac{1}{d-2},\label{e.ap1product}
\end{align}
then $\vec \alpha_{p\rightarrow1}^{(d)}$ is a principal vertex of the tower polytope in the $d$-dimensional theory and its deletion results in failure of the ESC.

If the $(p-1)$-brane is a principal norm candidate, then \eqref{e.ap1length} is satisfied, so let us investigate when \eqref{e.ap1product} is satisfied.

To do this, we must identify the possible origins for the second tower $\vec \alpha_1'^{(d)}$ and then translate the conditions \eqref{e.ap1length} and \eqref{e.ap1product} into statements in the $D$-dimensional theory. The tower $\vec \alpha_1'^{(d)}$ is either KK-like or a string oscillator. If it is KK-like, then it can come from brane winding modes from the $D$-dimensional theory, particle towers from the $D$-dimensional theory, KK-modes, KK-monopole towers if $d=4$, or exotic towers if $d=3$. If it is a string oscillator, then it can come from strings or wrapped branes from the $D$-dimensional theory, KK-monopoles if $d=5$, or exotic strings if $d=4$.

I will examine the case where the other tower is KK-like, but not string oscillator like. This is because---with the exception of 10d theories---a refinement of the ESC in \cite{Klaewer:2020lfg} has been proposed that all infinite distance emergent string limits are accompanied by KK-modes at the string scale. Thus, to check whether the wrapped $(p-1)$-brane is principal, in the $d<10$ case it suffices to only check \eqref{e.ap1product} with KK-like towers $\vec\alpha_1'^{(d)}$. 

Consider first the case where $\vec \alpha_1'^{(d)}$ is a KK-tower. In this case, \eqref{e.ap1product} is automatically satisfied by KK-modes, because the modes have only radion components, and these components are positive, whereas $\vec \alpha_{P\rightarrow1}^{(d)}$ has non-positive radion components.

Consider next KK-monopoles. These have $\alpha$-vectors whose lengths are given by
\begin{align}
	|\alpha_\text{KK mon}|^2=\frac{D_\text{mon}-2}{(D_\text{mon}-d)(d-2)}\leq \frac{d-1}{d-2},
\end{align}
for some integer $D_\text{mon}> d$. But, these monopoles do not have components in the $D$-dimensional theory's moduli space. Thus, KK-monopoles have dot products with the fully wrapped $(p-1)$ brane satisfying
\begin{align}
	\vec \alpha_{p\rightarrow 1}^{(d)}\cdot \vec \alpha_\text{KK mon}\leq \frac{d-3}{d-2}\sqrt{\frac{(D-d)(D_\text{mon}-2)}{(D-2)(D_\text{mon}-d)}}\leq \frac{d-3}{d-2}\sqrt{\frac{(D-d)(d-1)}{(D-2)}}.
\end{align}
This product is maximized when $D=d+1$, providing
\begin{align}
	\vec \alpha_{p\rightarrow 1}^{(d)}\cdot \vec \alpha_\text{KK mon}\leq  \frac{d-3}{d-2}. \label{e.ap1dotKKmon}
\end{align}

Since KK-monopoles have spacetime dimension $d-3$, they can only provide KK-like towers when the spacetime dimension is 4d. In 4d, \eqref{e.ap1dotKKmon} is
\begin{align}
	\vec \alpha_{p\rightarrow 1}^{(4)}\cdot \vec \alpha_\text{KK mon}\leq  \frac12. 
\end{align}
This satisfies, albeit barely and with possible saturation, \eqref{e.ap1product}. Thus, KK-mode oscillators will not affect a higher-dimensional brane's principality.

Let us next investigate exotic branes when $d=3$, which themselves can be the towers. Suppose we have a direction in the $D$-dimensional theory's $\phi$ direction that decompactifies $n_\phi$ dimensions. Then, upon compactification to the $d$-dimensional theory, we can have an exotic brane with a radion lattice component of
\begin{align}
	P_\rho=D+n_\phi-2.
\end{align}
Meanwhile, however, the exotic brane, in the $\phi$-direction, can be at most at the $P_\phi=p$ lattice site. In 3d, this means that it has a $\phi$ component of at most
\begin{align}
	\frac{1(D-2)-1(d-2)}{\sqrt{(D-d)(D-2)(d-2)}}=\sqrt{\frac{D-3}{D-2}}.
\end{align}
Meanwhile, the $\alpha$-vector of the fully wrapped $(p-1)$-brane has only components in the $\phi$ direction, and has a length of at most $\sqrt{\frac{D-1}{D-2}}$. Thus, in 3d,
\begin{align}
	\alpha_{p\rightarrow 1}\cdot \vec \alpha_\text{exotic}\leq \sqrt{\frac{D-3}{D-2}}\sqrt{\frac{D-1}{D-2}}=\frac{\sqrt{(D-3)(D-1)}}{D-2}<1.
\end{align}
Thus, in 3d, exotic branes do not affect the principal product condition with fully wrapped branes.

This leaves only the case where $\alpha_1'$ is of a wrapped brane or particle tower from the $D$-dimensional theory. We can restrict our analysis to the situations where $\vec \alpha_1'$ satisfies the principal norm condition \eqref{e.principal_norm}, since only principal towers can save the ESC in the lower-dimensional theory if the wrapped $(p-1)$-brane is deleted.

Consider the case where $\vec \alpha_1'$ is a wrapped $(q-1)$-brane from the $D$-dimensional theory, where $p\geq q$. The relevant moduli in the lower-dimensional theory are given by
\begin{align}
	(\vec \phi_D,\rho),
\end{align}
where $\vec \phi_D$ are the moduli of the $D$-dimensional theory and $\rho$ controls the volume of the $(p-1)$-torus. With respect to these moduli, the $\alpha$-vectors of the particles from the wrapped $(p-1)$ and $(q-1)$ branes are given by
\begin{align}
\begin{aligned}
	\vec \alpha_{p\rightarrow1}^{(d)}&=\vec \alpha_p^{(D)}+\frac{D-2-p(D-p-1)}{\sqrt{(p-1)(D-2)(D-p-1)}} \hat \rho,\\
	\vec \alpha_{q\rightarrow1}^{(d)}&=\vec \alpha_q^{(D)}+\frac{D-2-q(D-p-1)}{\sqrt{(p-1)(D-2)(D-p-1)}} \hat \rho.
\end{aligned}
\label{e.redap1aq1alphas}
\end{align}
By \eqref{e.redap1aq1alphas}, their dot products satisfy
\begin{align}
	\vec \alpha_{p\rightarrow 1}^{(d)}\cdot \vec \alpha_{p\rightarrow 1}^{(d)}=\vec \alpha_{p}^{(D)}\cdot \vec \alpha_{q}^{(D)}+\frac{1}{D-p-1}+q-1-\frac{pq}{D-2}.
\end{align}
Or,
\begin{align}
	\vec \alpha_{p}^{(D)}\cdot \vec \alpha_{q}^{(D)}=\vec \alpha_{p\rightarrow 1}^{(d)}\cdot \vec \alpha_{q\rightarrow 1}^{(d)}-\frac{1}{D-p-1}+1-q+\frac{pq}{D-2}.
\end{align}

Then the \emph{\textbf{principal product condition}} is that (dropping superscript $D$'s),
\begin{align}
	\boxed{\vec \alpha_{p}\cdot \vec \alpha_{q}\leq 1-q+\frac{pq}{D-2}},
\end{align}
for every other $(q-1)$-brane satisfying the principal norm condition, where $q\leq p$. This is equation \eqref{e.principal_product} from the Introduction.

\subsubsection{Lattice formulas}

One can express the principality conditions \eqref{e.principal_norm} and \eqref{e.principal_product} in terms of the lattice formula \eqref{e.lattice}. Inserting the lattice formula into \eqref{e.lattice} the principality conditions \eqref{e.principal_norm} and \eqref{e.principal_product} produces
\begin{align}
	\frac 1n+1-p=\sum_i  \frac {(P_i-p)^2}{D_i-d}-\frac{P ^2}{D -2},\qquad 
	1-q\geq \sum_i  \frac {(P_i-p)(Q_i-q)}{D_i-d}-\frac{P Q }{D -2} ,
\end{align}
where $q\leq p$.

\subsection{Principal brane lattices}
Since principal branes become principal vertices upon toroidal reduction to lower-dimensions, they provide winding mode towers in the lower dimensions that are T-dual to KK-modes, and thus other branes satisfy the radion lattice conditions with these winding modes. These conditions can be uplifted to dot products of $\alpha$-vectors in the higher-dimensional theory. As a result, dot products of other branes with principal branes take a simpler form than \eqref{e.lattice_product}. 

Consider $(p-1)$ and $(q-1)$-branes in a $D$-dimensional theory with $\alpha$-vectors of $\vec \alpha_p^{(D)}$ and $\vec \alpha_q^{(D)}$. Suppose we compactify this theory to a $d$-dimensional theory on a $(p-1)$-torus and have the $(p-1)$-brane wrap the torus, but have the $(q-1)$-brane not wrap the torus.\footnote{In the case where the unwrapped $(q-1)$-brane becomes too large in spacetime dimension, one could instead consider a similar argument, but with the $(q-1)$-brane wrapping the torus. In either case, the resulting formula below will be unaffected.} Suppose that the wrapped $(p-1)$-brane winding modes are dual to KK-modes, and the brane is principal tower in the $d$-dimensional theory.

The wrapped $(p-1)$-branes and unwrapped $(q-1)$-branes have $\alpha$-vectors in the $d$-dimensional theory of
\begin{align}
	\vec \alpha_{p,1}^{(d)}&=\frac{D-2-p(d-2)}{\sqrt{(D-d)(D-2)(d-2)}} \hat \rho +\vec \alpha_{p}^{(D)},\\
	\vec \alpha_{q,q}^{(d)}&=\frac{q(D-2)-q(d-2)}{\sqrt{(D-d)(D-2)(d-2)}} \hat \rho +\vec \alpha_{q}^{(D)}=q\sqrt{\frac{D-d}{(D-2)(d-2)}} \hat \rho +\vec \alpha_{q}^{(D)},
\end{align}
where $\rho$ is the modulus controlling the volume of the $(p-1)$-torus. Taking dot products,
\begin{align}
	\vec \alpha_{p,1}^{(d)}\cdot \vec \alpha_{q,q}^{(d)}&=\frac{(D-2)-p(d-2)}{(D-2)(d-2)} q+\vec \alpha_{p}^{(D)}\cdot \vec \alpha_{q}^{(D)}.
\end{align}
Meanwhile, since $\vec \alpha_{p,1}^{(d)}$ is, in some duality frame, a KK-mode corresponding to decompactification to $D_p$ dimensions, the radion lattice conditions \eqref{e.radion_lattice} imply that
\begin{align}
	\vec \alpha_{p,1}\cdot \vec \alpha_{q,q}^{(d)}=\frac{q(D_p-2)-Q(d-2)}{(D_p-d)(d-2)}.
\end{align}
Thus, we derive a lattice equation for principal branes:
\begin{align}
	\vec \alpha_{p}^{(D)}\cdot \vec \alpha_{q}^{(D)}=\frac{Q-q }{D-D_p-p+1}+\frac{p q}{D-2}.\label{e.principal_brane_lattice}
\end{align}
For the case where the $(p-1)$-brane wraps to provide a winding mode whose KK-mode interpretation involves decompactification of one dimension, then $D-D_p-p+1=1$, and thus \eqref{e.principal_brane_lattice} becomes
\begin{align}
	\vec \alpha_{p}^{(D)}\cdot \vec \alpha_{q}^{(D)}=Q-q +\frac{p q}{D-2}.
\end{align}
This matches the dot-product taxonomy formula discovered in  \cite{Etheredge:2024amg} and mentioned in the Introduction,
\begin{align}
	\vec\alpha_p\cdot \vec \alpha_q=a_{ij}+\delta_{p,q}+1-q+\frac
	{pq}{D-2},
\end{align}
with $a_{ij}=Q-q$. The reason this matches is because the $\frac 12$ BPS branes of M-theory have winding modes that are all KK-modes corresponding to decompactifications of 1 dimension \cite{Etheredge:2022opl}.

\subsection{Species polytopes and brane-species relationships}

To determine whether a lattice site $\vec \alpha_p$ for a $(p-1)$-brane site is principal, the principal product condition \eqref{e.principal_product} needs to be tested for every principal $(q-1)$-brane with $q\leq p$. For example, it requires testing this condition for $(q-1)$-branes that could be very heavy when the $(p-1)$-brane is light. It would be useful if one could only consider a subset of the principal $(q-1)$-branes to test \eqref{e.principal_product}. This will be the subject of future research, and in this section I discuss various lines of research that may be advantageous in the study of testing the principal product condition.

Given a principal norm candidate $(p-1)$-brane with an $\alpha$-vector $\vec \alpha_p$ such that $\vec\alpha_p/p$ (i.e. the $\alpha$-vector of the oscillator modes of the brane) is on the boundary of the species polytope, it would be interesting to know whether one can determine whether it is principal by testing the principal product conditions \eqref{e.principal_product} with only other branes that have $\alpha_p/p$ vectors also on the boundary of the species polytope.

It turns out that many, but not all, branes have $\vec\alpha_p/p$-vectors on the boundaries\footnote{Note that the $\vec \alpha_p/p$-vectors cannot be outside of the species polytope.} of the species polytopes. See Section \ref{s.classification} for many examples. It would be interesting to understand better why this is so often true.

These questions, if answered, could provide a route to finding a stronger version of the ESC that is preserved under dimensional reduction. It could also result in statements that do not require the study of all duality frames, but rather subsets of these frames.

\subsubsection{Principality and frame simplices}

It would be convenient if, given a principal norm candidate $\alpha$-vector $\alpha_p$ that points in the direction of some frame simplex $\mathcal F$, then the principal product condition \eqref{e.principal_product} must be tested with only principal $(q-1)$-branes that also point at the same frame simplex. If this were true, then one would need to test the principal product condition \eqref{e.principal_product} with only branes that become light in a specified way within the same duality frame, and thus reduce the set of $(q-1)$-branes one needs to consider. More precisely, suppose that $\vec \alpha_p$ points at a frame simplex $\mathcal F$ generated by vertices $\vec v_i$. Then, if we wanted to test \eqref{e.principal_product} with other $\alpha$-vectors $\vec \alpha_q$ that also pointed at $\mathcal F$, we would only need to test such vectors satisfying
\begin{align}
	\frac{Q}{D-2}\geq \frac{Q_i-q}{D_i-d}.
\end{align}
It would be interesting in future work to understand rigorously to what extent this holds.

\subsubsection{Species polytope}
Consider a frame, and suppose that $\vec v_i$ is a vertex of the frame simplex corresponding to decompactification to $D_i$ dimensions. It was shown in \cite{Etheredge:2024tok} that the tower-species pattern \cite{Castellano:2023jjt, Castellano:2023stg} relates this vertex $\vec v_i$ to the pericenter $\vec \lambda_i$ of the species polytope by 
\begin{align}
	\vec \lambda_i=\frac{\vec v_i}{(d-2)v_i^2}=\frac{D_i-d}{D_i-2}\vec v_i.
\end{align}
Suppose the $\vec \alpha_p/p$-vector for a $(p-1)$-brane satisfies 
\begin{align}
	\frac 1p\vec \alpha_p\cdot \vec \lambda_i=\vec \lambda_i^2,
\end{align}
i.e., it is on the species facet corresponding to $\vec v_i$. Then
\begin{align}
	\vec \alpha_p\cdot \hat  v_i=p\frac{D_i-d}{D_i-2}\sqrt{\frac{D_i-2}{(D_i-d)(d-2)}}= \frac{p(D_i-2)-p(d-2)}{\sqrt{(D_i-d)(D_i-2)(d-2)}},
\end{align}
i.e., $P_i=p$, or this corresponds to an unwrapped brane.  Thus, an unwrapped, non-particle brane has an $\vec \alpha_p/p$-vector that lies on the boundary of the species polytope. The existence of branes frequently lying on the boundary of the species polytope was noticed in \cite{Etheredge:2024amg}, but it remained somewhat mysterious. The condition $P_i=p$ provides a specific condition on the lattice that determines when this will happen.\footnote{I thank Jos\'e Calder\'on-Infante for discussions on this point.}

The length of a brane on the boundary of the species polytope, i.e. where $P_i-p=0$ for some $i$, satisfies
\begin{align}
	\alpha_p^2\geq p^2\frac{D_i-d}{(D_i-2)(d-2)}.
\end{align}
Comparing this length with the lower bound in the principal norm length \eqref{e.principal_norm} formula, $1-p+\frac{p^2}{d-2}$, gives
\begin{align}
	\vec \alpha_p^2\geq -1+p-\frac{p^2}{D_i-2}+1-p+\frac{p^2}{d-2}.\label{e.apPpbound}
\end{align}
Thus, a vector $\vec \alpha_p$ with $P_i=p$ automatically has a length long enough for it to potentially be a principal norm candidate if the right-hand side of \eqref{e.apPpbound} greater than $1-p+\frac{p^2}{d-2}$. That is, $\vec \alpha_p$ is long enough to be a principal norm candidate if
\begin{align}
	-1+p-\frac{p^2}{D_i-2}\geq 0,
\end{align}
and, if this is the case, then the $n$ in \eqref{e.principal_norm} satisfies $1/n\geq -1+p-\frac{p^2}{D_i-2}$. This happens when $D_i\geq 6$ and $p\leq d-3$.  Thus, when $D_i\geq 6$, a $(p-1\leq d-4)$-brane whose $\vec \alpha_p/p$-vector is on the boundary of the species polytope automatically is long-enough to satisfy the principal norm condition.

Let us next investigate the principal product condition. Suppose that $\vec \alpha_p/p$ satisfies $P_i=p$, and is thus on the boundary of the species polytope. First, consider the principal product condition with $\vec v_i$. This is
\begin{align}
	\vec \alpha_p\cdot \vec v_i=p\sqrt{\frac{D_i-d}{(D_i-2)(d-2)}} \sqrt{\frac{D_i-2}{(D_i-d)(d-2)}}=\frac{p}{d-2}.
\end{align}
Thus, this brane satisfies the principal brane condition with $\vec v_i$.

Consider next an $\alpha$-vector of some other $(q-1)$-brane satisfying $Q_j\geq q$. Then
\begin{align}
	\vec \alpha_p\cdot \vec \alpha_q&=\frac{pq}{d-2}-\frac{PQ}{D-2}+\sum_{j\neq i} \frac{(P_j-p)(Q_j-q)}{D_j-d}
\end{align} 
Now, for fixed $p$, $q$, $P_j$, and $Q_{j\neq i}$, this dot product is maximized when $Q$ is as small as possible, which occurs when $Q_i=q$. That means that if, for a $(p-1)$-brane with $P_i=p$, one is interested in testing whether this brane satisfies the principal product condition with a set of $(q-1)$-branes with fixed $Q_{j\neq i}$, one can determine that this set of branes does not fail the product condition if the $Q_i=q$ site does not fail the product condition. These results can likely be strengthened, and thus become more useful, as will be investigated in future work.

\subsection{Instantons \label{s.principal.instantons}}

The analysis of this paper has focused on particle towers and branes with spacetime dimensions $\geq 1$, but these results have consequences for instantons. Many of the formulas in this paper for $(p-1)$-branes also apply to instantons. For example, the lattice equations \eqref{e.lattice_form} apply to instantons, where $p=0$. 

In fact, some of this analysis mandates the existence of instantons. For instance, given a principal particle tower in a $d$-dimensional theory, compactification on a circle produces an instanton tower that comes from the particle towers. Thus, principality conditions can require the existence of instantons in lower dimensional theories. Also, sometimes in lower dimensional theories, one can determine that there must exist principal branes in higher dimensional theories, and the existence of these branes in higher-dimensional theories can often require the existence of instanton towers in the lower dimensional theories. This has potential relevance for the Axion WGC (see \cite{Etheredge:Bitowers}).

As an example, suppose one has a $d$-dimensional theory and a principal vector $\vec v_i$ corresponding to decompactification of 1 dimension, and suppose that there are particle towers with $P_i=1$. That is, suppose that there is a tower of particles in the $d$-dimensional theory that comes from a particle tower in the $(D_i=d+1)$-dimensional theory. Then this tower in the $D_i$-dimensional theory will produce a tower of instantons in the $d$-dimensional theory that come from the tower having worldlines along the $\vec v_i$ circle.

As discussed in \cite{Etheredge:2024amg}, it would be interesting to leverage this to produce a distance conjecture for instantons. It is possible that principality conditions, and other related results, may hold from a simple statement about instantons in lower dimensional theories.

It would be interesting to have a notion of principal instantons. For instance, the principal norm condition extrapolated to instantons takes the relatively simple, spacetime-dimension-independent form
\begin{align}
	\alpha^2=\frac 1n+1.
\end{align}
Additionally, the principal product conditions take the form
\begin{align}
	\vec \alpha_0\cdot \vec \alpha_{0}'\leq 1.
\end{align}
Just as the ESC and Sharpened DC holding in lower dimensions imply statements about particle towers and branes in higher-dimensions, statements about instantons in lower dimensions could maybe result in statements about particle towers and branes in higher-dimensional theories. Additionally, perhaps these instanton-statements could be made in even lower dimensions, such as 2d. For instance, many formulas for particle towers and branes diverge when $d=2$, but many formulas, such as the Brane DC bound in \cite{Etheredge:2024amg} when applied to instantons, do not diverge when $d=2$.

\section{Classification of 0d, 1d, and 2d moduli spaces in diverse dimensions \label{s.classification}}

The lattice formulas, together with the principality conditions, heavily constrain theories. In this section, I show that theories satisfying the assumptions of this paper\footnote{As a reminder, except when stated otherwise I am restricting my analysis to flat slices of moduli spaces, assuming the ESC holds, and assuming that there is no sliding (even after compactification or decompactification.)} form a finite set and are classifiable. I argue that 11d is the maximum spacetime dimension consistent with these assumptions, and I then classify all of the possible duality frames, lattices, principal branes, and tower/species polytopes of 0d, 1d, and 2d moduli spaces of theories with 3 to 11 spacetime dimensions. Remarkably, this classification reproduces the nontrivial particle and brane content of multiple different examples within the string landscape!

I begin by classifying the principal branes of 0d moduli spaces in diverse dimensions. This analysis requires that 11 is the maximum spacetime dimension a theory can have, as the ESC fails under circular reductions of higher-dimensional theories. My classification of 0d moduli spaces requires the existence of principal branes for 11d, 10d, 8d, 7d, and 6d theories, but not other spacetime dimensions. In particular, this classification requires 11-dimensional theories with 0d moduli spaces to have 2-branes and 5-branes, exactly matching M-theory. I also reproduce the brane content of IIB string theory after quotienting out by the dilaton.

I next classify the duality frames, principal branes, and tower polytopes in theories with 1d moduli spaces and 3 to 10 spacetime dimensions. To do this, I begin by listing all possible duality frames in theories with 3 to 10 spacetime dimensions (which are either emergent string limits or decompactification limits). For each duality frame, I then list all possible lattice sites satisfying the principal norm condition \eqref{e.principal_norm}. There can be choices in how the principal product condition \eqref{e.principal_product} are satisfied, and this results in choices for the allowable theories. I compute and list all of these consistent choices, and I also compute all of the principal branes in these theories. I compare parts of this classification with string landscape examples with 32, 16, and 8 supercharges. Heterotic string theory violates some of the assumptions behind this classification (such as no sliding), but I show that the known branes of heterotic string theory reside on my predicted lattices. If this classification holds with relaxed assumptions, then perhaps this analysis could predict the existence of branes with D-brane like tensions in heterotic string theory.  I also compare my classification with some examples of 5d $\mathcal N=1$ theories, and show that the branes in those examples match my classification. This also suggests that the assumptions behind my classification might be able to be relaxed, and my results might apply more generally to theories with 8 supercharges.

I next classify the duality frames, lattices, principal branes, and tower/species polytopes of theories with 2d moduli spaces and 3 to 9 spacetime dimensions. I begin by listing all of the duality frames. For each frame, I then list all possible lattice sites satisfying the principal norm condition. Not all principal norm candidates mutually satisfy the principal product condition, and thus different theories are labeled by choices that are made in satisfying them. For instance, for a stringy frame in 9d, I show that IIB string theory and its branes are reproduced by one choice, and that IIA and its branes are reproduced by another choice. I also discuss how the heterotic string theories, as well as Type I string theory, slide between choices, at least in a class of 9d examples.

In this section, when I list principal towers and branes, it is redundant to explicitly list both the string oscillator towers when there are principal strings. Thus, sometimes I will not list the oscillator towers, but have them implied by the explicitly listed principal strings. Additionally, in this classification, I have found some branes with $p>1$ and $n=\infty$ (i.e., their winding modes have $\alpha$-vectors with lengths of string oscillator modes). It is possible that the branes with this property are not needed for the ESC under dimensional reduction, but it is also possible that some not-yet-known principal requires them, as many of these types of branes appear in maximal supergravity. Since it is interesting that these types of branes are solutions to the principality equations, I have included these solutions in the analysis, tables, and figures below (except when stated otherwise).

\subsection{Maximal spacetime dimension is 11}

I begin by arguing that 11 is the maximum spacetime dimension consistent with the ESC and assuming that there is no sliding, either upon compactification or upon decompactification.

Each infinite distance limit is either an emergent string limit or a decompactification limit. In decompactification limits, the dimension of moduli space strictly decreases and the number of spacetime dimensions strictly increases. In each theory, we can find either emergent string limits, or decompactification limits to higher dimensions, and repeat this process indefinitely. Because the dimensions of non-compact moduli spaces decrease with each decompactification, eventually enough decompactifications will result in a theory with either only emergent string limits (like IIB string theory), or a theory with no moduli (like M-theory).

I now argue that theories with emergent string limits cannot occur in theories with more than 10 spacetime dimensions, and theories with no non-compact moduli cannot have more than 11 spacetime dimensions. Since all theories satisfying our assumptions decompactify to these types, theories cannot have more than 11 spacetime dimensions.

Consider first a stringy frame simplex with an oscillator vertex $\vec \alpha_\text{osc}$ of a $d$-dimensional theory. The Sharpened DC should hold in the direction opposite of $\vec \alpha_\text{osc}$. In order for this to work, this requires the existence of either a KK-tower or string oscillator tower with an $\alpha$-vector $\vec \alpha_\text{osc}'=\vec \alpha_1'$ satisfying
\begin{align}
		-\sqrt{\frac{d-1}{d-2}}\leq \vec \alpha_1'\cdot \hat \alpha_\text{osc}&\leq -\frac 1{\sqrt{d-2}},&\qquad \text{if $\vec \alpha_1'$ is a KK tower,}\label{e.KK_condition}\\
	\vec \alpha_1'\cdot \hat \alpha_\text{osc}&=-\frac1{\sqrt{d-2}},&\qquad \text{if $\vec \alpha_1'$ is a string oscillator tower.}\label{e.osc_saturation}
\end{align}
The reason why \eqref{e.osc_saturation} is a strict equality is because $\vec \alpha_\text{osc}'$ must have length exactly $1/\sqrt{d-2}$.

I now argue that equations \eqref{e.KK_condition} and \eqref{e.osc_saturation} are forbidden if $d\geq 11$. If $\vec \alpha_1'$ is from a KK-like tower, then the dilaton lattice condition \eqref{e.dilaton_lattice} implies
\begin{align}
	\vec \alpha_1'\cdot \hat \alpha_\text{osc}=\frac{1}{\sqrt{d-2}}-\frac{\sqrt{d-2}}{2}(P_\text{KK}-1),
\end{align}
where $P_\text{KK}$ is an integer. If $\vec \alpha_1'$ is from an oscillator tower $\vec \alpha_1'=\vec \alpha_\text{osc}'$, then the dilaton lattice condition \eqref{e.dilaton_lattice} implies
\begin{align}
	\vec \alpha_\text{osc}'\cdot \hat \alpha_\text{osc}=\frac{1}{\sqrt{d-2}}-\frac{\sqrt{d-2}}{4}(P_\text{osc}-2).
\end{align}
Now, if the spacetime dimension $d$ is greater than 10, then $\vec \alpha_1'$ satisfies, for both the case where $\vec \alpha_1'$ is a KK-like tower or $\vec \alpha_1'=\vec \alpha_\text{osc}'$ for an oscillator tower, the following properties:
\begin{align}
	\vec \alpha_1'\cdot \hat \alpha_\text{osc}\qquad &  \begin{matrix}
	<-\sqrt{\frac {d-1}{d-2}},&\qquad \text{if }P_\text{KK}-1>1,\\
	>-\frac 1{\sqrt{d-2}},&\qquad \text{if }P_\text{KK}-1\leq 1, 
	\end{matrix}\label{e.KK_ineq}
		\\
	\vec \alpha_\text{osc}'\cdot \hat \alpha_\text{osc}\qquad &\begin{matrix}
		=-\frac 1{\sqrt{d-2}},&\qquad \text{if }P_\text{osc}-2=\frac{8}{d-2},\\
		\neq -\frac 1{\sqrt{d-2}},&\qquad \text{if }P_\text{osc}-2\neq \frac{8}{d-2}.
	\end{matrix}\label{e.osc_ineq}
\end{align}
For $d=11$, \eqref{e.KK_condition} cannot be satisfied by any $P$, due to \eqref{e.KK_ineq}. Next, note that in \eqref{e.osc_ineq}, $P_\text{osc}$ is required to be an integer for equality, but this requires $\frac 8{d-2}$, which is only possible for $d=3$, 4, 6, and 10. Thus, it is not possible for the Sharpened DC to be satisfied with the required bounds \eqref{e.KK_condition} and \eqref{e.osc_saturation} for $d>10$. Thus, for theories with emergent string limits, the dimensions of spacetime must be lower than 11.

Consider next a $D$-dimensional theory with no moduli. Upon reduction on a circle to a $(d=D-1)$-dimensional theory, this theory will produce one non-compact modulus, the radion $\rho$. Assuming $D>5$, in the small-circle limit, the ESC mandates that the tower of light particles is either a tower of winding modes or string oscillators (from wrapped 2-branes). In the case where the string oscillators are the leading tower, from the above argument that emergent string limits happen in theories with at most 10 spacetime dimensions, $D\leq 10$. Suppose instead that the leading tower consists of string winding modes, and thus in a decompactification limit, the string winding modes have $\alpha$-vectors of
\begin{align}
	\vec \alpha_{2\rightarrow 1}=-\frac{d-3}{\sqrt{(d-1)(d-2)}}\hat \rho.
\end{align}
Having this tower be a decompactification limit to a $D'$-dimensional theory, we must have that
\begin{align}
	\frac{d-3}{\sqrt{(d-1)(d-2)}}=\sqrt{\frac{D'-2}{(D'-d)(d-2)}},
\end{align}
for some $D'>d$. Solving this provides
\begin{align}
	D'=\frac{d-1}{d-5}+d.
\end{align}
$D'$ is only an integer for $d=9$, 7, and 6, where respectively $D=11$, 7, and 6. Thus, we can restrict our analysis to theories with at most 11 spacetime dimensions.

Note that this analysis implies that the ESC does not hold in bosonic string theory---putting 26-dimensional bosonic string theory on a circle and taking the small radius limit results in ESC failure!

\subsection{0d moduli spaces \label{s.classification.2d}}

For a 0d moduli space, the principal norm condition \eqref{e.principal_norm} is just
\begin{align}
	\alpha_p^2=\frac 1n+1-p+\frac{p}{d-2}=0.
\end{align}
Thus, any two principal norm candidates satisfy
\begin{align}
	\vec \alpha_p\cdot \vec \alpha_q=0.
\end{align}
However, whether they satisfy the product condition \eqref{e.principal_product} is not automatic. This is because, for some values of $p$ and $q$, it can be the case that $1-q+\frac {pq}{d-2}$ is negative.

We begin with $d=11$ and consider the principal norm condition \eqref{e.principal_norm}, which is
\begin{align}
	\alpha_p^2=\frac 1n+1-p+\frac{p}{9}=0.
\end{align}
This is only satisfied for 2-branes and 5-branes, where $p=3$ and $6$, and in these cases $n=1$. Since there are no principal norm candidates for $p$ less than 3, the 2-branes are principal. Also, the 5-branes satisfy the principal product condition \eqref{e.principal_product} with the 2-branes, and thus they are principal. This matches M-theory!

For 10d, the principal norm condition \eqref{e.principal_norm} is that
\begin{align}
	(\vec \alpha_P^{(D)})^2=1+\frac 1n-p+\frac{p^2}{8}=0,
\end{align}
The only branes that can satisfy this principal norm condition are 1-branes, 3-branes, and 5-branes, where respectively $n=2$, $1$, and $2$. The 3-branes satisfy the product conditions with the 1-branes, and the 5-branes satisfy the product conditions with the 1-branes and 3-branes. Thus, 1-branes, 3-branes, and 5-branes are principal. This reproduces exactly the branes of IIB string theory (where the axiodilaton is quotiented out).

One can continue this for lower dimensions. The complete classification of principal branes are listed in Table \ref{t.0dpn}.

\begin{table}[H]
$$\begin{array}{c|c|c|c|c|c|c|c|c|c|c|}
d&p^n \\ \hline
11&3^1,6^1\\\hline
10&2^2,4^1,6^2\\\hline
9&\\\hline
8&2^3,3^2,4^3\\\hline
7&2^5,3^5\\\hline
6&2^\infty \\\hline
5&\\\hline
4&\\\hline
3&\\\hline
\end{array}$$ 
\caption{Principal branes of 0d moduli spaces in diverse dimensions. For each principal brane, $p$ refers to the spacetime dimension of the brane, and $n$ labels the number of dimensions that decompactify from the KK-modes that come from the winding modes of the brane.} \label{t.0dpn}
\end{table}

As a consequence of Table \ref{t.0dpn}, there are no principal branes of 0d moduli spaces in spacetime dimensions 9, 5, 4, and 3.

\subsection{1d moduli spaces \label{s.classification.1d}}

I now classify 1d moduli spaces in diverse dimensions. I begin by listing all duality frames. From each frame, I then calculate all principal norm candidates. Not all principal norm candidates mutually satisfy the principal product conditions, and I show that choices must be made, and I list all choices. This reproduces various examples from the landscape. For instance, in 10d, the resulting choices reproduce exactly the brane content of IIA and IIB string theory. I show that the known branes in 10d heterotic and Type I string theory reside on these principal lattice points. I also show that 5d some $\mathcal N=1$ sugra examples are reproduced by this classification. These examples with less supersymmetry suggest that perhaps the assumptions entering this classification could be relaxed. If so, it would be interesting if non-BPS branes in heterotic string theory could be predicted.

In 1d moduli spaces, each limit is either a decompactification to a $D\leq 11$-dimensional theory, or it is an emergent string limit, and thus each frame simplex is either a KK-mode of decompactification to a $D\leq 11$ dimensional theory, or it is a string oscillator. The relevant duality frames are depicted in Table 1.

\begin{table}[H]
$$\begin{array}{c|c|c|c|c|c|c|c|c|c|c|}
d&(D-d) \\ \hline
10&(1),(\infty)\\\hline
9&(1),(2),(\infty)\\\hline
8&(1),(2),(3),(\infty) \\\hline
7&(1),(2),(3),(4),(\infty) \\\hline
6&(1),(2),(3),(4),(5),(\infty) \\\hline
5&(1),(2),(3),(4),(5),(6),(\infty) \\\hline
4&(1),(2),(3),(4),(5),(6),(7),(\infty) \\\hline
3&(1),(2),(3),(4),(5),(6),(7),(8),(\infty) \\\hline
\end{array}$$ 
\caption{Duality frames of 1d moduli spaces. These are labeled by number of dimensions decompactified, which is either  a positive integer below $11-d$, or $\infty$ in the case of a stringy frame.} \label{t.1d_frames}
\end{table}

I now derive all consistent choices of principal branes, tower polytopes, and lattices, in all dimensions. By the tower-species pattern \cite{Castellano:2023jjt, Castellano:2023stg}, species polytopes are implied by the tower polytopes.

\subsubsection{10d}

The solutions to the  principal norm condition \eqref{e.principal_norm} are listed in Table \ref{t.1d_5d_norm} for all of the allowed frames of 1d moduli spaces of 10d theories.

\begin{table}[H]
$$\begin{array}{c|c|c|c|c|c|c|c|c|c|c|}
(D-d)&p_P^n \\ \hline
(1)&1_0^1,2_3^1,3_3^1,5_6^1,6_6^1,7_9^1\\\hline
(\infty)&1_1^{\infty },1_2^1,2_2^1,2_3^1,3_4^1,4_5^1,5_6^1,6_7^1,6_8^1,7_8^1,7_9^{\infty },8_9^1,8_{11}^1\\\hline
\end{array}$$ 
\caption{Principal norm candidates $p^n_P$, with respect to frames $(D-d)$, for 1d moduli spaces of 10d theories. $p$ refers to the spacetime dimension of the brane, $P$ refers to the coordinates of the brane with respect to the frame, and $n$ labels the number of dimensions that decompactify from the KK-modes that come from the winding modes of the brane. } \label{t.1d_10d_norm}
\end{table}

Consider first the $(1)$-frame. The frame simplex is generated by a KK-mode decompactifying 1-dimension to a 11d theory,
\begin{align}
	\vec v_1\equiv \frac 3{2\sqrt 2}\hat \rho,\label{e.10d_frame_KK}
\end{align}
where $\rho$ is the radion.

Consider first the principal norm candidate $1_0^1$ in Table \ref{t.1d_10d_norm}. Inserting this into the lattice formula \eqref{e.lattice} with frame simplex \eqref{e.10d_frame_KK}, the resulting $\alpha$-vector is
\begin{align}
	\vec \alpha(1_0^1)=\frac 3{2\sqrt 2}\hat \rho.
\end{align}
This is just the original KK-mode tower. There are no other principal norm candidates of particle towers that could violate the formula \eqref{e.principal_product}, so this is a necessary tower for the ESC to hold. Similarly, the oscillator modes of $2_3^1$, which are $1_{3/2}^\infty$, are principal towers.

Next consider the principal norm candidate $2_3^1$ in Table \ref{t.1d_10d_norm}, corresponding to a wrapped 2-brane from the 11d theory. By \eqref{e.lattice}, this has an $\alpha$-vector of
\begin{align}
	\vec \alpha(2_3^1)=-\frac 1{\sqrt 2}\rho.
\end{align}
The products of $\vec \alpha(2_3^1)$ with $\vec \alpha(1_0^1)$ and with $\vec \alpha(1_{3/2}^\infty)$ satisfy the principal product condition \eqref{e.principal_product}, and thus, since there are no other strings in the 10d theory, the $2_3$ string is principal and thus required. So far, this matches exactly the particle tower content of both 10d IIA string theory and heterotic $\mathrm {E_8\times E_8}$ string theory! That is, having one decompactification dimension forces the particle-tower and string content to reproduce exactly these two string theories. Note also that this analysis eliminates the case considered in 10d in \cite{Etheredge:2024tok} where both directions of moduli space are decompactification limits. It was not clear in \cite{Etheredge:2024tok} how to eliminate that case, but this analysis eliminates that case.

Consider next the $3_3^1$ brane, which is an unwrapped 2-brane. This has
\begin{align}
	\vec \alpha(3_3^1)=\frac{1}{2\sqrt 2}\hat \rho,
\end{align}
and this brane satisfies all of the principal product conditions with the other principal branes of spacetime dimension below 3. Thus this brane is principal and required. This brane is just like the D2 brane of IIA string theory.

The principal norm candidates $5_6^1$, $6_6^1$, and $7_9^1$, also satisfy the principal product conditions with branes of lower spacetime dimension, and thus all of these are principal (and thus required to be present by the ESC).

\begin{figure}
\centering
\begin{subfigure}{.49\linewidth}
\centering
\includegraphics[scale=.8]{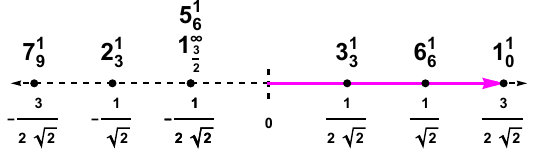}
\caption{$(1)$-frame, generated by $1_{0}^1$. }
\label{f.10d_1dclass_1}
\end{subfigure}
\begin{subfigure}{.49\linewidth}
\centering
\includegraphics[scale=.8]{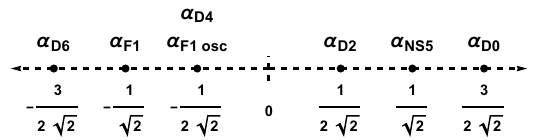}
\caption{Lattice of $\alpha$-vectors for $\frac 12$-BPS branes in 10d IIA string theory.}\label{f.10d_IIA}
\end{subfigure}
\begin{subfigure}{.49\linewidth}
\centering
\includegraphics[scale=.8]{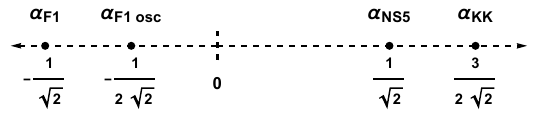}
\caption{Radion and dilaton lattices of known branes of 10d heterotic $\mathrm{E_8\times E_8}$ string theory.}\label{f.10d_HE}
\end{subfigure}
\caption{ Figure \ref{f.10d_1dclass_1} lists the principal $\alpha$-vectors, labeled by $p_{P}^n$, for 1d moduli spaces of 10d theories with a 1-dimension decompactification limit. $p$ refers to the spacetime dimension of the brane, $P$ refers to the coordinates of the brane with respect to the frame, and $n$ labels the number of dimensions that decompactify from the KK-modes that come from the winding modes of the brane. The magenta arrows point to the frame simplex vectors. This reproduces exactly the brane content of 10d IIA string theory, depicted in Figure \ref{f.10d_IIA}. The known branes in $\mathrm{E_8\times E_8}$ heterotic string theory satisfy the lattice conditions as well, and are depicted in Figure  \ref{f.10d_HE}. }
\label{f.10d_1}
\end{figure}

Remarkably, this analysis reproduces exactly the $\frac 12$-BPS brane structure of IIA string theory: the $1_0^1$, $2_3^1$, $3_3^1$, $5_6^1$, $6_6^1$, and $7_9^1$ branes have exactly the same $\alpha$-vectors of, respectively, D0-branes, fundamental strings, D2-branes, D4-branes, NS5-branes, and D6-branes---see Figure \ref{f.10d_1}!

Currently, heterotic $\mathrm{E_8\times E_8}$ string theory is only known to possess the $1_0^1$, $2_3^1$, and $6_6^1$ branes (i.e. the KK-mode in the M-theory limit, the fundamental string, and the NS5-brane), and they satisfy the lattice conditions as well, see Figure \ref{f.10d_HE}. If this analysis fully applies to $\mathrm{E_8\times E_8}$ heterotic string theory, then heterotic string theory requires a full set of D-even branes, just like IIA string theory. However, due to the presence of sliding \cite{Etheredge:2023odp} in heterotic reductions on tori, it is possible that this analysis does not entirely apply to the heterotic case.

Consider next the $(\infty)$-frame, generated by string oscillators. Here, let us use the frame simplex\footnote{The minus sign convention in \eqref{e.10d_frame_ES} is so that the results will be easier to compare with string theory examples.}
\begin{align}
	\vec v_\infty\equiv -\frac{1}{2\sqrt 2}\hat \phi. \label{e.10d_frame_ES}
\end{align}

The stringy frame simplex is generated by the point $1_1^\infty$ in Table \ref{t.1d_10d_norm}, and by evaluating this coordinate in the lattice formula \eqref{e.lattice} with frame simplex \eqref{e.10d_frame_ES}, we obtain an $\alpha$-vector of 
\begin{align}
	\vec \alpha(1_1^\infty )=-\frac{1}{2\sqrt 2}\hat \phi.
\end{align}

Next, we have several other principal norm candidates for towers. Table \ref{t.1d_10d_norm} lists $p_P^n$'s of $1_2^1$, and the strings $2_2^1$ and $2_3^1$ have oscillator towers with coordinates $1_1^\infty$ and $1_{3/2}^\infty$. The tower $1_2^1$, and the oscillator towers $1_1^\infty$ and $1_{3/2}^\infty$ of $2_2^1$ and $2_3^1$ have $\alpha$-vectors of, respectively,
\begin{align}
	\vec \alpha(1_2^1)=\frac 3{2\sqrt 2}\hat \phi,\qquad \vec\alpha (1_1^\infty)=-\frac 1{2\sqrt 2}\hat \phi,\qquad \vec\alpha (1_{3/2}^\infty )=\frac 1{2\sqrt 2}\hat \phi.
\end{align}
The oscillator tower of $2_2^2$ is just the generator $1_1^\infty$ of the frame simplex, so we already assume we have this tower. That means that the only new principal norm candidates are the $1_2^1$ tower and the oscillators $1_{3/2}^\infty$ of $2_3^1$. These two candidates satisfy the principal product condition \eqref{e.principal_product} with $1_1^\infty $, but they do not satisfy the principal product condition with each other, since
\begin{align}
	\vec \alpha(1_2^1)\cdot \vec \alpha(1_{3/2}^\infty)=\frac 3{8}>\frac 18.
\end{align}

Thus, a choice between two options must be made! That is, given an emergent string limit and the list of principal norm candidates of particle towers that follow, there are two ways in which the principal product conditions can be satisfied:
\begin{itemize}
	\item either there is another decompactification limit, where the tower of $1_2^1$ KK-modes is chosen, and the theory resembles IIA and Heterotic $\mathrm{E_8\times E_8}$ string theory,
	\item or there is another emergent string limit, where the tower of oscillators $1_{3/2}^\infty$ from $2_3^1$ strings is chosen, and thus the theory resembles Type IIB, heterotic SO(32), and Type I string theory).
\end{itemize}

In each case, the choice of 	$1_2^1$ or $1_{3/2}^\infty$ makes the $+\phi$ direction respectively either a geometric or stringy frame, and this introduces a new set of radion/dilaton lattice conditions that must also be satisfied from this new frame. For the $1_2^1$ choice, the lattice conditions are the ones found in the previous classification of geometric frames of decompactification to 11d. Meanwhile, the stringy $1_{3/2}^\infty$ choice needs to be symmetric along $\phi\rightarrow -\phi$, since both limits are emergent string limits, and thus the classification behaves the same way in both directions. With these restrictions, we find that the following sets of branes all mutually satisfy the principal product conditions \eqref{e.principal_product} in each choice
\begin{align}
	(\infty):\qquad \begin{matrix}\text{IIA-like:}&\qquad 1_1^\infty,1_2^1,2_2^1,3_4^1,5_6^1,6_8^1,7_8^1\\
	\text{IIB-like:}&\qquad  1_1^{\infty },1_{3/2}^{\infty },2_2^1,2_3^1,4_5^1,6_7^1,6_8^1,8_9^1,8_{11}^1
	\end{matrix}.\label{e.10d_1d_i}
\end{align} 
The choice with $1_2^1$ reproduces exactly the behavior of IIA string theory, where the $1_1^\infty $, $1_2^1$, $2_2^1$, $3_4^1$, $5_6^1$, $6_8^1$, and $7_8^1$ states respectively correspond to the fundamental string oscillators, D0-branes, D2-branes, D4-branes, NS5-branes, and D6-branes. The choice with $2_2^1$ reproduces exactly the same branes of IIB string theory, where the $1_1^\infty $, $2_2^1$, $2_3^1$, $4_5^1$, $6_7^1$, $8_9^1$, $8_{11}^1$ states respectively correspond to the fundamental string oscillators, fundamental strings, D1-branes, D3-branes, D5-branes, NS5-branes, D7-branes, and NS7-branes. 

If heterotic SO(32) and Type I string theory are governed by these rules, then they require a set of branes like D-odd branes, just like IIB string theory. But, due to sliding in heterotic compactifications, it is not the case that these assumptions apply. Interestingly, the instanton proposed in \cite{Alvarez-Garcia:2024vnr} is on this lattice as well. As discussed in Section \ref{s.principal.instantons}, principal branes sometimes require the existence of lower dimensional branes, such as instantons, and it is likely the instanton of \cite{Alvarez-Garcia:2024vnr} can be motivated this way. This will be the subject of future work.

\begin{figure}
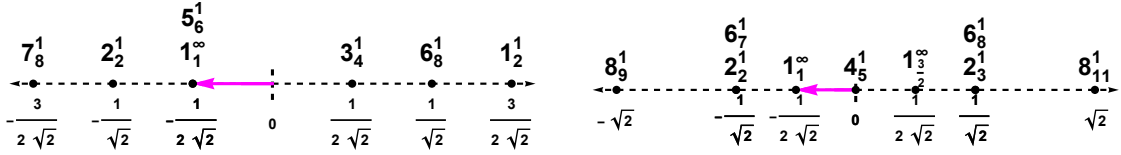
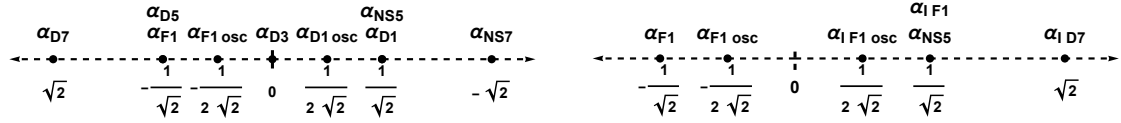

\centering
\begin{subfigure}{.49\linewidth}
\centering
\includegraphics[scale=.8]{figs/1d/classification/10d_i1.pdf}
\caption{$(\infty)$-frame, generated by $1_{1}^\infty$, and a decompactification limit $1_{3}^1$.} \label{f.10d_1dclass_i1}
\end{subfigure}
\begin{subfigure}{.49\linewidth}
\centering
\includegraphics[scale=.8]{figs/1d/classification/10d_i.pdf}
\caption{$(\infty)$-frame, generated by $1_{1}^\infty$, and another emergent string limit $1_{3/2}^\infty$. }
\label{f.10d_1dclass_i}
\end{subfigure}
\begin{subfigure}{.49\linewidth}
\centering
\includegraphics[scale=.8]{figs/1d/10d_IIB.pdf}
\caption{Lattice of $\alpha$-vectors for $\frac 12$-BPS branes in 10d IIB string theory.}\label{f.10d_IIB}
\end{subfigure}
\begin{subfigure}{.49\linewidth}
\centering
\includegraphics[scale=.8]{figs/1d/10d_HO.pdf}
\caption{Radion and dilaton lattices of known branes of 10d heterotic SO(32) string theory. }\label{f.10d_HO}
\end{subfigure}
\caption{Figures \ref{f.10d_1dclass_i1} and \ref{f.10d_1dclass_i} list the principal $\alpha$-vectors, labeled by $p_{P}^n$, for 1d moduli spaces of 10d theories with stringy frame simplices $1_1^\infty$ listed in \eqref{e.10d_1d_i}. $p$ refers to the spacetime dimension of the brane, $P$ refers to the coordinates of the brane with respect to the frame, and $n$ labels the number of dimensions that decompactify from the KK-modes that come from the winding modes of the brane. The magenta arrows point to the frame simplex vector $1_1^\infty$. Figure \ref{f.10d_1dclass_i1} reproduces exactly the brane content of the geometric frame analysis and 10d IIA string theory (and the known branes of $\mathrm{E_8\times E_8}$ heterotic string theory), depicted in Figure \ref{f.10d_1}. Figure \ref{f.10d_1dclass_i1} reproduces the brane content of IIB string theory, depicted in Figure \ref{f.10d_IIB}. Figure \ref{f.10d_1dclass_i1} also governs known branes in SO(32) heterotic string theory, which are depicted in Figure  \ref{f.10d_HO}. }
\label{f.10d_i}
\end{figure}

In summary, for 10d theories with one modulus in 10d theory, there are just two distinct theories, whose principal branes are listed in Table \ref{t.1d_10d_theories}. Remarkably, these reproduce the brane content of IIA and IIB string theory: One has geometric and stringy frames $(1)$ and $(\infty)$, reproducing exactly the $\frac 12$-BPS brane content of IIA string theory, depicted in Figure \ref{f.10d_1dclass_1}. The other theory has two stringy $(\infty)$-frames, reproducing exactly the $\frac 12$-BPS brane content of IIB string theory, depicted in Figure \ref{f.10d_1dclass_i}. Furthermore, the known branes in the heterotic and Type 1 string theories (some of which are non-BPS) also live at these principal lattice sites, and it would be interesting to know whether heterotic and Type I string theory possess non-BPS branes that fully populate these principal lattice sites.

\begin{table}[H]
$$\begin{array}{c|c|c|c|c|c|c|c|c|c|c|}
(D-d)& p_P^n\\ \hline
(1)& 1_0^1,1_{3/2}^{\infty },2_3^1,3_3^1,5_6^1,6_6^1,7_9^1\\\hline
(\infty)&
	1_1^{\infty },1_{3/2}^{\infty },2_2^1,2_3^1,4_5^1,6_7^1,6_8^1,8_9^1,8_{11}^1\\\hline
\end{array}$$ 
\caption{ 10d theories with 1d moduli spaces, labeled by their frame simplices and which principal branes $p_P^n$ they possess. $p$ refers to the spacetime dimension of the brane, $P$ refers to the coordinates of the brane with respect to the frame, and $n$ labels the number of dimensions that decompactify from the KK-modes that come from the winding modes of the brane.    These reproduce IIA and IIB string theory. These are depicted in Figures \ref{f.10d_1dclass_1} and \ref{f.10d_1dclass_i}.} \label{t.1d_10d_theories}
\end{table}

\subsubsection{9d}

The solutions to the  principal norm condition \eqref{e.principal_norm} are listed in Table \ref{t.1d_9d_norm} for all of the allowed frames of 1d moduli spaces of 9d theories.

\begin{table}[H]
$$\begin{array}{c|c|c|c|c|c|c|c|c|c|c|}
(D-d)&p_P^n \\ \hline
(1)&1_0^1,1_2^2,2_2^2,3_4^1,4_4^1,5_6^2,6_6^2,6_8^1\\\hline
(2)&1_0^2,1_2^{18},1_3^1,2_3^2,3_3^1,4_6^1,5_6^2,6_6^1,6_7^{18},6_9^2\\\hline
(\infty)&1_1^{\infty },2_2^1,5_7^1,6_8^{\infty }\\\hline
\end{array}$$ 
\caption{Principal norm candidates $p^n_P$, with respect to frames $(D-d)$, for 1d moduli spaces of 9d theories. $p$ refers to the spacetime dimension of the brane, $P$ refers to the coordinates of the brane with respect to the frame, and $n$ labels the number of dimensions that decompactify from the KK-modes that come from the winding modes of the brane.} \label{t.1d_9d_norm}
\end{table}

Consider the $(1)$-frame. Here the frame is generated by the $1_0^1$ KK-modes,
\begin{align}
	\vec \alpha(1_0^1)=\sqrt{\frac{8}{7}}\hat \rho.
\end{align}
The $1_2^2$ tower satisfies the principal product condition \eqref{e.principal_product} with $1_0^1$, but the $2_2^2$ string's oscillators do not. Thus, the $1_2^2$ tower is principal, but the $2_2^2$ string's oscillators are not principal. The $2_2^2$ string also does not satisfy the principality condition with the $1_0^1$ tower. Next, consider the $3_4^1$ 2-brane. This satisfies the principal product conditions, and thus is required. The $4_4^1$ brane does not satisfy the principal product conditions. The $5_6^2$ brane satisfies the conditions. The $6_8^1$ brane also satisfies the condition. Thus, the principal branes of $(1)$-frame are listed in Table \ref{t.1d_9d_theories} and depicted in Figure \ref{f.10d_1dclass_1}. The $(2)$-frame reproduces the $(1)$-frame. The $(\infty)$-frame does not provide a valid convex hull of particle towers satisfying the Sharpened DC.

\begin{table}[H]
$$\begin{array}{c|c|c|c|c|c|c|c|c|c|c|}
(D-d)& p_P^n\\ \hline
(1)& 1_0^1,1_2^2,2_2^2,3_4^1,4_4^1,5_6^2,6_6^2,6_8^1\\\hline
\end{array}$$ 
\caption{9d theories with 1d moduli spaces, labeled by their frame simplices and which principal branes $p_P^n$ they possess. $p$ refers to the spacetime dimension of the brane, $P$ refers to the coordinates of the brane with respect to the frame, and $n$ labels the number of dimensions that decompactify from the KK-modes that come from the winding modes of the brane. This reproduces the radion slice of IIB string theory on a circle. These are depicted in Figure \ref{f.9d_1d_class}.} \label{t.1d_9d_theories}
\end{table}

\begin{figure}
\centering
\begin{subfigure}{.49\linewidth}
\centering
\includegraphics[scale=.8]{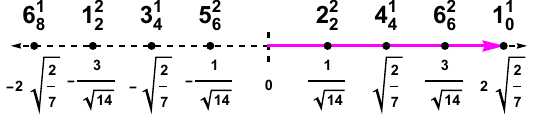}
\end{subfigure}\caption{Principal $\alpha$-vectors, labeled by $p_{P}^n$, for the 1d moduli space of a 9d theory. $p$ refers to the spacetime dimension of the brane, $P$ refers to the coordinates of the brane with respect to the frame, and $n$ labels the number of dimensions that decompactify from the KK-modes that come from the winding modes of the brane. The magenta arrow points to the frame simplex vector, generated by $1_0^1$, which generates the $(1)$-frame.  This reproduces exactly the radion slice of IIB string theory on a circle, where the $1_0^1$, $1_2^2$, $3_4^1$, $5_6^2$, and $6_6^2$ points respectively correspond to the KK-mode, wrapped strings, wrapped D3-branes, wrapped 5-branes, and unwrapped 5-branes. This also corresponds to the radion slice of M-theory on a 2-torus. 
}
\label{f.9d_1d_class}
\end{figure}

This exactly reproduces the $\frac 12$-BPS brane content of the radion slice of IIB string theory on a circle, where the $1_0^1$, $1_2^2$, $3_4^1$, $5_6^2$, and $6_6^2$ points respectively correspond to the KK-modes, wrapped strings, wrapped D3-branes, wrapped 5-branes, and unwrapped 5-branes! These principal branes are depicted in Figure \ref{f.9d_1d_class}.

\subsubsection{8d}

The solutions to the  principal norm condition \eqref{e.principal_norm} are listed in Table \ref{t.1d_8d_norm} for all of the allowed frames of 1d moduli spaces of 8d theories.

\begin{table}[H]
$$\begin{array}{c|c|c|c|c|c|c|c|c|c|c|}
(D-d)&p_P^n \\ \hline
(1)&1_0^1,5_7^1\\\hline
(2)&1_0^2,1_2^{\infty },2_2^2,2_4^1,3_4^2,4_4^1,4_6^2,5_6^{\infty },5_8^2,6_6^2,6_{10}^2\\\hline 
(3)&1_0^3,1_3^3,2_3^3,3_3^1,3_6^1,4_6^3,5_6^3,5_9^3,6_6^1,6_{12}^1\\\hline
(\infty) & 1_1^{\infty },1_2^2,2_2^1,2_3^2,3_4^2,4_5^2,4_6^1,5_6^2,5_7^{\infty },6_7^2,6_9^2\\\hline
\end{array}$$ 
\caption{Principal norm candidates $p^n_P$, with respect to frames $(D-d)$, for 1d moduli spaces of 8d theories. $p$ refers to the spacetime dimension of the brane, $P$ refers to the coordinates of the brane with respect to the frame, and $n$ labels the number of dimensions that decompactify from the KK-modes that come from the winding modes of the brane.} \label{t.1d_8d_norm}
\end{table}

\begin{figure}
\centering
\begin{subfigure}{.49\linewidth}
\centering
\includegraphics[scale=.8]{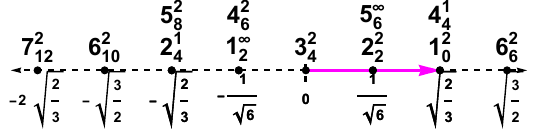}
\caption{$(2)$-frame, generated by $1_0^2$.}
\label{f.8d_1d_class_2}
\end{subfigure}
\begin{subfigure}{.49\linewidth}
\centering
\includegraphics[scale=.8]{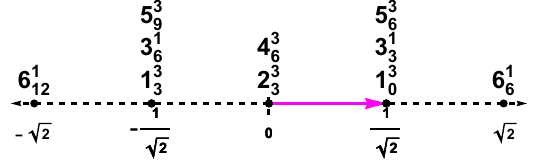}
\caption{$(3)$-frame, generated by $1_0^3$.}
\label{f.8d_1d_class_3}
\end{subfigure}
\caption{Principal $\alpha$-vectors, labeled by $p_{P}^n$, for 1d moduli spaces of 8d theories. $p$ refers to the spacetime dimension of the brane, $P$ refers to the coordinates of the brane with respect to the frame, and $n$ labels the number of dimensions that decompactify from the KK-modes that come from the winding modes of the brane. The magenta arrows point to the frame simplex vectors. }
\label{f.8d_1d_class}
\end{figure}

The $(1)$-frame does not have enough principal towers to satisfy the Sharpened DC. The $(2)$-frame uniquely produces the following principal branes $p_P^n$,
\begin{align}
	(2):\qquad 1_0^2,1_2^{\infty },2_2^2,2_4^1,3_4^2,4_4^1,4_6^2,5_6^{\infty },5_8^2,6_6^2,6_{10}^2,7_{12}^2,\label{e.8d_1d_2}
\end{align}
and they are depicted in Figure \ref{f.8d_1d_class_2}. The $(3)$-frame uniquely produces the following principal branes $p_P^n$,
\begin{align}
	(3):\qquad 1_0^3,1_3^3,2_3^3,3_3^1,3_6^1,4_6^3,5_6^3,5_9^3,6_6^1,6_{12}^1,\label{e.8d_1d_3}
\end{align}
and they are depicted in Figure \ref{f.8d_1d_class_3}. The $(\infty)$-frame produces a frame with a $(2)$-frame. Thus there are only two theories, listed in Table \ref{t.1d_8d_theories} and depicted in Figure \ref{f.8d_1d_class}.

\begin{table}[H]
$$\begin{array}{c|c|c|c|c|c|c|c|c|c|c|}
(D-d)& p_P^n\\ \hline
(2)& 1_0^2,1_2^{\infty },2_2^2,2_4^1,3_4^2,4_4^1,4_6^2,5_6^{\infty },5_8^2,6_6^2,6_{10}^2,7_{12}^2\\\hline
(3)& 1_0^3,1_3^3,2_3^3,3_3^1,3_6^1,4_6^3,5_6^3,5_9^3,6_6^1,6_{12}^1\\\hline
\end{array}$$ 
\caption{8d theories with 1d moduli spaces, labeled by their frame simplices and which principal branes $p_P^n$ they possess. $p$ refers to the spacetime dimension of the brane, $P$ refers to the coordinates of the brane with respect to the frame, and $n$ labels the number of dimensions that decompactify from the KK-modes that come from the winding modes of the brane. These are depicted in Figure \ref{f.8d_1d_class}. } \label{t.1d_8d_theories}
\end{table}

\subsubsection{7d}
The solutions to the principal norm condition \eqref{e.principal_norm} are listed in Table \ref{t.1d_7d_norm} for all of the allowed frames of 1d moduli spaces of 7d theories.

\begin{table}[H]
$$\begin{array}{c|c|c|c|c|c|c|c|c|c|c|}
(D-d)&p_P^n \\ \hline
(1)&1_0^1,1_2^3,2_2^3,2_3^2,3_3^2,3_4^3,4_4^3,4_6^1\\\hline
(2)&1_0^2,4_7^2\\\hline
(3)&1_0^3,1_4^1,2_2^2,2_4^3,3_4^3,3_6^2,4_4^1,4_8^3\\\hline
(4)&1_0^4,1_3^{\infty },2_3^4,2_6^1,3_3^1,3_6^4,4_6^{\infty },4_9^4,5_6^4,5_{12}^4\\\hline
(\infty)&1_1^{\infty },1_2^4,2_2^1,2_3^4,3_4^4,3_5^1,4_5^4,4_6^{\infty },5_6^4,5_8^4\\\hline
\end{array}$$ 
\caption{Principal norm candidates $p^n_P$, with respect to frames $(D-d)$, for 1d moduli spaces of 7d theories. $p$ refers to the spacetime dimension of the brane, $P$ refers to the coordinates of the brane with respect to the frame, and $n$ labels the number of dimensions that decompactify from the KK-modes that come from the winding modes of the brane.} \label{t.1d_7d_norm}
\end{table}

\begin{figure}
\centering
\begin{subfigure}{.49\linewidth}
\centering
\includegraphics[scale=.8]{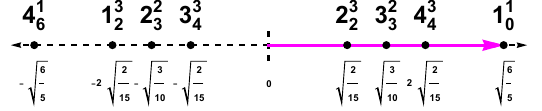}
\caption{$(1)$-frame, generated by $1_0^1$.}
\label{f.7d_1d_class_1}
\end{subfigure}
\begin{subfigure}{.49\linewidth}
\centering
\includegraphics[scale=.8]{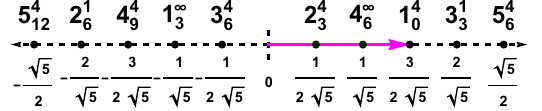}
\caption{$(4)$-frame, generated by $1_0^4$.}
\label{f.7d_1d_class_4}
\end{subfigure}
\caption{Principal $\alpha$-vectors, labeled by $p_{P}^n$, for 1d moduli spaces of 7d theories. $p$ refers to the spacetime dimension of the brane, $P$ refers to the coordinates of the brane with respect to the frame, and $n$ labels the number of dimensions that decompactify from the KK-modes that come from the winding modes of the brane. The magenta arrows point to the frame simplex vectors.  }
\label{f.7d_1d_class}
\end{figure}

The $(1)$-frame uniquely produces the following principal branes $p_P^n$,
\begin{align}
	(1):\qquad 1_0^1,1_2^3,2_2^3,2_3^2,3_3^2,3_4^3,4_4^3,4_6^1\label{e.7d_1d_1},
\end{align}
and they are depicted in Figure \ref{f.7d_1d_class_1}. The $(2)$-frame is not allowed. The $(3)$-frame reproduces the $(1)$-frame. The $(4)$-frame uniquely produces the following principal branes $p_P^n$,
\begin{align}
	(4):\qquad 1_0^4,1_3^{\infty },2_3^4,2_6^1,3_3^1,3_6^4,4_6^{\infty },4_9^4,5_6^4,5_{12}^4,\label{e.7d_1d_4}
\end{align}
and they are depicted in Figure \ref{f.7d_1d_class_4}. The $(\infty)$-frame reproduces the $(4)$-frame. Thus, there are only two theories, listed in Table \ref{t.1d_7d_theories}, and depicted in Figure \ref{f.7d_1d_class}.

\begin{table}[H]
$$\begin{array}{c|c|c|c|c|c|c|c|c|c|c|}
(D-d)&p_P^n \\ \hline
(1)& 1_0^1,1_2^3,2_2^3,2_3^2,3_3^2,3_4^3,4_4^3,4_6^1\\\hline
(4)& 1_0^4,1_3^{\infty },2_3^4,2_6^1,3_3^1,3_6^4,4_6^{\infty },4_9^4,5_6^4,5_{12}^4\\\hline
\end{array}$$ 
\caption{7d theories with 1d moduli spaces, labeled by their frame simplices and which principal branes $p_P^n$ they possess. $p$ refers to the spacetime dimension of the brane, $P$ refers to the coordinates of the brane with respect to the frame, and $n$ labels the number of dimensions that decompactify from the KK-modes that come from the winding modes of the brane. These are depicted in Figure \ref{f.7d_1d_class}. } \label{t.1d_7d_theories}
\end{table}

\subsubsection{6d}

The solutions to the  principal norm condition \eqref{e.principal_norm} are listed in Table \ref{t.1d_6d_norm} for all of the allowed frames of 1d moduli spaces of 6d theories.

\begin{table}[H]
$$\begin{array}{c|c|c|c|c|c|c|c|c|c|c|}
(D-d)&p_P^n \\ \hline
(1)&1_0^1,1_2^5,2_2^5,2_3^5,3_3^5,3_5^1\\\hline
(2)&1_0^2,1_3^2,2_2^3,2_3^{\infty },2_4^3,3_3^2,3_6^2,4_4^3,4_8^3\\\hline
(3)&1_0^3,1_3^{21},2_3^{21},2_4^{21},3_4^{21},3_7^3\\\hline
(4)&1_0^4,1_4^4,2_2^2,2_3^8,2_4^{\infty },2_5^8,2_6^2,3_4^4,3_8^4,4_4^1,4_5^8,4_{11}^8,4_{12}^1\\\hline
(5)&1_0^5,1_4^{45},1_6^1,2_3^5,2_4^{45},2_5^{45},2_6^5,3_3^1,3_5^{45},3_9^5\\\hline
(\infty)& 1_1^{\infty },1_2^{\infty },2_2^1,2_3^{\infty },2_4^1,3_4^{\infty },3_5^{\infty },4_5^{\infty },4_7^{\infty }\\\hline
\end{array}$$ 
\caption{Principal norm candidates $p^n_P$, with respect to frames $(D-d)$, for 1d moduli spaces of 6d theories. $p$ refers to the spacetime dimension of the brane, $P$ refers to the coordinates of the brane with respect to the frame, and $n$ labels the number of dimensions that decompactify from the KK-modes that come from the winding modes of the brane.} \label{t.1d_6d_norm}
\end{table}

\begin{figure}
\centering
\begin{subfigure}{.49\linewidth}
\centering
\includegraphics[scale=.8]{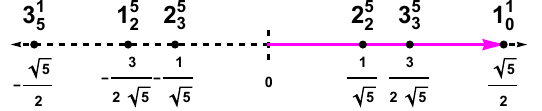}
\caption{$(1)$-frame, generated by $1_0^1$.}
\label{f.6d_1dclass_1}
\end{subfigure}
\begin{subfigure}{.49\linewidth}
\centering
\includegraphics[scale=.8]{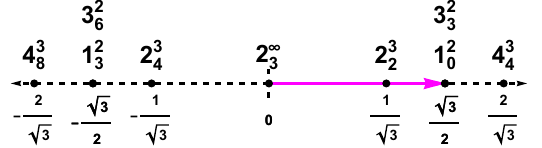}
\caption{$(2)$-frame, generated by $1_0^2$.}
\label{f.6d_1dclass_2}
\end{subfigure}
\begin{subfigure}{.49\linewidth}
\centering
\includegraphics[scale=.8]{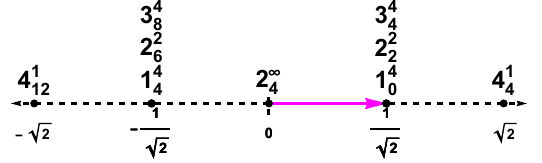}
\caption{$(4)$-frame, generated by $1_0^4$.}
\label{f.6d_1dclass_4}
\end{subfigure}
\begin{subfigure}{.49\linewidth}
\centering
\includegraphics[scale=.8]{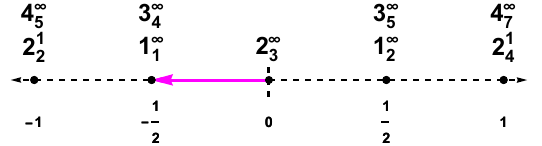}
\caption{$(\infty )$-frame, generated by $1_0^\infty$.}
\label{f.6d_1dclass_i}
\end{subfigure}
\caption{Principal $\alpha$-vectors, labeled by $p_{P}^n$, for 1d moduli spaces of 6d theories. $p$ refers to the spacetime dimension of the brane, $P$ refers to the coordinates of the brane with respect to the frame, and $n$ labels the number of dimensions that decompactify from the KK-modes that come from the winding modes of the brane. The magenta arrows point to the frame simplex vectors.}
\label{f.6d_1dclass}
\end{figure}

The $(1)$-frame uniquely produces the following principal branes $p_P^n$,
\begin{align}
	(1):\qquad 1_0^4,1_3^{\infty },2_3^4,2_6^1,3_3^1,3_6^4,4_6^{\infty },4_9^4,5_6^4,5_{12}^4, \label{e.6d_1d_1}
\end{align}
and they are depicted in Figure \ref{f.6d_1dclass_1}. The $(2)$-frame uniquely produces the following principal branes $p_P^n$,
\begin{align}
	(2):\qquad 1_0^4,1_3^{\infty },2_3^4,2_6^1,3_3^1,3_6^4,4_6^{\infty },4_9^4,5_6^4,5_{12}^4,
	\label{e.6d_1d_1}
\end{align}
and they are depicted in Figure \ref{f.6d_1dclass_2}. The $(3)$-frame does not provide a legal theory.  The $(4)$-frame produces multiple choices of principal strings, where $P$ could be $P\in \{2, 3,4,5,6\}$, but these strings do not all mutually satisfy the principal product conditions with each other. However, since this is a $(4)$-frame, the theory must be able to decompactify to a 10d theory with a 0d moduli space, and, by the earlier classification of 10d theories with 0d moduli spaces, there are no principal 2 and 4-branes in such a 10d theory. Thus, the branes $2_3^8$, $2_5^8$, and $4_5^8$ are forbidden in the 6d theory. As a result, the $(4)$-frame uniquely produces the following principal branes $p_P^n$,
\begin{align}
	(4):\qquad 1_0^4,1_4^4,2_2^2,2_4^{\infty },2_6^2,3_4^4,3_8^4,4_4^1,4_{12}^1,
	\label{e.6d_1d_4}
\end{align}
and they are depicted in Figure \ref{f.6d_1dclass_4}. The $(5)$-frame reproduces the $(1)$-frame. The $(\infty )$-frame uniquely produces the following principal branes $p_P^n$,
\begin{align}
	(\infty ):\qquad 1_1^{\infty },1_2^{\infty },2_2^1,2_3^{\infty },2_4^1,3_4^{\infty },3_5^{\infty },4_5^{\infty },4_7^{\infty },
	\label{e.6d_1d_i}
\end{align}
and they are depicted in Figure \ref{f.6d_1dclass_i}. Thus, there are only four theories, listed in Table \ref{t.1d_6d_theories}, and depicted in Figure \ref{f.6d_1dclass}.

\begin{table}[H]
$$\begin{array}{c|c|c|c|c|c|c|c|c|c|c|}
(D-d)&p_P^n \\ \hline
(1)& 1_0^4,1_3^{\infty },2_3^4,2_6^1,3_3^1,3_6^4,4_6^{\infty },4_9^4,5_6^4,5_{12}^4\\\hline
(2)& 1_0^4,1_3^{\infty },2_3^4,2_6^1,3_3^1,3_6^4,4_6^{\infty },4_9^4,5_6^4,5_{12}^4\\\hline
(4)& 1_0^4,1_4^4,2_2^2,2_4^{\infty },2_6^2,3_4^4,3_8^4,4_4^1,4_{12}^1\\\hline
(\infty) & 1_1^{\infty },1_2^{\infty },2_2^1,2_3^{\infty },2_4^1,3_4^{\infty },3_5^{\infty },4_5^{\infty },4_7^{\infty }\\\hline
\end{array}$$ 
\caption{ 6d theories with 1d moduli spaces, labeled by their frame simplices and which principal branes $p_P^n$ they possess. $p$ refers to the spacetime dimension of the brane, $P$ refers to the coordinates of the brane with respect to the frame, and $n$ labels the number of dimensions that decompactify from the KK-modes that come from the winding modes of the brane.    These are depicted in Figure \ref{f.6d_1dclass}.} \label{t.1d_6d_theories}
\end{table}

It would be interesting if this classification also describes the particle-tower and brane content of 6d string island theories.

\subsubsection{5d \label{s.classification.1d.5d}}

The solutions to the  principal norm condition \eqref{e.principal_norm} are listed in Table \ref{t.1d_5d_norm} for all of the allowed frames of 1d moduli spaces of 5d theories.

\begin{table}[H]
$$\begin{array}{c|c|c|c|c|c|c|c|c|c|c|}
(D-d)& p_P^n\\ \hline
(1)&1_0^1,1_2^{\infty },2_2^{\infty },2_4^1\\\hline
(2)&1_0^2,1_3^5,2_2^5,2_5^2,3_3^5,3_7^5\\\hline
(3)&1_0^3,1_4^3,2_2^3,2_6^3,3_3^2,3_9^2\\\hline
(4)&1_0^4,2_7^4\\\hline
(5)&1_0^5,1_6^2,2_2^2,2_8^5,3_4^5,3_{12}^5\\\hline
(6)&1_0^6,1_6^6,2_3^6,2_9^6,3_3^1,3_{15}^1\\\hline
(\infty)&1_1^{\infty },1_3^1,2_2^1,2_4^{\infty }\\\hline
\end{array}$$ 
\caption{Principal norm candidates $p^n_P$, with respect to frames $(D-d)$, for 1d moduli spaces of 5d theories. $p$ refers to the spacetime dimension of the brane, $P$ refers to the coordinates of the brane with respect to the frame, and $n$ labels the number of dimensions that decompactify from the KK-modes that come from the winding modes of the brane.} \label{t.1d_5d_norm}
\end{table}

\begin{figure}
\centering
\begin{subfigure}{.49\linewidth}
\centering
\includegraphics[scale=.8]{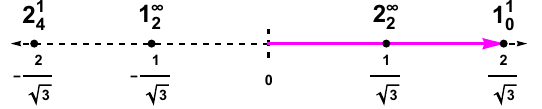}
\caption{$(1)$-frame, generated by $1_0^1$.}
\label{f.5d_1dclass_1}
\end{subfigure}
\begin{subfigure}{.49\linewidth}
\centering
\includegraphics[scale=.8]{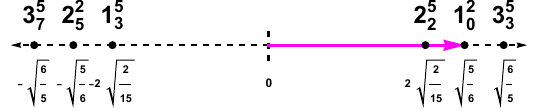}
\caption{$(2)$-frame, generated by $1_0^2$.}
\label{f.5d_1dclass_2}
\end{subfigure}
\begin{subfigure}{.49\linewidth}
\centering
\includegraphics[scale=.8]{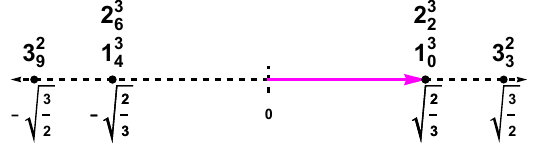}
\caption{$(3)$-frame, generated by $1_0^3$.}
\label{f.5d_1dclass_3}
\end{subfigure}
\begin{subfigure}{.49\linewidth}
\centering
\includegraphics[scale=.8]{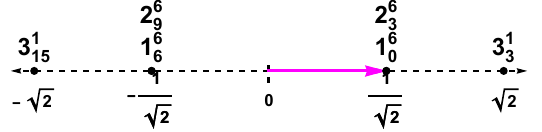}
\caption{$(6)$-frame, generated by $1_0^6$.}
\label{f.5d_1dclass_6}
\end{subfigure}
\caption{Principal $\alpha$-vectors, labeled by $p_{P}^n$, for 1d moduli spaces of 5d theories. $p$ refers to the spacetime dimension of the brane, $P$ refers to the coordinates of the brane with respect to the frame, and $n$ labels the number of dimensions that decompactify from the KK-modes that come from the winding modes of the brane. The magenta arrows point to the frame simplex vectors.}
\label{f.5d_1dclass}
\end{figure}

The $(1)$-frame uniquely produces the following principal branes $p_P^n$,
\begin{align}
	(1 ):\qquad 1_0^1,1_2^{\infty },2_2^{\infty },2_4^1,
	\label{e.5d_1d_1}
\end{align}
and they are depicted in Figure \ref{f.4d_1dclass_1}. The $(2)$-frame uniquely produces the following principal branes $p_P^n$,
\begin{align}
	(2 ):\qquad 1_0^2,1_3^5,2_2^5,2_5^2,3_3^5,3_7^5,
	\label{e.5d_1d_2}
\end{align}
and they are depicted in Figure \ref{f.4d_1dclass_2}. The $(3)$-frame uniquely produces the following principal branes $p_P^n$,
\begin{align}
	(3):\qquad 1_0^3,1_4^3,2_2^3,2_6^3,3_3^2,3_9^2,
	\label{e.5d_1d_3}
\end{align}
and they are depicted in Figure \ref{f.4d_1dclass_3}. The $(4)$-frame is not allowed. The $(5)$-frame reproduces the $(2)$-frame. The $(6)$-frame uniquely produces the following principal branes $p_P^n$,
\begin{align}
	(6):\qquad 1_0^6,1_6^6,2_3^6,2_9^6,3_3^1,3_{15}^1,
	\label{e.5d_1d_6}
\end{align}
and they are depicted in Figure \ref{f.4d_1dclass_6}. The $(\infty)$-frame reproduces the $(1)$-frame. Thus, there are only four theories, listed in Table \ref{t.1d_5d_theories}, and depicted in Figure \ref{f.5d_1dclass}.

\begin{table}[H]
$$\begin{array}{c|c|c|c|c|c|c|c|c|c|c|}
(D-d)&p_P^n \\ \hline
(1)& 1_0^1,1_2^{\infty },2_2^{\infty },2_4^1\\\hline
(2)& 1_0^2,1_3^5,2_2^5,2_5^2,3_3^5,3_7^5\\\hline
(3)& 1_0^3,1_4^3,2_2^3,2_6^3,3_3^2,3_9^2\\\hline
(6) & 1_0^6,1_6^6,2_3^6,2_9^6,3_3^1,3_{15}^1\\\hline
\end{array}$$ 
\caption{5d theories with 1d moduli spaces, labeled by their frame simplices and which principal branes $p_P^n$ they possess. $p$ refers to the spacetime dimension of the brane, $P$ refers to the coordinates of the brane with respect to the frame, and $n$ labels the number of dimensions that decompactify from the KK-modes that come from the winding modes of the brane.    These are depicted in Figure \ref{f.5d_1dclass}.} \label{t.1d_5d_theories}
\end{table}

As calculated in Appendix \ref{s.5dN1}, the 5d $\mathcal N=1$ symmetric flop and GMSV conifold examples are described by Figure \ref{f.5d_1dclass_1}. Thus, there exist landscape examples with only 8 supercharges described by this classification. Since these theories do not satisfy the assumptions used in this classification, but match the classification exactly, this suggests that the assumptions entering this classification might be able to be relaxed with similar results still following.

\subsubsection{4d}

The solutions to the  principal norm condition \eqref{e.principal_norm} are listed in Table \ref{t.1d_4d_norm} for all of the allowed frames of 1d moduli spaces of 4d theories.

\begin{table}[H]
$$\begin{array}{c|c|c|c|c|c|c|c|c|c|c|}
(D-d)& p_P^n\\ \hline
(1)&1_0^1,1_3^1\\\hline
(2)&1_0^2,1_4^2,2_2^{\infty },2_6^{\infty }\\\hline
(3)&1_0^3,1_5^3,2_2^5,2_8^5\\\hline
(4)&1_0^4,1_6^4,2_2^3,2_{10}^3\\\hline
(5)&1_0^5,1_7^5\\\hline
(6)&1_0^6,1_8^6,1_{10}^1,2_2^2,2_3^{24},2_{13}^{24},2_{14}^2\\\hline
(7)&1_0^7,1_9^7,2_3^7,2_{15}^7\\\hline
(\infty)& 1_1^{\infty },1_3^{\infty },2_2^1,2_6^1\\\hline
\end{array}$$ 
\caption{Principal norm candidates $p^n_P$, with respect to frames $(D-d)$, for 1d moduli spaces of 4d theories. $p$ refers to the spacetime dimension of the brane, $P$ refers to the coordinates of the brane with respect to the frame, and $n$ labels the number of dimensions that decompactify from the KK-modes that come from the winding modes of the brane.} \label{t.1d_4d_norm}
\end{table}

Many of the principal norm candidates in \ref{t.1d_4d_norm} are exotic. If the only way for exotic branes to exist are through the mechanisms described in Section \ref{s.lattice.exotic}, then this requires that the 1d moduli spaces investigated here are in fact slices/projections of higher-dimensional moduli spaces. This will be relevant in the application to the Dark Dimension Scenario in Section \ref{s.dark_dimension}. It would be interesting to know whether some of the large values of $P$ on these exotic branes forbid these branes. If so then this could exclude some of the towers and branes presented here.

\begin{figure}
\centering
\begin{subfigure}{.49\linewidth}
\centering
\includegraphics[scale=.8]{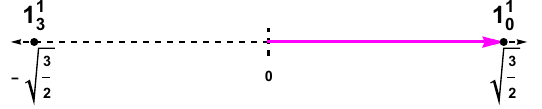}
\caption{$(1)$-frame, generated by $1_0^1$.}
\label{f.4d_1dclass_1}
\end{subfigure}
\begin{subfigure}{.49\linewidth}
\centering
\includegraphics[scale=.8]{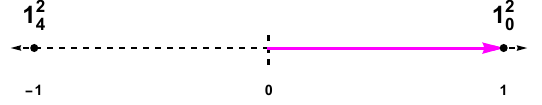}
\caption{$(2)$-frame, generated by $1_0^2$.  }
\label{f.4d_1dclass_2}
\end{subfigure}
\begin{subfigure}{.49\linewidth}
\centering
\includegraphics[scale=.8]{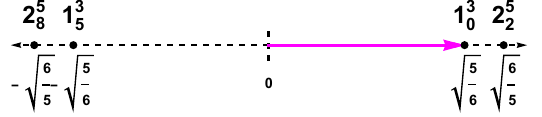}
\caption{$(3)$-frame, generated by $1_0^3$.   }
\label{f.4d_1dclass_3}
\end{subfigure}
\begin{subfigure}{.49\linewidth}
\centering
\includegraphics[scale=.8]{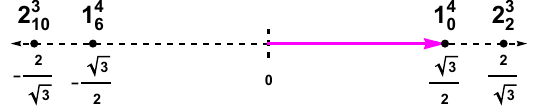}
\caption{$(4)$-frame, generated by $1_0^4$.   }
\label{f.4d_1dclass_4}
\end{subfigure}
\begin{subfigure}{.49\linewidth}
\centering
\includegraphics[scale=.8]{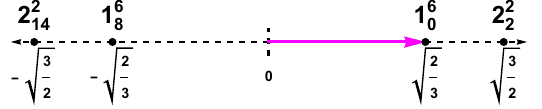}
\caption{$(6)$-frame, generated by $1_0^6$.   }
\label{f.4d_1dclass_6}
\end{subfigure}
\begin{subfigure}{.49\linewidth}
\centering
\includegraphics[scale=.8]{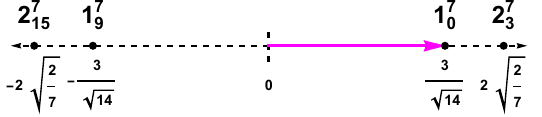}
\caption{$(7)$-frame, generated by $1_0^7$.   }
\label{f.4d_1dclass_7}
\end{subfigure}
\begin{subfigure}{.49\linewidth}
\centering
\includegraphics[scale=.8]{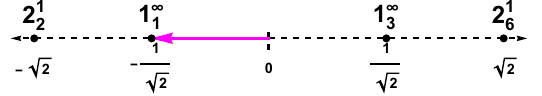}
\caption{$(\infty)$-frame, generated by $1_0^\infty $.   }
\label{f.4d_1dclass_i}
\end{subfigure}
\caption{Principal $\alpha$-vectors, labeled by $p_{P}^n$, for 1d moduli spaces of 4d theories. $p$ refers to the spacetime dimension of the brane, $P$ refers to the coordinates of the brane with respect to the frame, and $n$ labels the number of dimensions that decompactify from the KK-modes that come from the winding modes of the brane. The magenta arrows point to the frame simplex vectors.}
\label{f.4d_1dclass}
\end{figure}

The $(1)$-frame uniquely produces the following principal branes $p_P^n$,
\begin{align}
	(1):\qquad 1_0^1,1_3^1,
	\label{e.4d_1d_1}
\end{align}
and they are depicted in Figure \ref{f.4d_1dclass_1}.\footnote{Note that this theory's particle tower $1_3^1$ has an exotic value of $P$.} The $(2)$-frame uniquely produces the following principal branes $p_P^n$,
\begin{align}
	(2):\qquad 1_0^2,1_4^2,
	\label{e.4d_1d_2}
\end{align}
and they are depicted in Figure \ref{f.4d_1dclass_2}.\footnote{ I have removed $2_2^\infty$ and $2_6^\infty$, since they wrap to produce oscillator-length winding modes. I will do similar omissions in subsequent examples.} The $(3)$-frame uniquely produces the following principal branes $p_P^n$,
\begin{align}
	(3):\qquad 1_0^3,1_5^3,2_2^5,2_8^5,
	\label{e.4d_1d_3}
\end{align}
and they are depicted in Figure \ref{f.4d_1dclass_3}. The $(4)$-frame uniquely produces the following principal branes $p_P^n$,
\begin{align}
	(4):\qquad 1_0^4,1_6^4,2_2^3,2_{10}^3,
	\label{e.4d_1d_4}
\end{align}
and they are depicted in Figure \ref{f.4d_1dclass_4}.  The $(5)$-frame does not satisfy the Brane DC. After eliminating strings that cannot exist in 10d and requiring that the theory must be symmetric since both limits decompactify to 10d, the $(6)$-frame produces the following principal branes $p_P^n$,
\begin{align}
	(6):\qquad 1_0^6,1_8^6,2_2^2,2_{14}^2,
	\label{e.4d_1d_6}
\end{align}
and they are depicted in Figure \ref{f.4d_1dclass_6}. The $(7)$-frame uniquely produces the following principal branes $p_P^n$,
\begin{align}
	(7):\qquad 1_0^7,1_9^7,2_3^7,2_{15}^7,
	\label{e.4d_1d_7}
\end{align}
and they are depicted in Figure \ref{f.4d_1dclass_7}. The $(\infty )$-frame uniquely produces the following principal branes $p_P^n$,
\begin{align}
	(\infty ):\qquad 1_1^{\infty },1_3^{\infty },2_2^1,2_6^1,
	\label{e.4d_1d_i}
\end{align}
and they are depicted in Figure \ref{f.4d_1dclass_i}. Thus, there are only seven theories, listed in Table \ref{t.1d_4d_theories}, and depicted in Figure \ref{f.4d_1dclass}.

\begin{table}[H]
$$\begin{array}{c|c|c|c|c|c|c|c|c|c|c|}
(D-d)& p_P^n \\ \hline
(1)& 1_0^1,1_3^1\\\hline
(2)& 1_0^2,1_4^2\\\hline
(3)& 1_0^3,1_5^3,2_2^5,2_8^5\\\hline
(4)& 1_0^4,1_6^4,2_2^3,2_{10}^3\\\hline
(6)& 1_0^6,1_8^6,2_2^2,2_{14}^2\\\hline
(7)& 1_0^7,1_9^7,2_3^7,2_{15}^7\\\hline
(\infty) & 1_1^{\infty },1_3^{\infty },2_2^1,2_6^1\\\hline
\end{array}$$ 
\caption{ 4d theories with 1d moduli spaces, labeled by their frame simplices and which principal branes $p_P^n$ they possess. $p$ refers to the spacetime dimension of the brane, $P$ refers to the coordinates of the brane with respect to the frame, and $n$ labels the number of dimensions that decompactify from the KK-modes that come from the winding modes of the brane. These are depicted in Figure \ref{f.4d_1dclass}.} \label{t.1d_4d_theories}
\end{table}

\subsubsection{3d}

The solutions to the  principal norm condition \eqref{e.principal_norm} are listed in Table \ref{t.1d_3d_norm} for all of the allowed frames of 1d moduli spaces of 3d theories.

\begin{table}[H]
$$\begin{array}{c|c|c|c|c|c|c|c|c|c|c|}
(D-d)& p_P^n \\\hline
(1)&1_0^1,1_4^1\\\hline
(2)&1_0^2,1_6^2\\\hline
(3)&1_0^3,1_8^3\\\hline
(4)&1_0^4,1_{10}^4\\\hline
(5)&1_0^5,1_{12}^5\\\hline
(6)&1_0^6,1_{14}^6\\\hline
(7)&1_0^7,1_{16}^7\\\hline
(8)&1_0^8,1_{18}^8\\\hline
(\infty)&1_1^{\infty },1_5^{\infty }\\\hline
\end{array}$$ 
\caption{Principal norm candidate particle towers $1^n_P$, with respect to frames $(D-d)$, for 1d moduli spaces of 3d theories. $1$ refers to the spacetime dimension of the particle tower, $P$ refers to the lattice coordinates of the particle tower with respect to the frame, and $n$ labels the number of dimensions that the tower decompactifies (or $\infty$ for string oscillators).} \label{t.1d_3d_norm}
\end{table}

All of the principal norm candidate towers $1_P$ of geometric frames with $P>0$ are exotic branes. This is because a $(P-1)$-brane, upon compactification to 3d by wrapping a $(P-1)$-torus, cannot produce a particle tower that becomes light in the limit that the torus becomes small, due to radion lattice formula \eqref{e.radion_lattice} with $d=3$. The requirement of these exotic branes can motivate the existence of extra moduli in the higher-dimensional theories, as will be relevant to the Dark Dimension Scenario in Section \ref{s.dark_dimension}.

\begin{figure}
\centering
\begin{subfigure}{.49\linewidth}
\centering
\includegraphics[scale=.8]{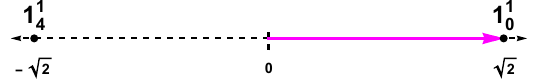}
\caption{$(1)$-frame, generated by $1_0^1$.   }
\label{f.3d_1dclass_1}
\end{subfigure}
\begin{subfigure}{.49\linewidth}
\centering
\includegraphics[scale=.8]{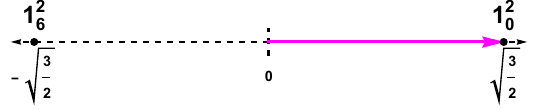}
\caption{$(2)$-frame, generated by $1_0^2$.   }
\label{f.3d_1dclass_2}
\end{subfigure}
\begin{subfigure}{.49\linewidth}
\centering
\includegraphics[scale=.8]{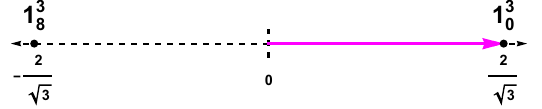}
\caption{$(3)$-frame, generated by $1_0^3$.   }
\label{f.3d_1dclass_3}
\end{subfigure}
\begin{subfigure}{.49\linewidth}
\centering
\includegraphics[scale=.8]{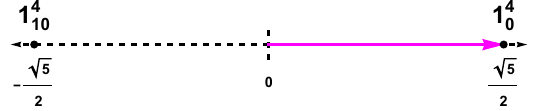}
\caption{$(4)$-frame, generated by $1_0^4$.   }
\label{f.3d_1dclass_4}
\end{subfigure}
\begin{subfigure}{.49\linewidth}
\centering
\includegraphics[scale=.8]{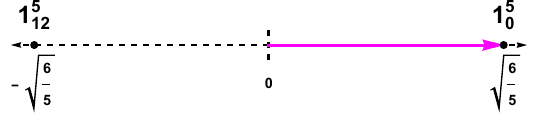}
\caption{$(5)$-frame, generated by $1_0^5$.   }
\label{f.3d_1dclass_5}
\end{subfigure}
\begin{subfigure}{.49\linewidth}
\centering
\includegraphics[scale=.8]{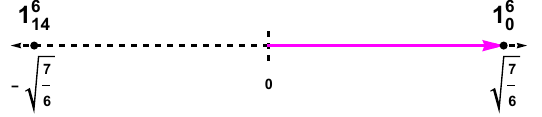}
\caption{$(6)$-frame, generated by $1_0^6$.   }
\label{f.3d_1dclass_6}
\end{subfigure}
\begin{subfigure}{.49\linewidth}
\centering
\includegraphics[scale=.8]{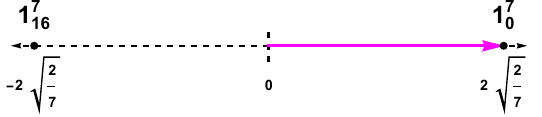}
\caption{$(7)$-frame, generated by $1_0^7$.   }
\label{f.3d_1dclass_7}
\end{subfigure}
\begin{subfigure}{.49\linewidth}
\centering
\includegraphics[scale=.8]{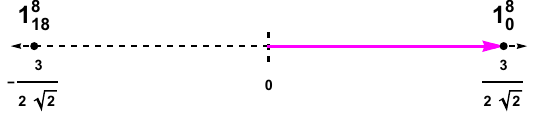}
\caption{$(8)$-frame, generated by $1_0^8$.   }
\label{f.3d_1dclass_8}
\end{subfigure}
\begin{subfigure}{.49\linewidth}
\centering
\includegraphics[scale=.8]{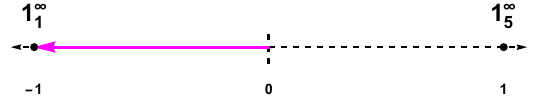}
\caption{$(\infty)$-frame, generated by $1_0^\infty $.   }
\label{f.3d_1dclass_i}
\end{subfigure}
\caption{Principal $\alpha$-vectors, labeled by $p_{P}^n$, for 1d moduli spaces of 3d theories. $p$ refers to the spacetime dimension of the brane, $P$ refers to the coordinates of the brane with respect to the frame, and $n$ labels the number of dimensions that decompactify from the KK-modes that come from the winding modes of the brane. The magenta arrows point to the frame simplex vectors.}
\label{f.3d_1dclass}
\end{figure}

The $(1)$-frame uniquely produces the following principal branes $p_P^n$,
\begin{align}
	(1 ):\qquad 1_0^1,1_4^1,
	\label{e.3d_1d_1}
\end{align}
and they are depicted in Figure \ref{f.3d_1dclass_1}. The $(2)$-frame uniquely produces the following principal branes $p_P^n$,
\begin{align}
	(2 ):\qquad 1_0^2,1_6^2,
	\label{e.3d_1d_2}
\end{align}
and they are depicted in Figure \ref{f.3d_1dclass_2}. The $(3)$-frame uniquely produces the following principal branes $p_P^n$,
\begin{align}
	(3):\qquad 1_0^3,1_8^3,
	\label{e.3d_1d_3}
\end{align}
and they are depicted in Figure \ref{f.3d_1dclass_3}. The $(4)$-frame uniquely produces the following principal branes $p_P^n$,
\begin{align}
	(4):\qquad 1_0^4,1_{10}^4,
	\label{e.3d_1d_4}
\end{align}
and they are depicted in Figure \ref{f.3d_1dclass_4}. The $(5)$-frame uniquely produces the following principal branes $p_P^n$,
\begin{align}
	(5):\qquad 1_0^5,1_{12}^5,
	\label{e.3d_1d_5}
\end{align}
and they are depicted in Figure \ref{f.3d_1dclass_5}. The $(6)$-frame uniquely produces the following principal branes $p_P^n$,
\begin{align}
	(6):\qquad 1_0^6,1_{14}^6,
	\label{e.3d_1d_6}
\end{align}
and they are depicted in Figure \ref{f.3d_1dclass_6}. The $(7)$-frame uniquely produces the following principal branes $p_P^n$,
\begin{align}
	(7):\qquad 1_0^7,1_{16}^7,
	\label{e.3d_1d_7}
\end{align}
and they are depicted in Figure \ref{f.3d_1dclass_7}. The $(8)$-frame uniquely produces the following principal branes $p_P^n$,\footnote{A $(1)$-frame could be produced, but it would be inconsistent with the previous $(1)$-frame analysis.}
\begin{align}
	(8):\qquad 1_0^8,1_{18}^8,
	\label{e.3d_1d_8}
\end{align}
and they are depicted in Figure \ref{f.3d_1dclass_8}. The $(\infty)$-frame uniquely produces the following principal branes $p_P^n$,\footnote{A $(1)$-frame could be produced, but but it would be inconsistent with the previous $(1)$-frame analysis.}:
\begin{align}
	(\infty ):\qquad 1_1^{\infty },1_5^{\infty },
	\label{e.3d_1d_i }
\end{align}
and they are depicted in Figure \ref{f.3d_1dclass_i}. Thus, there are only nine theories, listed in Table \ref{t.1d_3d_theories}, and depicted in Figure \ref{f.3d_1dclass}.

\begin{table}[H]
$$\begin{array}{c|c|c|c|c|c|c|c|c|c|c|}
(D-d)& p_P^n\\ \hline
(1)& 1_0^1,1_4^1\\\hline
(2)& 1_0^2,1_6^2\\\hline
(3)& 1_0^3,1_8^3\\\hline
(4)& 1_0^4,1_{10}^4\\\hline
(5)& 1_0^5,1_{12}^5\\\hline
(6)& 1_0^6,1_{14}^6\\\hline
(7)& 1_0^7,1_{16}^7\\\hline
(8)& 1_0^8,1_{18}^8\\\hline
(\infty) & 1_1^{\infty },1_3^{\infty },2_2^1,2_6^1\\\hline
\end{array}$$ 
\caption{ 3d theories with 1d moduli spaces, labeled by their frame simplices and which principal branes $p_P^n$ they possess. $p$ refers to the spacetime dimension of the brane, $P$ refers to the coordinates of the brane with respect to the frame, and $n$ labels the number of dimensions that decompactify from the KK-modes that come from the winding modes of the brane.    These are depicted in Figure \ref{f.3d_1dclass}.} \label{t.1d_3d_theories}
\end{table}

It is interesting to note that, in decompactification limits to 11d and 10d, the $\alpha$-vectors have familiar lengths such as $\frac 3{2\sqrt 2}$ and $2\sqrt{\frac 27}$ in, for example, Figures \ref{f.3d_1dclass_8} and \ref{f.3d_1dclass_8}. Recall that KK-modes from compactification of 11d to 10d, and 10d to 9d, have these lengths. The reason for this is because the KK-mode length-formula,
\begin{align}
	|\vec \alpha_\text{KK}|=\sqrt{\frac{D-2}{(D-d)(d-2)}},
\end{align}
is invariant under
\begin{align}
	d\leftrightarrow D-d+2.
\end{align}
It would be interesting if there were a deeper explanation for this symmetry of the formula between low and high dimensions.

\subsection{2d moduli spaces \label{s.classification.2d}}

I now list all frames of flat 2d moduli spaces where there is no sliding, in all spacetimes between 9d and 3d. With these, I list all possible lattice sites where the principal norm condition \eqref{e.principal_norm} is satisfied. I then solve the principal product conditions \eqref{e.principal_product}, and find all consistent sets of principal branes with respect to each duality frame. This classifies all principal branes, lattices, and tower/species polytopes of these theories.

 Some duality frames are prohibited from this analysis. Also, the resulting tower polytopes extend the 2d classification of \cite{Etheredge:2024tok} to 5d, 4d, and 3d theories, and also provides a rationale excluding some of the 2d tower polygons of 6d theories found in \cite{Etheredge:2024tok}. If sliding occurs between principal lattice sites, then this analysis also classifies all examples of this kind of sliding of 2d moduli spaces.

Each frame is specified by two vertices that form a frame simplex, where each generator is either a string oscillator mode or a KK-mode involving decompactification to at most 11-dimensions. In a $d$-dimensional theory, a geometric frame must be generated by a pair of KK-vertices such that the number of dimensions decompactified by each vertex does not sum to a number that exceeds $11-d$. This means that geometric frames cannot exist in theories of greater than 9 spacetime dimensions. Stringy frames can also be prohibited in theories of greater than 9 spacetime dimensions. This is because one cannot have 1d moduli spaces in 11d (by the analysis of the previous subsections), and thus 2d stringy frames are excluded in 10d (since decompactification along the KK-vertex direction to 11d would still leave a 11d theory with a dilaton). Thus, flat, noncompact, 2-dimensional moduli spaces exist in only 9-dimensions and below. The allowed frames are listed in Table \ref{t.2d_frames}.

\begin{table}[H]
$$\begin{array}{c|c|c|c|c|c|c|c|c|c|c|}
d& (D_1-d,D_2-d)\\ \hline
9&(1,1),(1,\infty)\\\hline
8&(1,1),(1,2),(1,\infty),(2,\infty)\\\hline
7&(1,1),(1,2),(1,3),(2,2),(1,\infty),(2,\infty),(3,\infty)\\\hline
6&(1,1),(1,2),(1,3),(1,4),(2,2),(2,3),(1,\infty),(2,\infty),(3,\infty),(4,\infty)\\\hline
5&\begin{matrix}
	(1,1),(1,2),(1,3),(1,4),(1,5),(2,2),(2,3),(2,4),(3,3)\\(1,\infty),(2,\infty),(3,3),(3,\infty),(4,\infty),(5,\infty)
\end{matrix}\\\hline
4&\begin{matrix}
	(1,1),(1,2),(1,3),(1,4),(1,5),(1,6),(2,2),(2,3),(2,4),(2,5),\\(3,3),(3,4),(1,\infty),(2,\infty),(3,\infty),(4,\infty),(5,\infty),(6,\infty)
\end{matrix}\\\hline
3&\begin{matrix}
	(1,1),(1,2),(1,3),(1,4),(1,5),(1,6),(1,7),(2,2),(2,3),(2,4),(2,5),(2,6),\\(3,3),(3,4),(3,5),(4,4),(1,\infty),(2,\infty),(3,\infty),(4,\infty),(5,\infty),(6,\infty),(7,\infty)
\end{matrix}\\\hline
\end{array}$$ 
\caption{Frame simplices for 2d moduli spaces. These simplices are labeled by the number of dimensions each vertex decompactifies, $(D_1-d,D_2-d)$, and string oscillators have $D_i-d=\infty$, by the radion-to-dilaton prescription.} \label{t.2d_frames}
\end{table}

I now classify all consistent choices of principal branes, tower/species polytopes, lattices, in all dimensions, for all of these frames.

\subsubsection{9d}

We begin with the 9d theory. From Table \ref{t.2d_frames}, there are just two frame simplices, which are the $(1,1)$ and $(1,\infty)$-frames, corresponding to either two KK-modes, or one KK-mode and one string oscillator mode. Given these two frames, the principal norm conditions can be solved by hand or with a computer program to give the principal norm candidates $p_{P_1,P_2}^n$, and they are listed in Table \ref{t.2d_9d_norm}. Here $p$ refers to the spacetime dimension of the brane, $P_1$ and $P_2$ refer to the coordinates of the brane with respect to the frame, and $n$ labels the number of dimensions that decompactify from the KK-modes that come from the winding modes of the brane.

\begin{table}[H]
$$\begin{array}{c|c|c|c|c|c|c|c|c|c|c|}
(D_1-d,D_2-d)&p_{P_1,P_2}^n \\ \hline
(1,1)&1_{0,1}^1,1_{1,0}^1,1_{2,2}^1,2_{2,3}^1,2_{3,2}^1,3_{3,3}^1,4_{5,5}^1,5_{5,6}^1,5_{6,5}^1,6_{6,6}^1,6_{7,8}^1,6_{8,7}^1,7_{7,9}^1,7_{9,7}^1\\\hline
(1,\infty)&\begin{matrix}
	1_{0,1}^1,1_{1,1}^{\infty },1_{1,2}^1,1_{2,1}^1,1_{2,2}^1,2_{2,2}^1,2_{2,3}^1,2_{3,3}^1,3_{3,4}^1,3_{4,4}^1,4_{4,5}^1,4_{5,5}^1,5_{5,6}^1,5_{6,6}^1,\\5_{6,7}^1,6_{6,7}^1,6_{6,8}^1,6_{7,7}^1,6_{7,8}^{\infty },6_{8,8}^1,7_{7,8}^1,7_{7,9}^{\infty },7_{8,8}^1,7_{8,10}^1,7_{9,9}^{\infty },7_{9,10}^1\end{matrix}\\\hline
\end{array}$$ 
\caption{Principal norm candidates $p^n_{P_1,P_2}$, with respect to frames $(D_1-d,D_2-d)$, for 2d moduli spaces of 9d theories. Here $p$ refers to the spacetime dimension of the brane, $P_1$ and $P_2$ refer to the coordinates of the brane with respect to the frame, and $n$ labels the number of dimensions of decompactification corresponding to the KK-mode that comes from the winding mode of the brane.} \label{t.2d_9d_norm}
\end{table}

Consider first the $(1,1)$-frame simplex generated by two KK-modes, where each KK-mode decompactifies to 10d. We can write the $\alpha$-vectors of these KK-modes using the basis
\begin{align}
	\vec v_1=\vec\alpha(1_{0,1}^1)=\sqrt{\frac87}\hat \rho,\qquad \vec v_2=\vec \alpha(1_{1,0}^1)=\frac 2{\sqrt{14}}\hat \rho +\frac{3}{2\sqrt 2}\hat \phi,
\end{align}
for some canonically normalized moduli $\hat \rho$ and $\hat \phi$ (which we will later recognize as being like the radion and dilaton of IIA on a circle). The superscripts $1$ on $1_{0,1}^1$ and $1_{1,0}^1$ are reminders that these KK-modes decompactify one dimension. As a check, the $1_{0,1}^1$ and $1_{1,0}^1$ towers satisfy the principal product condition \eqref{e.principal_product} with one another. Next, the $1_{2,2}^1$ tower and the oscillator towers from $2_{2,3}^1$ and $2_{3,2}^1$, which I denote $1_{1,3/2}^\infty$ and $1_{3/2,1}^\infty$, all satisfy the principal product conditions with $1_{0,1}^1$ and $1_{1,0}^1$. In fact, they all satisfy the product conditions with one another. This, as a result, produces exactly the same particle tower content as the radion-dilaton slice of IIA string theory on a circle (or, equivalently, the radion-radion slice of M-theory on a 2-torus, or the dilaton-radion slice of IIB on a circle)! That is, the $1_{0,1}^1$, $1_{1,0}^1$, $1_{2,2}^1$, $1_{1,3/2}^\infty $, and $1_{3/2,1}^\infty $ towers respectively correspond to, in the IIA on a circle perspective, KK-modes, D0-branes, fundamental string winding modes, fundamental string oscillator modes, and wrapped D2-brane oscillator modes.

In this example, all of the principal candidate branes in the $(1,1)$ column of Table \ref{t.2d_9d_norm} satisfy the principal product condition \eqref{e.principal_product} with each other, and are thus principal and thus required to exist. This matches exactly the radion-dilaton slice of IIA string theory on a circle (or, equivalently, the radion-radion slice of M-theory on a 2-torus, and also the dilaton-radion slice of IIB on a circle)!

\begin{table}[H]
$$\begin{array}{c|c|c|c|c|c|c|c|c|c|c|}
\text{principal branes}&\text{IIA on $S^1$} \\\hline
1_{0,1}^1&\text{KK mode}\\\hline 
1_{1,0}^1&\text{D0 brane}\\\hline
1_{2,2}^1&\text{fundamental string winding mode}\\\hline
2_{2,3}^1&\text{unwrapped fundamental string}\\\hline
2_{3,2}^1&\text{wrapped D2-brane}\\\hline
3_{3,3}^1&\text{unwrapped D2-brane}\\\hline
4_{5,5}^1&\text{wrapped D4-brane}\\\hline
5_{5,6}^1&\text{unwrapped D4-brane}\\\hline
5_{6,5}^1&\text{wrapped NS5-brane}\\\hline
6_{6,6}^1&\text{unwrapped NS5-brane}\\\hline
6_{7,8}^1&\text{wrapped D6-brane}\\\hline
6_{8,7}^1&\text{KK-monopole}\\\hline
7_{7,9}^1&\text{unwrapped D6-brane}\\\hline
7_{9,7}^1&\text{exotic 6-brane}\\\hline
\end{array}$$ 
\caption{Principal branes $p_{P_1,P_2}^n$ for the $(1,1)$-frame in 9d, and how they correspond to $\frac 12$-BPS branes of IIA string theory on a circle. Here $p$ refers to the spacetime dimension of the brane, $P_1$ and $P_2$ refer to the coordinates of the brane with respect to the frame, and $n$ labels the number of dimensions of decompactification corresponding to the KK-mode that comes from the winding mode of the brane.} \label{t.2d_11_princ}
\end{table}

\begin{figure}
\centering
\includegraphics{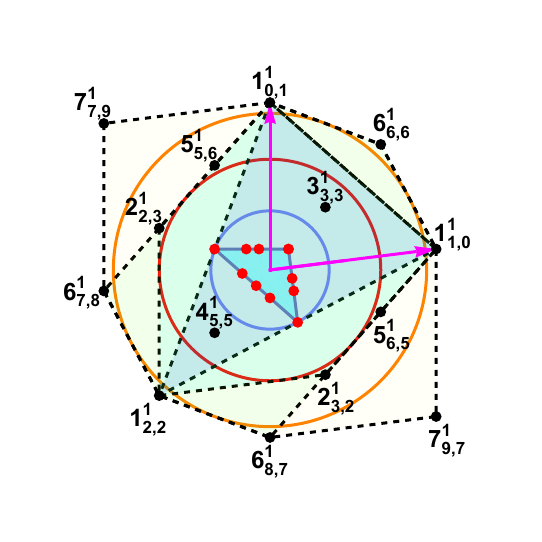}
\caption{Principal $\alpha$-vectors for the $(1,1)$-frame in 9d, labeled by $p_{P_1,P_2}^n$. Here $p$ refers to the spacetime dimension of the brane, $P_1$ and $P_2$ refer to the coordinates of the brane with respect to the frame, and $n$ labels the number of dimensions of decompactification corresponding to the KK-mode that comes from the winding mode of the brane. The magenta arrows point to the frame simplex generators $1_{0,1}^1$ and $1_{1,0}^1$. The circles of radius $1/\sqrt{d-2}$, $1/\sqrt{2}$, and $1$ are listed, and are contained by the relevant convex hulls, thus satisfying the Brane DC. The cyan polytope is the species polytope, and all of the $\vec\alpha_p/p$-vectors, labeled in red, lie on the boundary of the species polytope.}
\label{f.9d_class_11}
\end{figure}

These principal branes are plotted\footnote{Some of the string oscillator modes are not explicitly depicted, since they are implied from the strings which are plotted.} in Figure \ref{f.9d_class_11}. The species polytope is fixed by the tower polygon, and all of the $\alpha_p/p$-vectors lie on the boundary of the species polygon. The Brane DC \cite{Etheredge:2024amg} is also satisfied by this example.

It is worth emphasizing how extraordinary this is. By just starting with simple assumptions about the ESC and no sliding, we have completely recovered the detailed $\frac 12$-BPS brane content of IIA string theory on a circle!

Consider next the $(1,\infty)$-frame, corresponding to a stringy frame simplex generated by a KK-vertex and a string oscillator vertex. This frame simplex is generated by
\begin{align}
	\vec v_1=\vec\alpha(1_{0,1}^1)=\sqrt{\frac87}\hat \rho,\qquad \vec v_2=\vec \alpha(1_{1,1}^\infty)=\frac 1{2\sqrt{14}}\hat \rho -\frac{1}{2\sqrt 2}\hat \phi ,\label{e.10d_IIA_basis}
\end{align}
for some canonically normalized moduli $\hat \rho$ and $\hat \phi$ (which we will later recognize as behaving like the radion and dilaton of IIB on a circle). The superscripts $n=1$ and $n=\infty$ in \eqref{e.10d_IIA_basis} indicate that the $1_{1,0}^1$ results in decompactification of 1-dimension, and $1_{1,1}^\infty$ is a string oscillator (since, by the radion-to-dilaton prescription, emergent string limits formally behave in some ways like decompactification limits of infinitely many dimensions at once).  The $1_{0,1}^1$ and $1_{1,1}^\infty $ towers satisfy the principal product condition \eqref{e.principal_product} with one another, so this is a consistent frame simplex.

\begin{figure}
\centering
\includegraphics{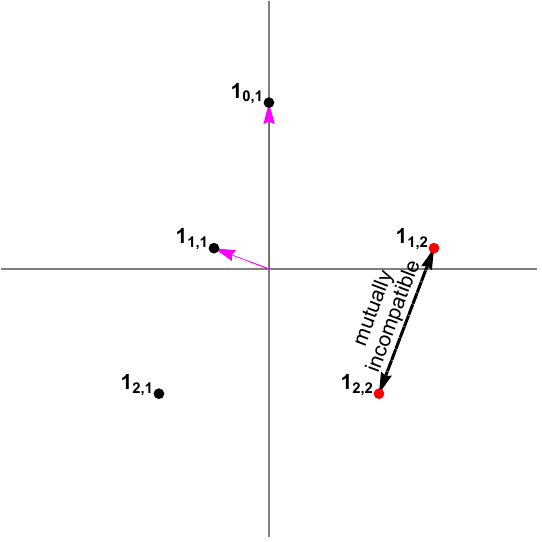}
\caption{Choices must be made for satisfying the principal product condition for the principal norm-candidate $\alpha$-vectors $p_{P_1,P_2}$ for the $(1,\infty)$-frame in 9d. The black dots are principal. The two red dots, $1_{1,2}^1$ and $1_{2,2}^1$, do not satisfy the principal product conditions with each other, and thus a choice must be made. Choosing the $1_{1,2}^1$ reproduces IIA string theory, which was depicted in Figure \ref{f.9d_class_11}, and also reproduces $\mathrm{E_8\times E_8}$ heterotic string theory on a circle with Wilson lines reducing the group to SO(16)$\times$SO(16). Choosing $1_{2,2}^1$ produces IIB string theory on a circle, and also heterotic SO(32) on a circle with Wilson lines reducing the group to SO(16)$\times$SO(16). Heterotic SO(32) and $\mathrm{E_8\times E_8}$ on a circle without Wilson lines involve sliding between $1_{1,2}^1$ and $1_{2,2}^1$.}
\label{f.9d_stringy_choices}
\end{figure}

Next, the other towers that satisfy the principal norm condition are the $1_{1,2}^1$, $1_{2,1}^1$, $1_{2,2}^1$ towers, and the oscillator modes of the $2_{2,2}^1 $, $2_{2,3}^1$, and $2_{3,3}^1$ strings, whose oscillators I denote $1_{1,1}^\infty $, $1_{1,3/2}^\infty$, and $1_{3/2,3/2}^\infty $. Despite the fact that all of the towers $1_{1,2}^1$, $1_{2,1}^1$, $1_{2,2}^1$, $1_{1,3/2}^\infty $, and $1_{3/2,3/2}^\infty $ satisfy the principal product condition \eqref{e.principal_product} with the $1_{0,1}^1$ and $1_{1,1}^\infty $ towers, they do not all satisfy the product conditions with each other. For instance, the $1_{1,2}^1$ and the $1_{2,2}^1$ states do not satisfy the principal product condition \eqref{e.principal_product} with each other, see Figure \ref{f.9d_stringy_choices} . Thus, choices must be made to find maximal sets of principal norm candidates that mutually satisfy the product condition, and all of these choices must contain $1_{0,1}^1$, $1_{1,1}^\infty $. All of these choices must also include the $1_{2,1}^1$ tower, since it satisfies the product condition with all of the other towers.

One choice is the set of towers $1_{0,1}^1$, $1_{1,1}^\infty $, $1_{1,3/2}^\infty $, $1_{2,1}^1$, $1_{2,2}^1$, and the other choice is the set of towers $1_{0,1}^1$, $1_{1,1}^\infty $, $1_{1,2}^1$, $1_{3/2,3/2}^\infty $, $1_{2,1}^1$. The choice with the point $1_{2,2}^1$ produces IIB string theory on a circle, and also heterotic SO(32) on a circle with Wilson lines reducing the group to SO(16)$\times$SO(16). The choice with the point $1_{1,2}^1$ reproduces IIA string theory, which was depicted in Figure \ref{f.9d_class_11}, and also reproduces $\mathrm{E_8\times E_8}$ heterotic string theory on a circle with Wilson lines reducing the group to SO(16)$\times$SO(16). The cases of heterotic SO(32) and $\mathrm{E_8\times E_8}$ on a circle involve sliding between the $1_{1,2}$ and $1_{2,2}$ points. 

\begin{figure}
\centering
\includegraphics{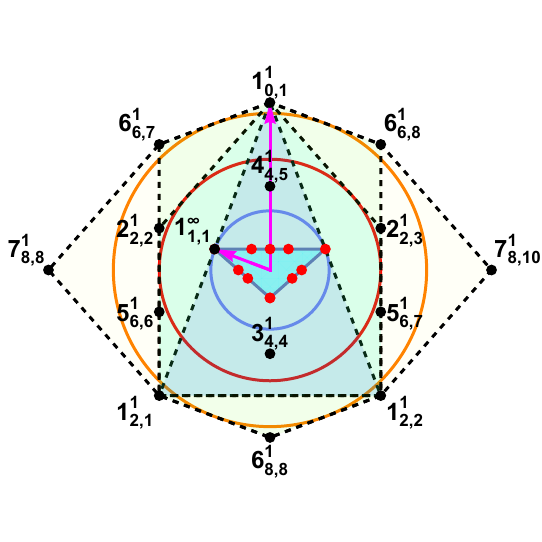}
\caption{Principal $\alpha$-vectors for the $(1,\infty )$-frame in 9d, labeled by $p_{P_1,P_2}^n$. $p$ refers to the spacetime dimension of the brane, $P_1$ and $P_2$ refer to the coordinates of the brane with respect to the frame, and $n$ labels the number of dimensions that decompactify from the KK-modes that come from the winding modes of the brane. The magenta arrows point to the frame simplex generators $1_{0,1}^1$ and $1_{1,1}^\infty $. The circles of radius $1/\sqrt{d-2}$, $1/\sqrt{2}$, and $1$ are listed, and are contained by the relevant convex hulls, thus satisfying the Brane DC. The cyan polytope is the species polytope, and all of the $\vec\alpha_p/p$-principal branes, labeled in red, lie on the boundary of the species polytope.}
\label{f.9d_class_1i}
\end{figure}

For each choice of principal towers, one can consider next principal norm candidates of branes in Table \ref{t.2d_9d_norm} and test the principal product condition \eqref{e.principal_product}. In the end, the $(1,\infty)$-frame allows for only two sets of principal branes,
\begin{align}
	\text{IIA-like:}&\qquad \begin{matrix}1_{0,1},1_{1,1},1_{1,2},1_{3/2,3/2},1_{2,1},2_{2,2},2_{3,3},3_{3,4},4_{5,5},\\5_{5,6},5_{6,7},6_{6,8},6_{7,7},6_{8,8},7_{7,8},7_{9,10}\end{matrix},\label{e.9d_str_IIA}\\
	\text{IIB-like:}&\qquad \begin{matrix}1_{0,1},1_{1,1},1_{1,3/2},1_{2,1},1_{2,2},2_{2,2},2_{2,3},3_{4,4},4_{4,5},\\5_{6,6},4_{6,7},6_{6,7},6_{6,8},6_{7,8},6_{8,8},7_{8,8},7_{8,10}\end{matrix}.\label{e.9d_str_IIB}
\end{align}
The first choice reproduces the principal branes determined earlier with respect to the $(1,1)$-frame, and thus behaves like IIA on a circle, and reproduces the branes in Figure \ref{f.9d_class_11} (albeit with respect to a different frame). The second choice produces a set of branes exactly like IIB string theory on a circle, and the $\alpha$-vectors are depicted in Figure \ref{f.9d_class_1i}.

The two theories \eqref{e.9d_str_IIA} and \eqref{e.9d_str_IIB} are actually equivalent. Figures \ref{f.9d_class_11} and \ref{f.9d_class_1i} are rotations of each other. Thus, there is a unique 9d theory, listed in Table \ref{t.2d_9d_theories} and depicted in both Figures \ref{f.9d_class_11} and \ref{f.9d_class_1i}. This analysis also fixes the species polytope, depicted in cyan in both plots. All of the principal branes have $\alpha_p/p$-vectors that lie on the boundary of the species polytope.

\begin{table}[H]
$$\begin{array}{c|c|c|c|c|c|c|c|c|c|c|}
(D_1-d,D_2-d)& p_{P_1,P_2}^n\\ \hline
(1,1)& 1_0^1,1_2^2,2_2^2,3_4^1,4_4^1,5_6^2,6_6^2,6_8^1\\\hline
\end{array}$$ 
\caption{The unique 9d theory with a 2d moduli space, labeled by its frame and which principal branes $p_{P_1,P_2}^n$ it possesses. $p$ refers to the spacetime dimension of the brane, $P_1$ and $P_2$ refer to the coordinates of the brane with respect to the frame, and $n$ labels the number of dimensions that decompactify from the KK-modes that come from the winding modes of the brane. This reproduces IIA and IIB string theory on a circle. The branes are depicted in Figure \ref{f.9d_class_11}.} \label{t.2d_9d_theories}
\end{table}

We can compare these theories with $\mathrm{E_8\times E_8}$ or SO(32) heterotic string theory on a circle with a Wilson line that yields an SO(16)$\times$SO(16) gauge group. As discussed in \cite{Etheredge:2024amg}, the known branes are not enough to satisfy the codimension-2 Brane DC, and new non-BPS branes are needed. It is known from \cite{Etheredge:2023odp} that, at least the particle towers, in all of the duality frames of this theory do not slide (but it is not known whether the higher-dimensional branes slide under dimensional reduction). If this ESC analysis applies to this theory, then heterotic SO(32) string theory needs a set of branes with D-odd-brane tensions, just like those of IIB string theory, and $\mathrm {E_8\times E_8}$ requires a set of branes with D-even-brane tensions, just like in IIA string theory. Additionally, since SO(32) theory would require a D-brane content like IIB's, it is natural to suspect that the argument could be extrapolated to argue that the SO(32) theory requires a D-brane like instanton, and this would further motivate the non-BPS heterotic instanton proposed in \cite{Alvarez-Garcia:2024vnr}. Examining further this heterotic example will be the subject of further work.

\subsubsection{8d}

The solutions to the  principal norm condition \eqref{e.principal_norm} are listed in Table \ref{t.2d_8d_norm} for all of the allowed frames of 2d moduli spaces of 8d theories.

\begin{table}[H]
$$\begin{array}{c|c|c|c|c|c|c|c|c|c|c|}
(D_1-d,D_2-d)& p_{P_1,P_2}^n\\ \hline
(1,1)&\begin{matrix}
	1_{0,1}^1,1_{1,0}^1,1_{1,2}^2,1_{2,1}^2,2_{2,2}^2,2_{3,3}^1,3_{3,4}^1,3_{4,3}^1,4_{4,4}^1,4_{5,5}^2,5_{5,6}^2,5_{6,5}^2,5_{6,7}^1,\\5_{7,6}^1,6_{6,6}^2,6_{6,8}^1,6_{8,6}^1,6_{8,8}^2
\end{matrix}\\\hline
(1,2)&\begin{matrix}
	1_{0,1}^1,1_{1,0}^2,1_{1,2}^{18},1_{1,3}^1,1_{2,2}^2,2_{2,3}^2,2_{3,2}^1,3_{3,3}^1,3_{4,5}^1,4_{4,6}^1,4_{5,5}^2,5_{5,6}^2,5_{6,5}^1,\\5_{6,6}^{18},5_{6,8}^2,5_{7,7}^1,6_{6,6}^1,6_{6,7}^{18},6_{6,9}^2,6_{8,7}^2,6_{8,9}^{18},6_{8,10}^1
\end{matrix}\\\hline
(1,\infty)&
	1_{0,1}^1,1_{1,1}^{\infty },1_{2,1}^1,2_{2,2}^1,4_{5,6}^1,5_{5,7}^1,5_{6,7}^{\infty },5_{7,7}^1,6_{6,8}^{\infty },6_{8,8}^{\infty }\\\hline
(2,\infty)&\begin{matrix}
	1_{0,1}^2,1_{1,1}^{\infty },1_{1,2}^1,1_{2,1}^2,1_{2,2}^2,1_{3,2}^1,2_{2,2}^1,2_{2,3}^1,2_{3,3}^2,\\2_{4,3}^1,3_{3,4}^1,3_{4,4}^2,3_{5,4}^1,4_{4,5}^1,4_{5,5}^2,4_{6,5}^1,4_{6,6}^1,5_{5,6}^1,5_{6,6}^2,5_{6,7}^2,5_{7,6}^1,\\5_{7,7}^{\infty },5_{8,7}^2,6_{6,7}^1,6_{6,8}^1,6_{7,7}^2,6_{8,7}^1,6_{8,9}^1,6_{9,9}^2,6_{10,8}^1,6_{10,9}^1
\end{matrix}\\\hline
\end{array}$$ 
\caption{Principal norm candidates $p^n_{P_1,P_2}$, with respect to frames $(D_1-d,D_2-d)$, for 2d moduli spaces of 8d theories. $p$ refers to the spacetime dimension of the brane, $P_1$ and $P_2$ refer to the coordinates of the brane with respect to the frame, and $n$ labels the number of dimensions that decompactify from the KK-modes that come from the winding modes of the brane.} \label{t.2d_8d_norm}
\end{table}

\begin{figure}
\centering
\begin{subfigure}{.49\linewidth}
\centering
\includegraphics[scale=.75]{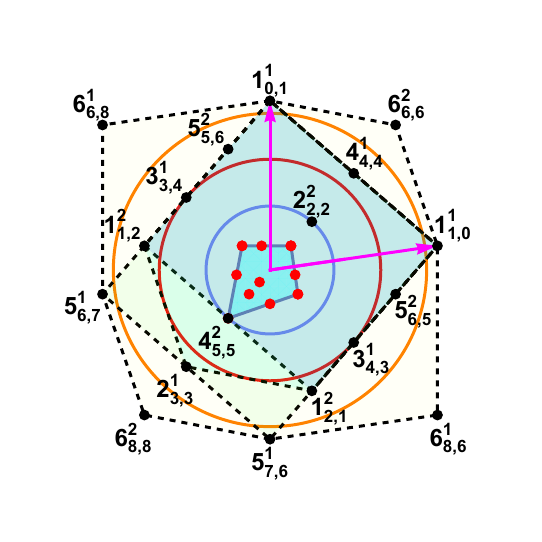}
\caption{$(1,1)$-frame generated by $1_{0,1}^1$ and $1_{1,0}^1$. }
\label{f.8d_class_11}
\end{subfigure}
\begin{subfigure}{.49\linewidth}
\centering
\includegraphics[scale=.75]{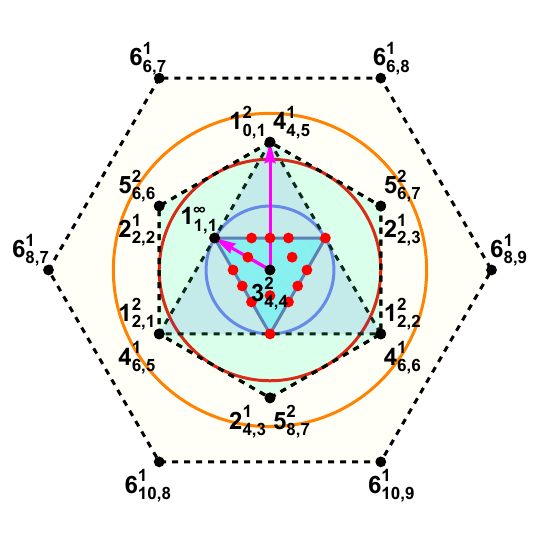}
\caption{$(2,\infty)$-frame generated by $1_{0,1}^2$ and $1_{1,1}^\infty $.}
\label{f.8d_class_2i}
\end{subfigure}
\caption{Principal $\alpha$-vectors for 2d frames in 8d, labeled by $p_{P_1,P_2}^n$. $p$ refers to the spacetime dimension of the brane, $P_1$ and $P_2$ refer to the coordinates of the brane with respect to the frame, and $n$ labels the number of dimensions that decompactify from the KK-modes that come from the winding modes of the brane. The magenta arrows point to the frame simplex vectors. The circles of radius $1/\sqrt{d-2}$, $1/\sqrt{2}$, and $1$ are listed, and are contained by the relevant convex hulls, thus satisfying the brane DC. All of the $\vec\alpha_p/p$-vectors, labeled in red, lie either within or on the boundary of the species polytope, labeled in cyan. }
\label{f.8d_class}
\end{figure}

The $(1,1)$-frame uniquely produces the following principal branes $p_{P_1,P_2}^n$,
\begin{align}
	(1,1):\qquad \begin{matrix}
		1_{0,1}^1,1_{1,0}^1,1_{1,2}^2,1_{2,1}^2,2_{2,2}^2,2_{3,3}^1,3_{3,4}^1,3_{4,3}^1,4_{4,4}^1,4_{5,5}^2,5_{5,6}^2,\\5_{6,5}^2,5_{6,7}^1,5_{7,6}^1,6_{6,6}^2,6_{6,8}^1,6_{8,6}^1,6_{8,8}^2
	\end{matrix},\label{e.8d_class_11}
\end{align}
and they are depicted in Figure \ref{f.8d_class_11}. This matches the brane content of the volume-dilaton modulus of IIA string theory on a 2-torus. The $(1,2)$ frame produces two potential theories, where one is the one described with the $(1,1)$-frame, but the other theory can be eliminated since its tower polytope uses the $1_{1,2}^{18}$ entry in Table \ref{t.2d_8d_norm}, which would involve decompactification to 26 dimensions. The $(1,\infty)$-frame is prohibited, since all of the particles are co-linear. There exist exactly two choices of particle towers allowed with the $(2,\infty)$-frame. One of these choices reproduces the \eqref{e.8d_class_11} theory (see the $(2,\infty)$ facet in Figure \ref{f.8d_class_11}), so let us consider the novel choice. This novel choice has the particle towers creating a triangular convex hull, with emergent string limits at the center of each of the three faces. All principal norm candidates must be principal norm candidates with respect to all three facets, and this eliminates some states, such as $2_{3,3}^2$, $4_{5,5}^2$, $5_{6,5}^1$, and $5_{7,6}^2$. As a result, the $(2,\infty)$-frame produces a theory with the following principal branes $p_{P_1,P_2}^n$,
\begin{align}
	(2,\infty):\qquad \begin{matrix}
		1_{0,1}^2,1_{1,1}^{\infty },1_{2,1}^2,1_{2,2}^2,2_{2,2}^1,2_{2,3}^1,2_{4,3}^1,3_{4,4}^2,4_{4,5}^1,4_{6,5}^1,\\4_{6,6}^1,5_{6,6}^2,5_{6,7}^2,5_{8,7}^2,6_{6,7}^1,6_{6,8}^1,6_{8,7}^1,6_{8,9}^1,6_{10,8}^1,6_{10,9}^1
	\end{matrix},\label{e.8d_class_2i}
\end{align}
and they are depicted in Figure \ref{f.8d_class_2i}. This matches exactly brane content of the volume-dilaton modulus of IIB string theory on a 2-torus, (c.f., Figure 3(b) of \cite{Etheredge:2024amg}). Thus, there are only two unique theories consistent with this analysis, listed in Table \ref{t.2d_8d_theories} and depicted in Figure \ref{f.8d_class}. These exactly match the radion-dilaton profiles of IIA and IIB string theory on tori. The tower and species polygons also exactly match the analysis in \cite{Etheredge:2024tok}.

\begin{table}[H]
$$\begin{array}{c|c|c|c|c|c|c|c|c|c|c|}
(D_1-d,D_2-d)&p_{P_1,P_2}^n \\ \hline
(1,1)& \begin{matrix}
	1_{0,1}^1,1_{1,0}^1,1_{1,2}^2,1_{2,1}^2,2_{2,2}^2,2_{3,3}^1,3_{3,4}^1,3_{4,3}^1,4_{4,4}^1,4_{5,5}^2,5_{5,6}^2,5_{6,5}^2,5_{6,7}^1,\\5_{7,6}^1,6_{6,6}^2,6_{6,8}^1,6_{8,6}^1,6_{8,8}^2
\end{matrix}\\\hline
(2,\infty )& \begin{matrix}
		1_{0,1}^2,1_{1,1}^{\infty },1_{2,1}^2,1_{2,2}^2,2_{2,2}^1,2_{2,3}^1,2_{4,3}^1,3_{4,4}^2,4_{4,5}^1,4_{6,5}^1,\\4_{6,6}^1,5_{6,6}^2,5_{6,7}^2,5_{8,7}^2,6_{6,7}^1,6_{6,8}^1,6_{8,7}^1,6_{8,9}^1,6_{10,8}^1,6_{10,9}^1
	\end{matrix}\\\hline
\end{array}$$ 
\caption{ 8d theories with 2d moduli spaces, labeled by their frame simplices and which principal branes $p_{P_1,P_2}^n$ they possess. $p$ refers to the spacetime dimension of the brane, $P_1$ and $P_2$ refer to the coordinates of the brane with respect to the frame, and $n$ labels the number of dimensions that decompactify from the KK-modes that come from the winding modes of the brane.    The branes are depicted in Figure \ref{f.8d_class}.} \label{t.2d_8d_theories}
\end{table}

\subsubsection{7d}

The solutions to the  principal norm condition \eqref{e.principal_norm} are listed in Table \ref{t.2d_6d_norm} for all of the allowed frames of 2d moduli spaces of 6d theories.

\clearpage
\begin{table}[H]
$$\begin{array}{c|c|c|c|c|c|c|c|c|c|c|}
(D_1-d,D_2-d)&p_{P_1,P_2}^n \\ \hline
(1,1)&\begin{matrix}
	1_{0,1}^1,1_{1,0}^1,4_{5,6}^1,4_{6,5}^1,5_{5,7}^1,5_{7,5}^1
	\end{matrix}\\\hline 
(1,2)&\begin{matrix}
	1_{0,1}^1,1_{1,0}^2,1_{1,2}^{\infty },1_{2,1}^2,1_{2,3}^1,2_{2,2}^2,2_{2,4}^1,2_{3,3}^2,3_{3,4}^2,3_{4,3}^1,3_{4,5}^2,4_{4,4}^1,4_{4,6}^2,4_{5,5}^{\infty },\\4_{5,7}^2,4_{6,6}^1,5_{5,6}^{\infty },5_{5,8}^2,5_{6,5}^2,5_{6,9}^2,5_{7,6}^2,5_{7,8}^{\infty }
\end{matrix}\\\hline
(1,3)&\begin{matrix}
	1_{0,1}^1,1_{1,0}^3,1_{1,3}^3,1_{2,2}^3,2_{2,3}^3,2_{3,2}^1,2_{3,5}^1,3_{3,3}^1,3_{3,6}^1,3_{4,5}^3,4_{4,6}^3,\\4_{5,5}^3,4_{5,8}^3,4_{6,7}^1,5_{5,6}^3,5_{5,9}^3,5_{6,5}^1,5_{6,11}^1,5_{7,7}^3,5_{7,10}^3
\end{matrix}\\\hline
(2,2)&\begin{matrix}
	1_{0,1}^2,1_{1,0}^2,1_{1,2}^{18},1_{1,3}^1,1_{2,1}^{18},1_{2,2}^{\infty },1_{3,1}^1,2_{2,3}^2,2_{3,2}^2,2_{4,4}^1,3_{3,3}^1,3_{4,5}^2,3_{5,4}^2,4_{4,6}^1,\\4_{5,5}^{\infty },4_{5,6}^{18},4_{6,4}^1,4_{6,5}^{18},4_{6,7}^2,4_{7,6}^2,5_{5,6}^2,5_{6,5}^2,5_{6,8}^{\infty },5_{8,6}^{\infty },5_{8,9}^2,5_{9,8}^2
\end{matrix}\\\hline
(1,\infty)&\begin{matrix}
	1_{0,1}^1,1_{1,1}^{\infty },1_{1,2}^2,1_{2,1}^1,1_{2,2}^2,2_{2,2}^1,2_{2,3}^2,2_{3,3}^2,3_{3,4}^2,3_{4,4}^2,3_{4,5}^1,4_{4,5}^2,4_{4,6}^1,4_{5,5}^2,\\4_{5,6}^{\infty },4_{6,6}^1,5_{5,6}^2,5_{5,7}^{\infty },5_{6,6}^2,5_{6,8}^2,5_{7,7}^{\infty },5_{7,8}^2
\end{matrix}\\\hline
(2,\infty)&\begin{matrix}
	1_{0,1}^2,1_{1,1}^{\infty },1_{2,1}^2,1_{2,2}^4,2_{2,2}^1,2_{3,3}^4,3_{4,4}^4,3_{5,5}^1,4_{5,5}^4,4_{5,6}^2,4_{6,6}^{\infty },4_{7,6}^2,\\5_{5,7}^1,5_{6,6}^4,5_{8,8}^4,5_{9,7}^1
\end{matrix}\\\hline
(3,\infty)&\begin{matrix}
	1_{0,1}^3,1_{1,1}^{\infty },1_{1,2}^1,1_{2,1}^3,1_{2,2}^3,1_{3,2}^3,1_{4,2}^1,2_{2,2}^1,2_{2,3}^1,2_{3,3}^3,2_{4,3}^3,2_{5,3}^1,3_{3,4}^1,\\3_{4,4}^3,3_{5,4}^3,3_{6,4}^1,3_{6,5}^1,4_{4,5}^1,4_{5,5}^3,4_{6,5}^3,4_{6,6}^3,4_{7,5}^1,4_{7,6}^{\infty },\\4_{8,6}^3,5_{5,6}^1,5_{6,6}^3,5_{6,7}^3,5_{7,6}^3,5_{8,6}^1,5_{8,8}^1,5_{9,8}^3,5_{10,7}^3,5_{10,8}^3,5_{11,8}^1
\end{matrix}\\\hline
\end{array}$$ 
\caption{Principal norm candidates $p^n_{P_1,P_2}$, with respect to frames $(D_1-d,D_2-d)$, for 2d moduli spaces of 7d theories. $p$ refers to the spacetime dimension of the brane, $P_1$ and $P_2$ refer to the coordinates of the brane with respect to the frame, and $n$ labels the number of dimensions that decompactify from the KK-modes that come from the winding modes of the brane.} \label{t.2d_7d_norm}
\end{table}

\begin{figure}
\centering
\begin{subfigure}{.49\linewidth}
\centering
\includegraphics[scale=.75]{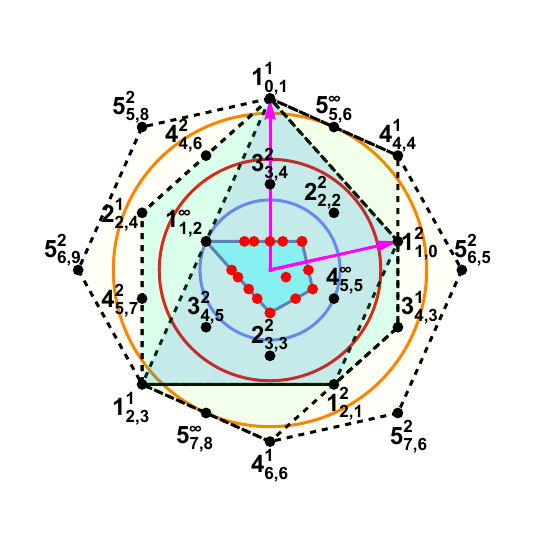}
\caption{$(1,2)$-frame, generated by $1_{0,1}^1$ and $1_{1,0}^2$. }
\label{f.7d_class_12}
\end{subfigure}
\begin{subfigure}{.49\linewidth}
\centering
\includegraphics[scale=.75]{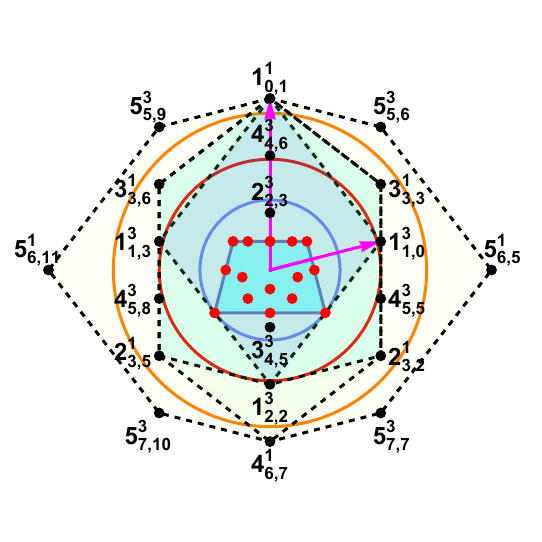}
\caption{$(1,3)$-frame, generated by $1_{0,1}^2$ and $1_{1,0}^3 $.}
\label{f.7d_class_13}
\end{subfigure}
\caption{Principal $\alpha$-vectors for 2d frames in 7d, labeled by $p_{P_1,P_2}^n$. $p$ refers to the spacetime dimension of the brane, $P_1$ and $P_2$ refer to the coordinates of the brane with respect to the frame, and $n$ labels the number of dimensions that decompactify from the KK-modes that come from the winding modes of the brane. The magenta arrows point to the frame simplex vectors. The circles of radius $1/\sqrt{d-2}$, $1/\sqrt{2}$, and $1$ are listed, and are contained by the relevant convex hulls, thus satisfying the Brane DC. All of the $\vec\alpha_p/p$-vectors, labeled in red, lie either within or on the boundary of the species polytope, labeled in cyan.}
\label{f.7d_class}
\end{figure}

With the $(1,1)$-frame, there are no allowed sets of principal towers that satisfy the Sharpened DC, since there are only two principal towers. The $(1,2)$-frame uniquely produces the following principal branes $p_{P_1,P_2}^n$,
\begin{align}
	(1,2):\qquad \begin{matrix}
		1_{0,1}^1,1_{1,0}^2,1_{1,2}^{\infty },1_{2,1}^2,1_{2,3}^1,2_{2,2}^2,2_{2,4}^1,2_{3,3}^2,3_{3,4}^2,3_{4,3}^1,3_{4,5}^2,4_{4,4}^1,4_{4,6}^2,\\4_{5,5}^{\infty },4_{5,7}^2,4_{6,6}^1,5_{5,6}^{\infty },5_{5,8}^2,5_{6,5}^2,5_{6,9}^2,5_{7,6}^2,5_{7,8}^{\infty }
	\end{matrix},\label{e.7d_class_12}
\end{align}
and they are depicted in Figure \ref{f.7d_class_12}. The $(1,4)$-frame uniquely produces the following principal branes $p_{P_1,P_2}^n$,
\begin{align}
	(1,3):\qquad \begin{matrix}
		1_{0,1}^1,1_{1,0}^3,1_{1,3}^3,1_{2,2}^3,2_{2,3}^3,2_{3,2}^1,2_{3,5}^1,3_{3,3}^1,3_{3,6}^1,3_{4,5}^3,4_{4,6}^3,\\4_{5,5}^3,4_{5,8}^3,4_{6,7}^1,5_{5,6}^3,5_{5,9}^3,5_{6,5}^1,5_{6,11}^1,5_{7,7}^3,5_{7,10}^3
	\end{matrix},\label{e.7d_class_13}
\end{align}
and they are depicted in Figure \ref{f.7d_class_13}. These towers and branes exactly match the radion-dilaton profile of IIB string theory on a 3-torus (c.f., Figure 4(a) of \cite{Etheredge:2024amg}). There is a unique theory with the $(2,2)$-frame, but it is the same as the theory found with the $(1,2)$-frame \eqref{e.7d_class_12} (the $(2,2)$ facet appears in the tower polygon depicted in Figure \ref{f.7d_class_12}). Consider next the $(1,\infty)$-frame. There is a unique theory with this frame, and it is the same as the theory found with the $(1,2)$-frame \eqref{e.7d_class_12} (the the $(1,\infty)$ facet appears in this tower polygon depicted in Figure \ref{f.7d_class_12}). The $(2,\infty)$ frame is eliminated because the Sharpened DC is not satisfied by principal towers that satisfy the product condition \eqref{e.principal_product}. Thus, there are only two theories consistent with this analysis, listed in Table \ref{t.2d_7d_theories} depicted in Figure \ref{f.7d_class}. 

\begin{table}[H]
$$\begin{array}{c|c|c|c|c|c|c|c|c|c|c|}
(D_1-d,D_2-d)&p_{P_1,P_2}^n \\ \hline
(1,2)& \begin{matrix}
	1_{0,1}^1,1_{1,0}^2,1_{1,2}^{\infty },1_{2,1}^2,1_{2,3}^1,2_{2,2}^2,2_{2,4}^1,2_{3,3}^2,3_{3,4}^2,3_{4,3}^1,3_{4,5}^2,4_{4,4}^1,4_{4,6}^2,\\4_{5,5}^{\infty },4_{5,7}^2,4_{6,6}^1,5_{5,6}^{\infty },5_{5,8}^2,5_{6,5}^2,5_{6,9}^2,5_{7,6}^2,5_{7,8}^{\infty }
\end{matrix}\\\hline
(1,3 )& \begin{matrix}
	1_{0,1}^1,1_{1,0}^3,1_{1,3}^3,1_{2,2}^3,2_{2,3}^3,2_{3,2}^1,2_{3,5}^1,3_{3,3}^1,3_{3,6}^1,3_{4,5}^3,4_{4,6}^3,4_{5,5}^3,4_{5,8}^3,\\4_{6,7}^1,5_{5,6}^3,5_{5,9}^3,5_{6,5}^1,5_{6,11}^1,5_{7,7}^3,5_{7,10}^3
	\end{matrix}\\\hline
\end{array}$$ 
\caption{ 7d theories with 2d moduli spaces, labeled by their frame simplices and which principal branes $p_{P_1,P_2}^n$ they possess. $p$ refers to the spacetime dimension of the brane, $P_1$ and $P_2$ refer to the coordinates of the brane with respect to the frame, and $n$ labels the number of dimensions that decompactify from the KK-modes that come from the winding modes of the brane.    The branes are depicted in Figure \ref{f.7d_class}.} \label{t.2d_7d_theories}
\end{table}

The tower polytopes and species scales here agree with those found in \cite{Etheredge:2024tok}, though the brane analysis is novel. As discussed in \cite{Etheredge:2024tok}, these polytopes appear in maximal supergravity.

\subsubsection{6d}

The solutions to the  principal norm condition \eqref{e.principal_norm} are listed in Table \ref{t.2d_6d_norm} for all of the allowed frames of 2d moduli spaces of 6d theories.

\clearpage
\begin{table}[H]
$$\begin{array}{c|c|c|c|c|c|c|c|c|c|c|}
(D_1-d,D_2-d)& p_{P_1,P_2}^n \\ \hline
(1,1)&\begin{matrix}
1_{0,1}^1,1_{1,0}^1,1_{1,2}^3,1_{2,1}^3,1_{2,2}^2,2_{2,2}^3,2_{2,3}^2,2_{3,2}^2,2_{3,3}^3,3_{3,3}^2,3_{3,4}^3,3_{4,3}^3,3_{4,5}^1,3_{5,4}^1,\\4_{4,4}^3,4_{4,6}^1,4_{6,4}^1,4_{6,6}^3
\end{matrix}\\\hline 
(1,2)&\begin{matrix}
	1_{0,1}^1,1_{1,0}^2,3_{4,6}^2,3_{5,5}^1,4_{4,7}^2,4_{6,5}^2
\end{matrix}\\\hline
(1,3)&\begin{matrix}
	1_{0,1}^1,1_{1,0}^3,1_{1,4}^1,1_{2,1}^2,1_{2,3}^3,2_{2,2}^2,2_{2,4}^3,2_{3,3}^3,2_{3,5}^2,\\3_{3,4}^3,3_{3,6}^2,3_{4,3}^1,3_{4,7}^3,3_{5,6}^1,4_{4,4}^1,4_{4,8}^3,4_{6,6}^3,4_{6,10}^1
\end{matrix}\\\hline
(1,4)&\begin{matrix}
	1_{0,1}^1,1_{1,0}^4,1_{1,3}^{\infty },1_{2,2}^4,1_{2,5}^1,2_{2,3}^4,2_{2,6}^1,2_{3,2}^1,2_{3,5}^4,3_{3,3}^1,3_{3,6}^4,3_{4,5}^{\infty },3_{4,8}^4,\\3_{5,7}^1,4_{4,6}^{\infty },4_{4,9}^4,4_{5,5}^4,4_{5,11}^4,4_{6,7}^4,4_{6,10}^{\infty }\end{matrix}\\\hline
(2,2)&\begin{matrix}
	1_{0,1}^2,1_{1,0}^2,1_{1,2}^{\infty },1_{2,1}^{\infty },1_{2,3}^2,1_{3,2}^2,2_{2,2}^2,2_{2,4}^1,2_{3,3}^{\infty },2_{4,2}^1,2_{4,4}^2,3_{3,4}^2,3_{4,3}^2,3_{4,5}^{\infty },\\3_{5,4}^{\infty },3_{5,6}^2,3_{6,5}^2,4_{4,4}^1,4_{4,6}^2,4_{5,7}^{\infty },4_{6,4}^2,4_{6,8}^2,4_{7,5}^{\infty },4_{8,6}^2,4_{8,8}^1
\end{matrix}\\\hline
(2,3)&\begin{matrix}
	1_{0,1}^2,1_{1,0}^3,1_{1,3}^3,1_{2,1}^{18},1_{2,3}^{18},1_{3,1}^1,1_{3,4}^1,2_{2,3}^3,2_{3,2}^2,2_{3,3}^{18},2_{3,4}^{18},2_{3,5}^2,2_{4,4}^3,\\3_{3,3}^1,3_{3,6}^1,3_{4,4}^{18},3_{4,6}^{18},3_{5,4}^3,3_{5,7}^3,3_{6,6}^2,4_{4,6}^3,4_{6,4}^1,4_{6,10}^1,4_{8,8}^3
\end{matrix}\\\hline
(1,\infty)&\begin{matrix}
	1_{0,1}^1,1_{1,1}^{\infty },1_{1,2}^4,1_{2,1}^1,1_{2,2}^4,2_{2,2}^1,2_{2,3}^4,2_{3,3}^4,2_{3,4}^1,3_{3,4}^4,3_{3,5}^1,3_{4,4}^4,3_{4,5}^{\infty },\\3_{5,5}^1,4_{4,5}^4,4_{4,6}^{\infty },4_{5,5}^4,4_{5,7}^4,4_{6,6}^{\infty },4_{6,7}^4
\end{matrix}\\\hline
(2,\infty)&\begin{matrix}
	1_{0,1}^2,1_{1,1}^{\infty },1_{1,2}^2,1_{2,1}^2,1_{2,2}^{\infty },1_{3,2}^2,2_{2,2}^1,2_{2,3}^2,2_{3,3}^{\infty },2_{4,3}^2,2_{4,4}^1,3_{3,4}^2,3_{4,4}^{\infty },\\3_{4,5}^2,3_{5,4}^2,3_{5,5}^{\infty },3_{6,5}^2,4_{4,5}^2,4_{4,6}^1,4_{5,5}^{\infty },4_{6,5}^2,4_{6,7}^2,4_{7,7}^{\infty },4_{8,6}^1,4_{8,7}^2
\end{matrix}\\\hline
(3,\infty)&\begin{matrix}
	1_{0,1}^3,1_{1,1}^{\infty },1_{2,1}^3,1_{2,2}^{12},1_{3,2}^{12},2_{2,2}^1,2_{3,3}^{12},2_{4,3}^{12},2_{5,4}^1,3_{4,4}^{12},3_{5,4}^{12},\\3_{5,5}^3,3_{6,5}^{\infty },3_{7,5}^3,4_{5,5}^{12},4_{5,6}^3,4_{6,5}^{12},4_{8,7}^{12},4_{9,6}^3,4_{9,7}^{12}
\end{matrix}\\\hline
(4,\infty)&\begin{matrix}
	1_{0,1}^4,1_{1,1}^{\infty },1_{1,2}^1,1_{2,1}^4,1_{2,2}^4,1_{3,1}^1,1_{3,2}^{\infty },1_{4,2}^4,1_{5,2}^1,2_{2,2}^1,2_{2,3}^1,2_{3,3}^4,2_{4,3}^{\infty },\\2_{5,3}^4,2_{6,3}^1,2_{6,4}^1,3_{3,4}^1,3_{4,4}^4,3_{5,4}^{\infty },3_{5,5}^1,3_{6,4}^4,3_{6,5}^4,3_{7,4}^1,3_{7,5}^{\infty },3_{8,5}^4,3_{9,5}^1,\\4_{4,5}^1,4_{5,5}^4,4_{6,5}^{\infty },4_{6,6}^{\infty },4_{7,5}^4,4_{8,5}^1,4_{8,7}^1,4_{9,7}^4,4_{10,6}^{\infty },4_{10,7}^{\infty },4_{11,7}^4,4_{12,7}^1
\end{matrix}\\\hline
\end{array}$$ 
\caption{Principal norm candidates $p^n_{P_1,P_2}$, with respect to frames $(D_1-d,D_2-d)$, for 2d moduli spaces of 6d theories. $p$ refers to the spacetime dimension of the brane, $P_1$ and $P_2$ refer to the coordinates of the brane with respect to the frame, and $n$ labels the number of dimensions that decompactify from the KK-modes that come from the winding modes of the brane.} \label{t.2d_6d_norm}
\end{table}

\begin{figure}
\centering
\begin{subfigure}{.49\linewidth}
\centering
\includegraphics[scale=.75]{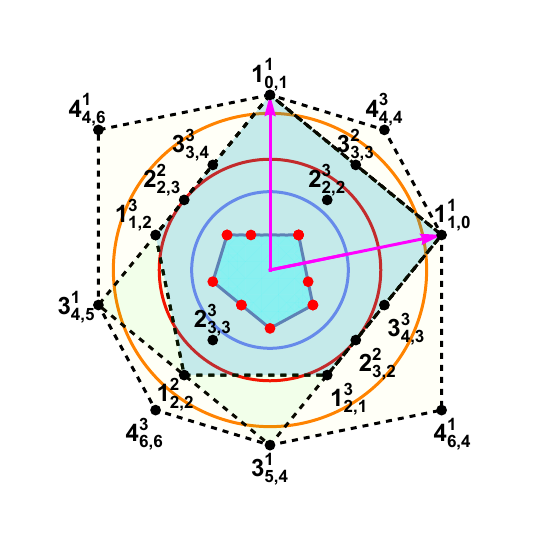}
\caption{$(1,1)$-frame, generated by $1_{0,1}^1$ and $1_{1,0}^1$. }
\label{f.6d_class_11}
\end{subfigure}
\begin{subfigure}{.49\linewidth}
\centering
\includegraphics[scale=.75]{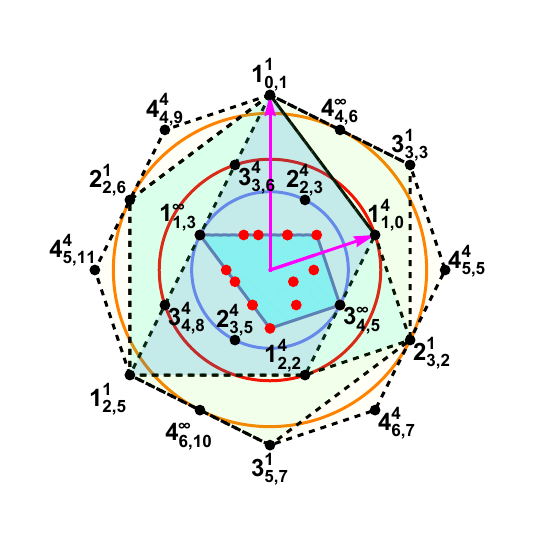}
\caption{$(1,4)$-frame, generated by $1_{0,1}^1$ and $1_{1,0}^4$. }
\label{f.6d_class_14}
\end{subfigure}
\begin{subfigure}{.49\linewidth}
\centering
\includegraphics[scale=.75]{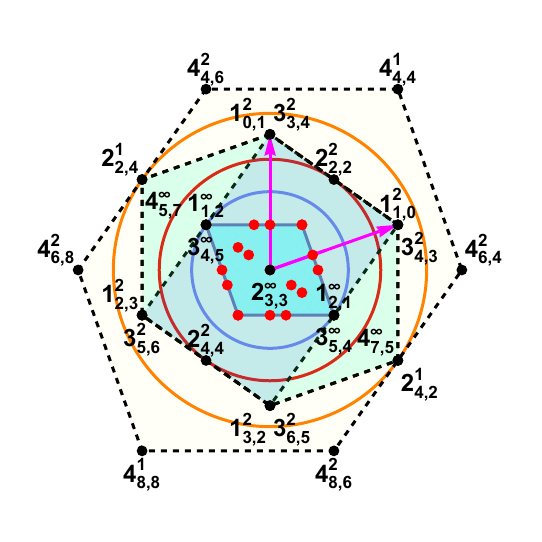}
\caption{$(2,2)$-frame, generated by $1_{0,1}^2$ and $1_{1,0}^2$. }
\label{f.6d_class_22}
\end{subfigure}
\begin{subfigure}{.49\linewidth}
\centering
\includegraphics[scale=.75]{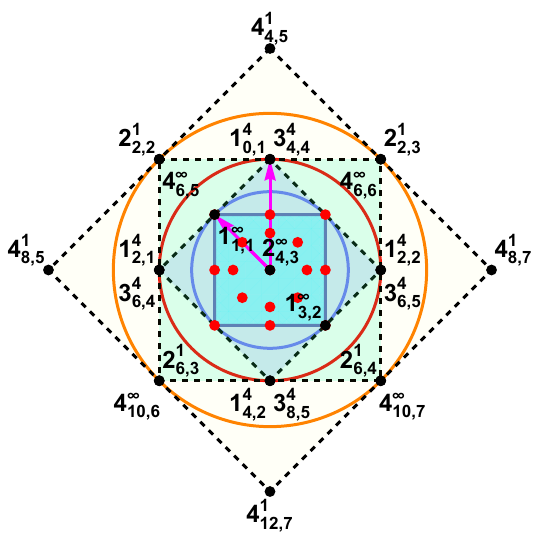}
\caption{$(4,\infty)$-frame, generated by $1_{0,1}^4$ and $1_{1,1}^\infty $.}
\label{f.6d_class_4i}
\end{subfigure}
\caption{Principal $\alpha$-vectors for 2d moduli spaces of 6d theories. The principal branes are labeled by $p_{P_1,P_2}^n$, with respect to the duality frames, whose frame simplices are generated by the magenta arrows. $p$ refers to the spacetime dimension of the brane, $P_1$ and $P_2$ refer to the coordinates of the brane with respect to the frame, and $n$ labels the number of dimensions that decompactify from the KK-modes that come from the winding modes of the brane. The circles of radii $1/\sqrt{d-2}$, $1/\sqrt{2}$, and $1$, are enclosed within the relevant convex hulls, thus satisfying the Sharpened DC and Brane DC. Many of the $\vec\alpha_p/p$-vectors, labeled in red, lie on the boundary of the species polytope, labeled in cyan.}
\label{f.6d_class}
\end{figure}

The $(1,1)$-frame uniquely produces the following principal branes $p_{P_1,P_2}^n$,
\begin{align}
	(1,1):\qquad \begin{matrix}
		1_{0,1}^1,1_{1,0}^1,1_{1,2}^3,1_{2,1}^3,1_{2,2}^2,2_{2,2}^3,2_{2,3}^2,2_{3,2}^2,2_{3,3}^3,3_{3,3}^2,3_{3,4}^3,\\3_{4,3}^3,3_{4,5}^1,3_{5,4}^1,4_{4,4}^3,4_{4,6}^1,4_{6,4}^1,4_{6,6}^3
	\end{matrix},\label{e.6d_class_11}
\end{align}
and they are depicted in Figure \ref{f.6d_class_11}. The $(1,2)$-frame does not have enough particle towers to satisfy the Sharpened DC, and thus is not allowed.  There is a unique tower polytope and set of principal branes associated with the $(1,3)$-frame, but it is the same theory found with the $(1,1)$-frame in \eqref{e.6d_class_11} and depicted in Figure \ref{f.6d_class_11}. The $(1,4)$-frame uniquely produces the following principal branes $p_{P_1,P_2}^n$,
\begin{align}
	(1,4):\qquad \begin{matrix}
		1_{0,1}^1,1_{1,0}^4,1_{1,3}^{\infty },1_{2,2}^4,1_{2,5}^1,2_{2,3}^4,2_{2,6}^1,2_{3,2}^1,2_{3,5}^4,3_{3,3}^1,3_{3,6}^4,3_{4,5}^{\infty },\\3_{4,8}^4,3_{5,7}^1,4_{4,6}^{\infty },4_{4,9}^4,4_{5,5}^4,4_{5,11}^4,4_{6,7}^4,4_{6,10}^{\infty }
	\end{matrix},\label{e.6d_class_14}
\end{align}
and they are depicted in Figure \ref{f.6d_class_14}.  The $(2,2)$-frame uniquely produces the following principal branes $p_{P_1,P_2}^n$,
\begin{align}
	(2,2):\qquad \begin{matrix}
	1_{0,1}^2,1_{1,0}^2,1_{1,2}^{\infty },1_{2,1}^{\infty },1_{2,3}^2,1_{3,2}^2,2_{2,2}^2,2_{2,4}^1,2_{3,3}^{\infty },2_{4,2}^1,2_{4,4}^2,3_{3,4}^2,3_{4,3}^2,3_{4,5}^{\infty },\\3_{5,4}^{\infty },3_{5,6}^2,3_{6,5}^2,4_{4,4}^1,4_{4,6}^2,4_{5,7}^{\infty },4_{6,4}^2,4_{6,8}^2,4_{7,5}^{\infty },4_{8,6}^2,4_{8,8}^1
	\end{matrix},\label{e.6d_class_22}
\end{align}
and they are depicted in Figure \ref{f.6d_class_22}. There is a unique tower polytope and set of principal branes associated with the $(2,3)$-frame, but it is the same theory found with the $(1,1)$-frame and depicted in Figure \ref{f.6d_class_11}. There is a unique tower polytope and set of principal branes associated with the $(1,\infty)$-frame, but it is the same theory found with the $(1,4)$-frame and depicted in Figure \ref{f.6d_class_14}. There is a unique tower polytope and set of principal branes associated with the $(2,\infty)$-frame, but it is the same theory found with the $(2,2)$-frame and depicted in Figure \ref{f.6d_class_22}. The $(3,\infty)$-frame does not allow for tower polytopes satisfying the Sharpened DC. The $(4,\infty)$ allows for two unique theories, one of which is the one with the $(1,4)$-frame in \eqref{e.6d_class_14}. The other is novel and consists of the following principal branes $p_{P_1,P_2}^n$,
\begin{align}
	(4,\infty ):\qquad \begin{matrix}1_{0,1}^4,1_{1,1}^{\infty },1_{2,1}^4,1_{2,2}^4,1_{3,2}^{\infty },1_{4,2}^4,2_{2,2}^1,2_{2,3}^1,2_{4,3}^{\infty },2_{6,3}^1,2_{6,4}^1,3_{4,4}^4,\\3_{6,4}^4,3_{6,5}^4,3_{8,5}^4,4_{4,5}^1,4_{6,5}^{\infty },4_{6,6}^{\infty },4_{8,5}^1,4_{8,7}^1,4_{10,6}^{\infty },4_{10,7}^{\infty },4_{12,7}^1
	\end{matrix},\label{e.6d_class_4i}
\end{align}
and they are depicted in Figure \ref{f.6d_class_4i}. This frame reproduces the dilaton-radion profile of IIB string theory on a 4-torus (c.f., Figure 6(b) of \cite{Etheredge:2024amg}). Thus, there are only four theories consistent with this analysis, listed in Table \ref{t.2d_7d_theories} and depicted in Figure \ref{f.6d_class}.


\begin{table}[H]
$$\begin{array}{c|c|c|c|c|c|c|c|c|c|c|}
(D_1-d,D_2-d)& p_{P_1,P_2}^n\\ \hline
(1,1)& \begin{matrix}
	1_{0,1}^1,1_{1,0}^1,1_{1,2}^3,1_{2,1}^3,1_{2,2}^2,2_{2,2}^3,2_{2,3}^2,2_{3,2}^2,2_{3,3}^3,3_{3,3}^2,3_{3,4}^3,\\3_{4,3}^3,3_{4,5}^1,3_{5,4}^1,4_{4,4}^3,4_{4,6}^1,4_{6,4}^1,4_{6,6}^3
	\end{matrix}\\\hline
(1,4)& \begin{matrix}
	1_{0,1}^1,1_{1,0}^4,1_{1,3}^{\infty },1_{2,2}^4,1_{2,5}^1,2_{2,3}^4,2_{2,6}^1,2_{3,2}^1,2_{3,5}^4,3_{3,3}^1,3_{3,6}^4,3_{4,5}^{\infty },\\3_{4,8}^4,3_{5,7}^1,4_{4,6}^{\infty },4_{4,9}^4,4_{5,5}^4,4_{5,11}^4,4_{6,7}^4,4_{6,10}^{\infty }
	\end{matrix}\\\hline
(2,2)& \begin{matrix}
	1_{0,1}^2,1_{1,0}^2,1_{1,2}^{\infty },1_{2,1}^{\infty },1_{2,3}^2,1_{3,2}^2,2_{2,2}^2,2_{2,4}^1,2_{3,3}^{\infty },2_{4,2}^1,2_{4,4}^2,3_{3,4}^2,3_{4,3}^2,3_{4,5}^{\infty },\\3_{5,4}^{\infty },3_{5,6}^2,3_{6,5}^2,4_{4,4}^1,4_{4,6}^2,4_{5,7}^{\infty },4_{6,4}^2,4_{6,8}^2,4_{7,5}^{\infty },4_{8,6}^2,4_{8,8}^1
	\end{matrix}\\\hline
(4,\infty)& \begin{matrix}
	1_{0,1}^4,1_{1,1}^{\infty },1_{2,1}^4,1_{2,2}^4,1_{3,2}^{\infty },1_{4,2}^4,2_{2,2}^1,2_{2,3}^1,2_{4,3}^{\infty },2_{6,3}^1,2_{6,4}^1,3_{4,4}^4,\\3_{6,4}^4,3_{6,5}^4,3_{8,5}^4,4_{4,5}^1,4_{6,5}^{\infty },4_{6,6}^{\infty },4_{8,5}^1,4_{8,7}^1,4_{10,6}^{\infty },4_{10,7}^{\infty },4_{12,7}^1
	\end{matrix}\\\hline
\end{array}$$ 
\caption{ 6d theories with 2d moduli spaces, labeled by their frame simplices and which principal branes $p_{P_1,P_2}^n$ they possess. $p$ refers to the spacetime dimension of the brane, $P_1$ and $P_2$ refer to the coordinates of the brane with respect to the frame, and $n$ labels the number of dimensions that decompactify from the KK-modes that come from the winding modes of the brane.    The branes are depicted in Figure \ref{f.6d_class}.} \label{t.2d_6d_theories}
\end{table}

These four tower polygons and species polygons appear in the 2d classification in Table 1 of \cite{Etheredge:2024tok}, where it was noted that all four of these tower polygons appear in maximal supergravity compactifications. However, in that paper, seven other polygons appeared, but those polygons did not appear in maximal supergravity. In that paper, it was not known how to eliminate those seven polygons. The analysis here, using lattices, eliminates those seven polygons.

\subsubsection{5d}

The solutions to the  principal norm condition \eqref{e.principal_norm} are listed in Table \ref{t.2d_5d_norm} for all of the allowed frames of 2d moduli spaces of 5d theories.
\clearpage
\begin{table}[H]
$$\begin{array}{c|c|c|c|c|c|c|c|c|c|c|}
(D_1-d,D_2-d)&p_{P_1,P_2}^{n} \\ \hline
(1,1)&\begin{matrix}
	1_{0,1}^1,1_{1,0}^1,1_{1,2}^5,1_{2,1}^5,1_{2,2}^5,2_{2,2}^5,2_{2,3}^5,2_{3,2}^5,2_{3,4}^1,2_{4,3}^1,3_{3,3}^5,3_{3,5}^1,3_{5,3}^1,3_{5,5}^5
\end{matrix}\\\hline 
(1,2)&\begin{matrix}
	1_{0,1}^1,1_{1,0}^2,1_{1,3}^2,1_{2,1}^3,1_{2,2}^{\infty },1_{2,3}^3,2_{2,2}^3,2_{2,3}^{\infty },2_{2,4}^3,2_{3,2}^2,2_{3,5}^2,2_{4,4}^1,\\3_{3,3}^2,3_{3,6}^2,3_{4,3}^3,3_{4,7}^3,3_{5,4}^2,3_{5,7}^2
\end{matrix}\\\hline
(1,3)&\begin{matrix}
	1_{0,1}^1,1_{1,0}^3,1_{1,3}^{21},1_{2,2}^{21},1_{2,3}^{21},2_{2,3}^{21},2_{2,4}^{21},2_{3,3}^{21},2_{3,6}^3,2_{4,5}^1,3_{3,4}^{21},3_{3,7}^3,3_{5,5}^3,3_{5,8}^{21}
\end{matrix}\\\hline
(1,4)&\begin{matrix}
	1_{0,1}^1,1_{1,0}^4,1_{1,4}^4,1_{2,1}^2,1_{2,2}^8,1_{2,3}^{\infty },1_{2,4}^8,1_{2,5}^2,2_{2,2}^2,2_{2,3}^8,2_{2,4}^{\infty },2_{2,5}^8,\\2_{2,6}^2,2_{3,3}^4,2_{3,7}^4,2_{4,6}^1,3_{3,4}^4,3_{3,8}^4,3_{4,3}^1,3_{4,4}^8,3_{4,10}^8,3_{4,11}^1,3_{5,6}^4,3_{5,10}^4
\end{matrix}\\\hline
(1,5)&\begin{matrix}
	1_{0,1}^1,1_{1,0}^5,1_{1,4}^{45},1_{1,6}^1,1_{2,2}^5,1_{2,3}^{45},1_{2,4}^{45},1_{2,5}^5,2_{2,3}^5,2_{2,4}^{45},2_{2,5}^{45},\\2_{2,6}^5,2_{3,2}^1,2_{3,4}^{45},2_{3,8}^5,2_{4,7}^1,3_{3,3}^1,3_{3,5}^{45},3_{3,9}^5,3_{5,7}^5,3_{5,11}^{45},3_{5,13}^1
\end{matrix}\\\hline
(2,2)&\begin{matrix}
	1_{0,1}^2,1_{1,0}^2,2_{4,5}^2,2_{5,4}^2,3_{4,6}^{\infty },3_{6,4}^{\infty }
\end{matrix}\\\hline
(2,3)&\begin{matrix}
	1_{0,1}^2,1_{1,0}^3,1_{1,4}^1,1_{2,1}^{\infty },1_{3,2}^3,1_{3,4}^2,2_{2,2}^2,2_{2,4}^3,2_{3,5}^{\infty },2_{4,2}^1,2_{4,6}^3,\\2_{5,5}^2,3_{3,4}^3,3_{3,6}^2,3_{4,3}^2,3_{6,9}^2,3_{7,6}^2,3_{7,8}^3
\end{matrix}\\\hline
(2,4)&\begin{matrix}
	1_{0,1}^2,1_{1,0}^4,1_{1,3}^{\infty },1_{2,1}^{18},1_{2,5}^2,1_{3,1}^1,1_{3,4}^4,2_{2,3}^4,2_{2,6}^1,2_{3,2}^2,2_{3,6}^{18},2_{4,4}^{\infty },\\2_{4,7}^4,2_{5,6}^2,3_{3,3}^1,3_{3,6}^4,3_{5,4}^4,3_{5,10}^4,3_{7,8}^4,3_{7,11}^1
\end{matrix}\\\hline
(3,3)&\begin{matrix}
	1_{0,1}^3,1_{1,0}^3,1_{1,3}^3,1_{3,1}^3,1_{3,4}^3,1_{4,3}^3,2_{2,3}^3,2_{3,2}^3,2_{3,5}^3,2_{5,3}^3,2_{5,6}^3,2_{6,5}^3,3_{3,3}^1,\\3_{3,6}^1,3_{6,3}^1,3_{6,9}^1,3_{9,6}^1,3_{9,9}^1
\end{matrix}\\\hline
(1,\infty)&\begin{matrix}
	1_{0,1}^1,1_{1,1}^{\infty },1_{1,2}^{\infty },1_{2,1}^1,1_{2,2}^{\infty },1_{2,3}^1,2_{2,2}^1,2_{2,3}^{\infty },2_{2,4}^1,2_{3,3}^{\infty },2_{3,4}^{\infty },2_{4,4}^1,\\3_{3,4}^{\infty },3_{3,5}^{\infty },3_{4,4}^{\infty },3_{4,6}^{\infty },3_{5,5}^{\infty },3_{5,6}^{\infty }
\end{matrix}\\\hline
(2,\infty)&\begin{matrix}
	1_{0,1}^2,1_{1,1}^{\infty },1_{1,2}^4,1_{2,1}^2,1_{3,2}^4,1_{3,3}^1,2_{2,2}^1,2_{2,3}^4,2_{3,4}^2,2_{4,3}^4,2_{4,4}^{\infty },2_{5,4}^2,\\3_{3,4}^4,3_{3,5}^1,3_{5,4}^4,3_{5,6}^4,3_{7,5}^1,3_{7,6}^4
\end{matrix}\\\hline
(3,\infty)&\begin{matrix}
	1_{0,1}^3,1_{1,1}^{\infty },1_{1,2}^2,1_{2,1}^3,1_{4,2}^2,1_{4,3}^1,2_{2,2}^1,2_{2,3}^2,2_{4,4}^3,2_{5,3}^2,2_{5,4}^{\infty },2_{6,4}^3,3_{3,4}^2,\\3_{4,5}^3,3_{6,4}^2,3_{6,6}^2,3_{8,5}^3,3_{9,6}^2
\end{matrix}\\\hline
(4,\infty)&\begin{matrix}
	1_{0,1}^4,1_{1,1}^{\infty },1_{2,1}^4,1_{2,2}^{\infty },1_{3,1}^1,1_{4,2}^{\infty },1_{5,3}^1,2_{2,2}^1,2_{3,3}^{\infty },2_{4,4}^1,2_{5,3}^{\infty },2_{5,4}^4,\\2_{6,4}^{\infty },2_{7,4}^4,2_{8,4}^1,3_{4,4}^{\infty },3_{5,5}^{\infty },3_{6,4}^{\infty },3_{8,6}^{\infty },3_{9,5}^{\infty },3_{10,6}^{\infty }
\end{matrix}\\\hline
(5,\infty)&\begin{matrix}
	1_{0,1}^5,1_{1,1}^{\infty },1_{1,2}^1,1_{2,1}^5,1_{2,2}^5,1_{5,2}^5,1_{6,2}^1,1_{6,3}^1,2_{2,2}^1,2_{2,3}^1,2_{3,3}^5,2_{6,3}^5,2_{6,4}^5,\\2_{7,3}^1,2_{7,4}^{\infty },2_{8,4}^5,3_{3,4}^1,3_{4,4}^5,3_{7,4}^5,3_{8,4}^1,3_{8,6}^1,3_{9,6}^5,3_{12,6}^5,3_{13,6}^1
\end{matrix}\\\hline
\end{array}$$ 
\caption{Principal norm candidates $p^n_{P_1,P_2}$, with respect to frames $(D_1-d,D_2-d)$, for 2d moduli spaces of 5d theories. $p$ refers to the spacetime dimension of the brane, $P_1$ and $P_2$ refer to the coordinates of the brane with respect to the frame, and $n$ labels the number of dimensions that decompactify from the KK-modes that come from the winding modes of the brane.} \label{t.2d_5d_norm}
\end{table}

\begin{figure}
\centering
\begin{subfigure}{.49\linewidth}
\centering
\includegraphics[scale=.75]{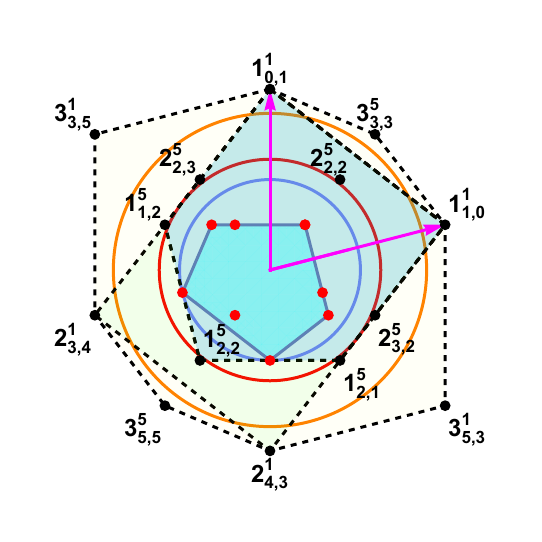}
\caption{$(1,1)$-frame, generated by $1_{0,1}^1$ and $1_{1,0}^1$. }
\label{f.5d_class_11}
\end{subfigure}
\begin{subfigure}{.49\linewidth}
\centering
\includegraphics[scale=.75]{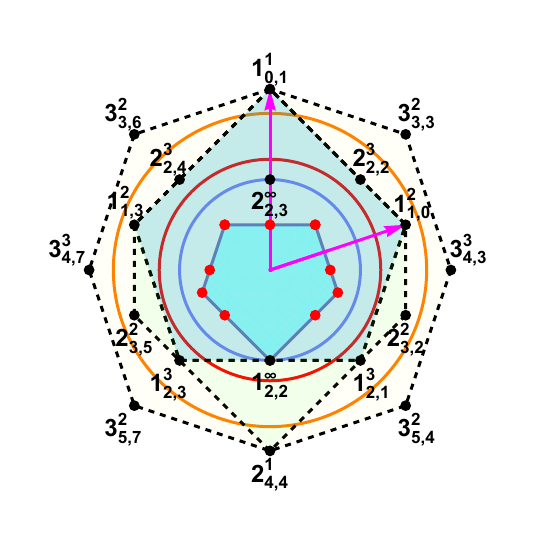}
\caption{$(1,2)$-frame, generated by $1_{0,1}^1$ and $1_{1,0}^2$. }
\label{f.5d_class_12}
\end{subfigure}
\begin{subfigure}{.49\linewidth}
\centering
\includegraphics[scale=.75]{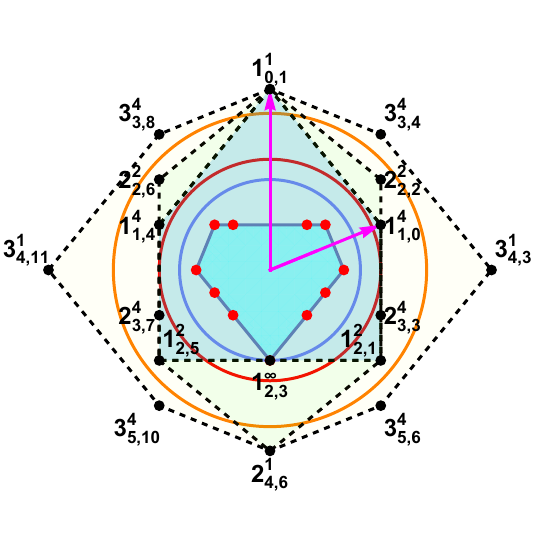}
\caption{$(1,4)$-frame, generated by $1_{0,1}^1$ and $1_{1,0}^4$.}
\label{f.5d_class_14}
\end{subfigure}
\begin{subfigure}{.49\linewidth}
\centering
\includegraphics[scale=.75]{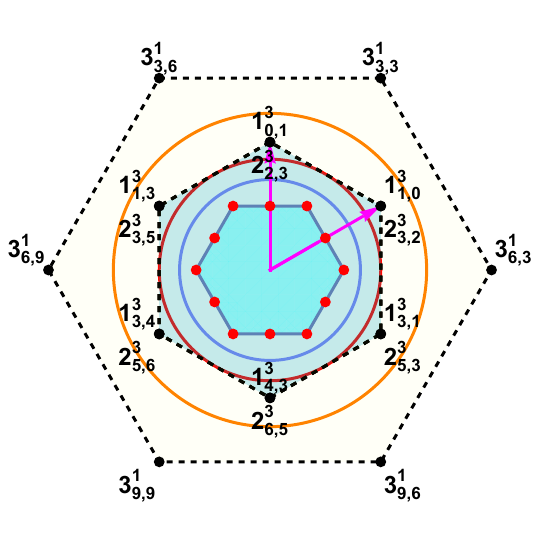}
\caption{$(3,3)$-frame, generated by $1_{0,1}^3$ and $1_{1,0}^3$.}
\label{f.5d_class_33}
\end{subfigure}
\caption{Principal $\alpha$-vectors for 2d moduli spaces of 5d theories. The principal branes are labeled  by $p_{P_1,P_2}^n$, with respect to the duality frames, whose frame simplices are generated by the magenta arrows. $p$ refers to the spacetime dimension of the brane, $P_1$ and $P_2$ refer to the coordinates of the brane with respect to the frame, and $n$ labels the number of dimensions that decompactify from the KK-modes that come from the winding modes of the brane. The circles of radii $1/\sqrt{d-2}$, $1/\sqrt{2}$, and $1$, are enclosed within the relevant convex hulls, thus satisfying the Sharpened DC and Brane DC. Most of the $\vec\alpha_p/p$-vectors, labeled in red, lie on the boundary of the species polytope, labeled in cyan.}
\label{f.5d_class_geom}
\end{figure}

\begin{figure}
\centering
\begin{subfigure}{.49\linewidth}
\centering
\includegraphics[scale=.75]{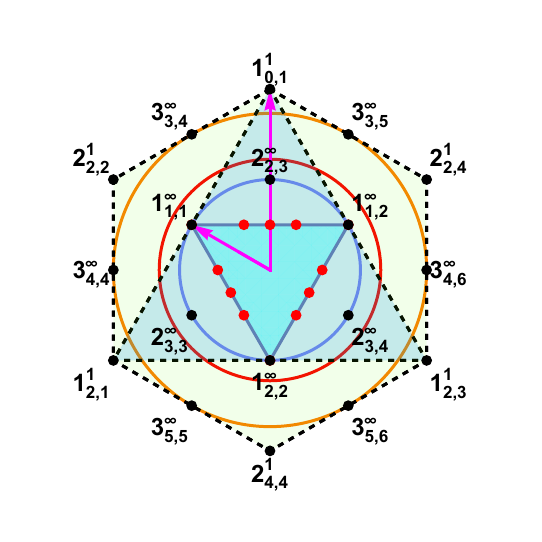}
\caption{$(1,\infty)$-frame, generated by $1_{0,1}^1$ and $1_{1,1}^\infty $.}
\label{f.5d_class_1i}
\end{subfigure}
\caption{Principal $\alpha$-vectors for 2d moduli spaces of 5d theories. The principal branes are labeled  by $p_{P_1,P_2}^n$, with respect to the duality frames, whose frame simplices are generated by the magenta arrows. $p$ refers to the spacetime dimension of the brane, $P_1$ and $P_2$ refer to the coordinates of the brane with respect to the frame, and $n$ labels the number of dimensions that decompactify from the KK-modes that come from the winding modes of the brane. The circles of radii $1/\sqrt{d-2}$, $1/\sqrt{2}$, and $1$, are enclosed within the relevant convex hulls, thus satisfying the Sharpened DC and Brane DC. Most of the $\vec\alpha_p/p$-vectors, labeled in red, lie on the boundary of the species polytope, labeled in cyan.}
\label{f.5d_class_str}
\end{figure}

The $(1,1)$-frame uniquely produces the following principal branes $p_{P_1,P_2}^n$,
\begin{align}
	(1,1):\qquad \begin{matrix}
		1_{0,1}^1,1_{1,0}^1,1_{1,2}^5,1_{2,1}^5,1_{2,2}^5,2_{2,2}^5,2_{2,3}^5,2_{3,2}^5,2_{3,4}^1,2_{4,3}^1,3_{3,3}^5,3_{3,5}^1,3_{5,3}^1,3_{5,5}^5
	\end{matrix},\label{e.5d_class_11}
\end{align}
and they are depicted in Figure \ref{f.5d_class_11}. This reproduces the dilaton-radion profile of IIB string theory on a 5-torus (c.f., Figure 4(c) of \cite{Etheredge:2024amg}). The $(1,2)$-frame uniquely produces the following principal branes $p_{P_1,P_2}^n$,
\begin{align}
	(1,2):\qquad \begin{matrix}
		1_{0,1}^1,1_{1,0}^2,1_{1,3}^2,1_{2,1}^3,1_{2,2}^{\infty },1_{2,3}^3,2_{2,2}^3,2_{2,3}^{\infty },2_{2,4}^3,2_{3,2}^2,2_{3,5}^2,\\2_{4,4}^1,3_{3,3}^2,3_{3,6}^2,3_{4,3}^3,3_{4,7}^3,3_{5,4}^2,3_{5,7}^2
	\end{matrix},\label{e.5d_class_12}
\end{align}
and they are depicted in Figure \ref{f.5d_class_12}. The $(1,3)$-frame does not produce anything allowed. The $(1,4)$-frame uniquely produces the following principal branes $p_{P_1,P_2}^n$,
\begin{align}
	(1,4):\qquad \begin{matrix}
		1_{0,1}^1,1_{1,0}^4,1_{1,4}^4,1_{2,1}^2,1_{2,3}^{\infty },1_{2,5}^2,2_{2,2}^2,2_{2,6}^2,2_{3,3}^4,2_{3,7}^4,2_{4,6}^1,3_{3,4}^4,\\3_{3,8}^4,3_{4,3}^1,3_{4,11}^1,3_{5,6}^4,3_{5,10}^4
	\end{matrix},\label{e.5d_class_14}
\end{align}
and they are depicted in Figure \ref{f.5d_class_14}. The $(1,5)$-frame reproduces the $(1,1)$-frame. The $(2,2)$-frame does not produce anything allowed. The $(2,3)$-frame reproduces the $(1,2)$-frame. The $(2,4)$-frame reproduces the $(1,4)$-frame.  The $(3,3)$-frame uniquely produces the following principal branes $p_{P_1,P_2}^n$,
\begin{align}
	(3,3):\qquad \begin{matrix}
		1_{0,1}^3,1_{1,0}^3,1_{1,3}^3,1_{3,1}^3,1_{3,4}^3,1_{4,3}^3,2_{2,3}^3,2_{3,2}^3,2_{3,5}^3,2_{5,3}^3,2_{5,6}^3,\\2_{6,5}^3,3_{3,3}^1,3_{3,6}^1,3_{6,3}^1,3_{6,9}^1,3_{9,6}^1,3_{9,9}^1
	\end{matrix},\label{e.5d_class_33}
\end{align}
and they are depicted in Figure \ref{f.5d_class_33}. The $(1,\infty)$-frame uniquely produces the following principal branes $p_{P_1,P_2}^n$,
\begin{align}
	(1,\infty ):\qquad \begin{matrix}1_{0,1}^1,1_{1,1}^{\infty },1_{1,2}^{\infty },1_{2,1}^1,1_{2,2}^{\infty },1_{2,3}^1,2_{2,2}^1,2_{2,3}^{\infty },2_{2,4}^1,2_{3,3}^{\infty },2_{3,4}^{\infty },\\2_{4,4}^1,3_{3,4}^{\infty },3_{3,5}^{\infty },3_{4,4}^{\infty },3_{4,6}^{\infty },3_{5,5}^{\infty },3_{5,6}^{\infty }
	\end{matrix},\label{e.5d_class_1i}
\end{align}
and they are depicted in Figure \ref{f.5d_class_1i}. The $(2,\infty)$ reproduces the $(1,4)$-frame. The $(3,\infty)$ reproduces the $(1,2)$-frame. The $(4,\infty)$ and $(5,\infty)$-frame only reproduce previous frames. Thus, there are only five theories consistent with this analysis, listed in Table \ref{t.2d_7d_theories} and depicted in Figures \ref{f.5d_class_geom} and  \ref{f.5d_class_str}.

\begin{table}[H]
$$\begin{array}{c|c|c|c|c|c|c|c|c|c|c|}
(D_1-d,D_2-d)&p_{P_1,P_2}^n \\ \hline
(1,1)& \begin{matrix}
	1_{0,1}^1,1_{1,0}^1,1_{1,2}^5,1_{2,1}^5,1_{2,2}^5,2_{2,2}^5,2_{2,3}^5,2_{3,2}^5,2_{3,4}^1,2_{4,3}^1,3_{3,3}^5,3_{3,5}^1,3_{5,3}^1,3_{5,5}^5
	\end{matrix}\\\hline
(1,2)& \begin{matrix}1_{0,1},1_{1,0},1_{1,3},1_{2,1},1_{2,2},1_{2,3},2_{2,2},2_{2,3},2_{2,4},2_{3,2},2_{3,5},2_{4,4},\\3_{3,3},3_{3,6},3_{4,3},3_{4,7},3_{5,4},3_{5,7}
	\end{matrix}\\\hline
(1,4)& \begin{matrix}
	1_{0,1}^1,1_{1,0}^4,1_{1,4}^4,1_{2,1}^2,1_{2,3}^{\infty },1_{2,5}^2,2_{2,2}^2,2_{2,6}^2,2_{3,3}^4,2_{3,7}^4,2_{4,6}^1,3_{3,4}^4,\\3_{3,8}^4,3_{4,3}^1,3_{4,11}^1,3_{5,6}^4,3_{5,10}^4
	\end{matrix}\\\hline
(3,3)& \begin{matrix}
	1_{0,1}^3,1_{1,0}^3,1_{1,3}^3,1_{3,1}^3,1_{3,4}^3,1_{4,3}^3,2_{2,3}^3,2_{3,2}^3,2_{3,5}^3,2_{5,3}^3,2_{5,6}^3,\\2_{6,5}^3,3_{3,3}^1,3_{3,6}^1,3_{6,3}^1,3_{6,9}^1,3_{9,6}^1,3_{9,9}^1
	\end{matrix}\\\hline
(1,\infty)& \begin{matrix}
	1_{0,1}^1,1_{1,1}^{\infty },1_{1,2}^{\infty },1_{2,1}^1,1_{2,2}^{\infty },1_{2,3}^1,2_{2,2}^1,2_{2,3}^{\infty },2_{2,4}^1,2_{3,3}^{\infty },\\2_{3,4}^{\infty },2_{4,4}^1,3_{3,4}^{\infty },3_{3,5}^{\infty },3_{4,4}^{\infty },3_{4,6}^{\infty },3_{5,5}^{\infty },3_{5,6}^{\infty }
	\end{matrix}\\\hline
\end{array}$$ 
\caption{5d theories with 2d moduli spaces, labeled by their frame simplices and which principal branes $p_{P_1,P_2}^n$ they possess. $p$ refers to the spacetime dimension of the brane, $P_1$ and $P_2$ refer to the coordinates of the brane with respect to the frame, and $n$ labels the number of dimensions that decompactify from the KK-modes that come from the winding modes of the brane.  The branes are depicted in Figures \ref{f.5d_class_geom} and  \ref{f.5d_class_str}.} \label{t.2d_5d_theories}
\end{table}

\subsubsection{4d}

The solutions to the  principal norm condition \eqref{e.principal_norm} are listed in Table \ref{t.2d_4d_norm} for all of the allowed frames of 2d moduli spaces of 4d theories.
\clearpage

\begin{table}[H]
$$\begin{array}{c|c|c|c|c|c|c|c|c|c|c|}
(D_1-d,D_2-d)&p_{P_1,P_2}^n \\ \hline
(1,1)&\begin{matrix}
	1_{0,1}^1,1_{1,0}^1,1_{1,2}^{\infty },1_{2,1}^{\infty },1_{2,3}^1,1_{3,2}^1,2_{2,2}^{\infty },2_{2,4}^1,2_{4,2}^1,2_{4,4}^{\infty }
\end{matrix}\\\hline 
(1,2)&\begin{matrix}
	1_{0,1}^1,1_{1,0}^2,1_{1,3}^5,1_{2,1}^5,1_{2,4}^2,1_{3,3}^1,2_{2,2}^5,2_{2,5}^2,2_{3,2}^5,2_{3,6}^5,2_{4,3}^2,2_{4,6}^5
\end{matrix}\\\hline
(1,3)&\begin{matrix}
	1_{0,1}^1,1_{1,0}^3,1_{1,4}^3,1_{2,1}^3,1_{2,5}^3,1_{3,4}^1,2_{2,2}^3,2_{2,6}^3,2_{3,2}^2,2_{3,8}^2,2_{4,4}^3,2_{4,8}^3
\end{matrix}\\\hline
(1,4)&\begin{matrix}
	1_{0,1}^1,1_{1,0}^4,1_{2,6}^4,1_{3,5}^1,2_{2,7}^4,2_{4,5}^4
\end{matrix}\\\hline
(1,5)&\begin{matrix}
	1_{0,1}^1,1_{1,0}^5,1_{1,6}^2,1_{2,1}^2,1_{2,7}^5,1_{3,6}^1,2_{2,2}^2,2_{2,8}^5,2_{3,3}^5,2_{3,11}^5,2_{4,6}^5,2_{4,12}^2
\end{matrix}\\\hline
(1,6)&\begin{matrix}
	1_{0,1}^1,1_{1,0}^6,1_{1,6}^6,1_{2,2}^6,1_{2,8}^6,1_{3,7}^1,2_{2,3}^6,2_{2,9}^6,2_{3,2}^1,2_{3,14}^1,2_{4,7}^6,2_{4,13}^6
\end{matrix}\\\hline
(2,2)&\begin{matrix}
	1_{0,1}^2,1_{1,0}^2,1_{1,3}^2,1_{3,1}^2,1_{3,4}^2,1_{4,3}^2,2_{2,2}^3,2_{2,3}^{\infty },2_{2,4}^3,2_{3,2}^{\infty },2_{3,5}^{\infty },2_{4,2}^3,2_{4,6}^3,\\2_{5,3}^{\infty },2_{5,6}^{\infty },2_{6,4}^3,2_{6,5}^{\infty },2_{6,6}^3
\end{matrix}\\\hline
(2,3)&\begin{matrix}
	1_{0,1}^2,1_{1,0}^3,1_{1,3}^{21},1_{3,2}^{21},1_{3,5}^3,1_{4,4}^2,2_{2,3}^{21},2_{2,4}^{21},2_{6,6}^{21},2_{6,7}^{21}
\end{matrix}\\\hline
(2,4)&\begin{matrix}
	1_{0,1}^2,1_{0,3}^1,1_{1,0}^4,1_{1,4}^4,1_{2,1}^{\infty },1_{2,5}^{\infty },1_{3,2}^4,1_{3,6}^4,1_{4,3}^1,1_{4,5}^2,1_{4,7}^1,2_{2,2}^2,2_{2,3}^8,\\2_{2,4}^{\infty },2_{2,5}^8,2_{2,6}^2,2_{4,2}^1,2_{4,3}^8,2_{4,9}^8,2_{4,10}^1,2_{6,6}^2,2_{6,7}^8,2_{6,8}^{\infty },2_{6,9}^8,2_{6,10}^2
\end{matrix}\\\hline
(2,5)&\begin{matrix}
	1_{0,1}^2,1_{1,0}^5,1_{1,4}^{45},1_{1,6}^1,1_{2,1}^{18},1_{2,6}^{18},1_{3,1}^1,1_{3,3}^{45},1_{3,7}^5,1_{4,6}^2,\\2_{2,3}^5,2_{2,4}^{45},2_{2,5}^{45},2_{2,6}^5,2_{3,2}^2,2_{5,12}^2,2_{6,8}^5,2_{6,9}^{45},2_{6,10}^{45},2_{6,11}^5
\end{matrix}\\\hline
(3,3)&\begin{matrix}
	1_{0,1}^3,1_{1,0}^3,1_{1,4}^1,1_{4,1}^1,1_{4,5}^3,1_{5,4}^3,2_{2,2}^2,2_{2,4}^3,2_{4,2}^3,2_{6,8}^3,2_{8,6}^3,2_{8,8}^2
\end{matrix}\\\hline
(3,4)&\begin{matrix}
	1_{0,1}^3,1_{1,0}^4,1_{1,3}^{\infty },1_{2,5}^3,1_{3,1}^3,1_{4,3}^{\infty },1_{4,6}^4,1_{5,5}^3,\\2_{2,3}^4,2_{2,6}^1,2_{3,2}^3,2_{5,3}^4,2_{5,9}^4,2_{7,10}^3,2_{8,6}^1,2_{8,9}^4
\end{matrix}\\\hline
(1,\infty)&\begin{matrix}
	1_{0,1}^1,1_{1,1}^{\infty },1_{1,3}^1,1_{2,1}^1,1_{2,3}^{\infty },1_{3,3}^1,2_{2,2}^1,2_{2,4}^{\infty },2_{4,4}^{\infty },2_{4,6}^1
\end{matrix}\\\hline
(2,\infty)&\begin{matrix}
	1_{0,1}^2,1_{1,1}^{\infty },1_{1,2}^{\infty },1_{2,1}^2,1_{2,3}^2,1_{3,2}^{\infty },1_{3,3}^{\infty },1_{4,3}^2,2_{2,2}^1,2_{2,3}^{\infty },2_{2,4}^1,\\2_{4,3}^{\infty },2_{4,5}^{\infty },2_{6,4}^1,2_{6,5}^{\infty },2_{6,6}^1
\end{matrix}\\\hline
(3,\infty)&\begin{matrix}
	1_{0,1}^3,1_{1,1}^{\infty },1_{1,2}^4,1_{2,1}^3,1_{3,3}^3,1_{4,2}^4,1_{4,3}^{\infty },1_{5,3}^3,\\2_{2,2}^1,2_{2,3}^4,2_{3,4}^3,2_{5,3}^4,2_{5,5}^4,2_{7,4}^3,2_{8,5}^4,2_{8,6}^1
\end{matrix}\\\hline
(4,\infty)&\begin{matrix}
	1_{0,1}^4,1_{1,1}^{\infty },1_{1,2}^2,1_{2,1}^4,1_{3,1}^1,1_{3,3}^1,1_{4,3}^4,1_{5,2}^2,1_{5,3}^{\infty },1_{6,3}^4,1_{7,3}^1,2_{2,2}^1,\\2_{2,3}^2,2_{4,4}^{\infty },2_{6,3}^2,2_{6,5}^2,2_{8,4}^{\infty },2_{10,5}^2,2_{10,6}^1
\end{matrix}\\\hline
(5,\infty)&\begin{matrix}
	1_{0,1}^5,1_{1,1}^{\infty },1_{2,1}^5,1_{5,3}^5,1_{6,3}^{\infty },1_{7,3}^5,2_{2,2}^1,2_{12,6}^1
\end{matrix}\\\hline
(6,\infty)&\begin{matrix}
	1_{0,1}^6,1_{1,1}^{\infty },1_{1,2}^1,1_{2,1}^6,1_{2,2}^6,1_{6,2}^6,1_{6,3}^6,1_{7,2}^1,1_{7,3}^{\infty },1_{8,3}^6,2_{2,2}^1,2_{2,3}^1,\\2_{3,3}^6,2_{5,4}^2,2_{7,3}^6,2_{8,3}^1,2_{8,5}^1,2_{9,5}^6,2_{11,4}^2,2_{13,5}^6,2_{14,5}^1,2_{14,6}^1
\end{matrix}\\\hline
\end{array}$$ 
\caption{Principal norm candidates $p^n_{P_1,P_2}$, with respect to frames $(D_1-d,D_2-d)$, for 2d moduli spaces of 4d theories. $p$ refers to the spacetime dimension of the brane, $P_1$ and $P_2$ refer to the coordinates of the brane with respect to the frame, and $n$ labels the number of dimensions that decompactify from the KK-modes that come from the winding modes of the brane.} \label{t.2d_4d_norm}
\end{table}

\begin{figure}[H]
\centering
\begin{subfigure}{.49\linewidth}
\centering
\includegraphics[scale=.75]{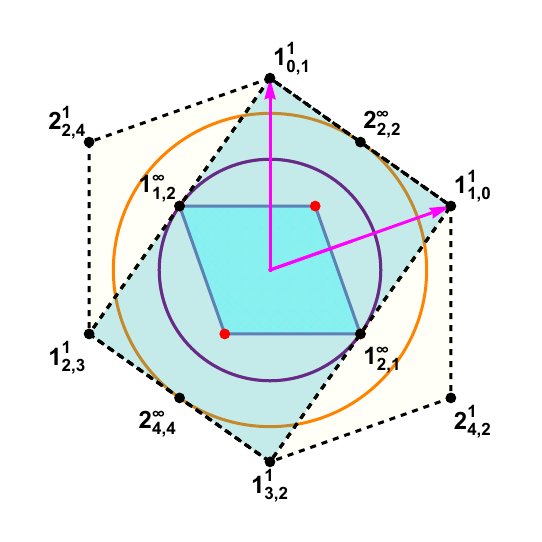}
\caption{$(1,1)$-frame, generated by $1_{0,1}^1$ and $1_{1,0}^1$. }
\label{f.4d_class_11}
\end{subfigure}
\begin{subfigure}{.49\linewidth}
\centering
\includegraphics[scale=.75]{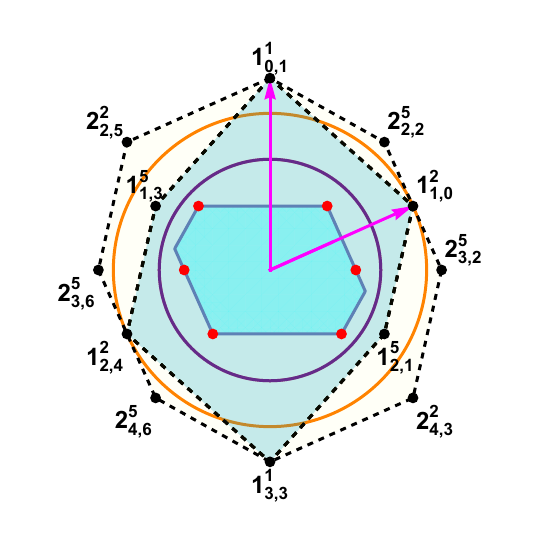}
\caption{$(1,2)$-frame, generated by $1_{0,1}^1$ and $1_{1,0}^2$. }
\label{f.4d_class_12}
\end{subfigure}
\begin{subfigure}{.49\linewidth}
\centering
\includegraphics[scale=.75]{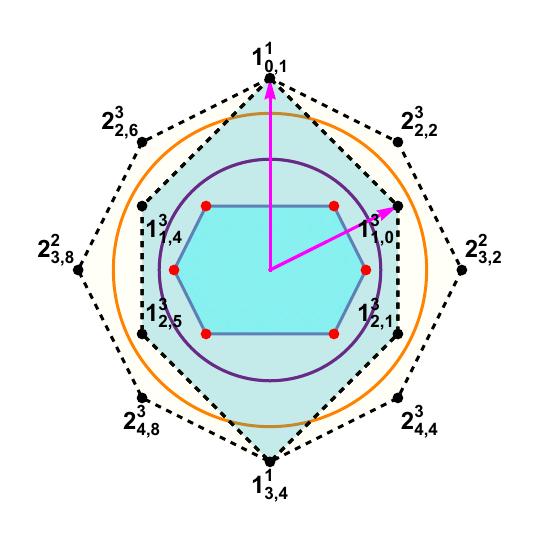}
\caption{$(1,3)$-frame, generated by $1_{0,1}^1$ and $1_{1,0}^3$. }
\label{f.4d_class_13}
\end{subfigure}
\begin{subfigure}{.49\linewidth}
\centering
\includegraphics[scale=.75]{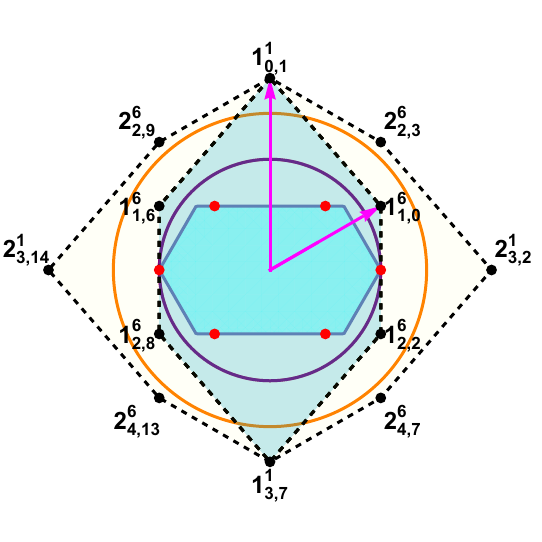}
\caption{$(1,6)$-frame, generated by $1_{0,1}^1$ and $1_{1,0}^6$. }
\label{f.4d_class_16}
\end{subfigure}
\caption{Principal $\alpha$-vectors for 2d moduli spaces of 4d theories. The principal branes are labeled  by $p_{P_1,P_2}^n$, with respect to the duality frames, whose frame simplices are generated by the magenta arrows. $p$ refers to the spacetime dimension of the brane, $P_1$ and $P_2$ refer to the coordinates of the brane with respect to the frame, and $n$ labels the number of dimensions that decompactify from the KK-modes that come from the winding modes of the brane. The circles of radii $1/\sqrt{d-2}$ and $1$, are enclosed within the relevant convex hulls, thus satisfying the Sharpened DC and Brane DC. All of the $\vec\alpha_p/p$-vectors, labeled in red, lie on the boundary of the species polytope, labeled in cyan.}
\label{f.4d_class_geo1}
\end{figure}

\begin{figure}[H]
\centering
\begin{subfigure}{.49\linewidth}
\centering
\includegraphics[scale=.75]{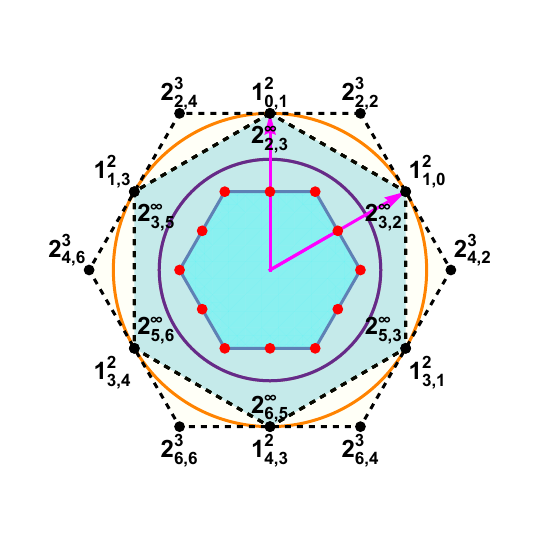}
\caption{$(2,2)$-frame, generated by $1_{0,1}^2$ and $1_{1,0}^2$. }
\label{f.4d_class_22}
\end{subfigure}
\begin{subfigure}{.49\linewidth}
\centering
\includegraphics[scale=.75]{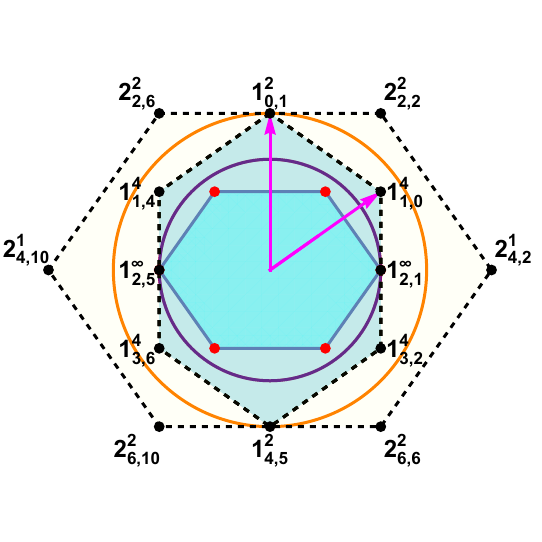}
\caption{$(2,4)$-frame, generated by $1_{0,1}^2$ and $1_{1,0}^4$. }
\label{f.4d_class_24}
\end{subfigure}
\begin{subfigure}{.49\linewidth}
\centering
\includegraphics[scale=.75]{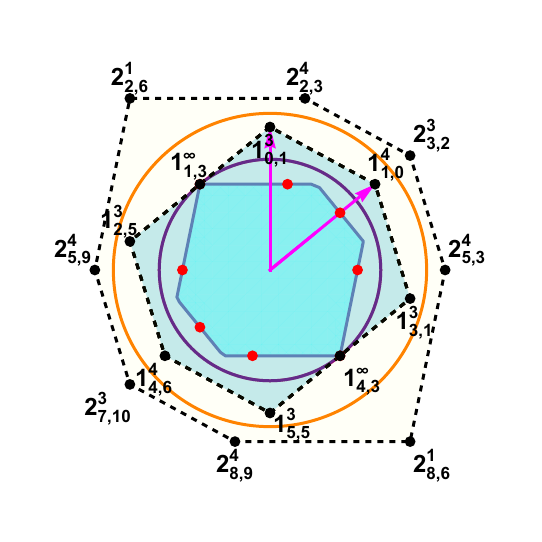}
\caption{$(3,4)$-frame, generated by $1_{0,1}^3$ and $1_{1,0}^4$. }
\label{f.4d_class_34}
\end{subfigure}
\caption{Principal $\alpha$-vectors for 2d moduli spaces of 4d theories. The principal branes are labeled  by $p_{P_1,P_2}^n$, with respect to the duality frames, whose frame simplices are generated by the magenta arrows. $p$ refers to the spacetime dimension of the brane, $P_1$ and $P_2$ refer to the coordinates of the brane with respect to the frame, and $n$ labels the number of dimensions that decompactify from the KK-modes that come from the winding modes of the brane. The circles of radii $1/\sqrt{d-2}$ and $1$, are enclosed within the relevant convex hulls, thus satisfying the Sharpened DC and Brane DC. All of the $\vec\alpha_p/p$-vectors, labeled in red, lie on the boundary of the species polytope, labeled in cyan.}
\label{f.4d_class_geo2}
\end{figure}

\begin{figure}[H]
\centering
\begin{subfigure}{.49\linewidth}
\centering
\includegraphics[scale=.75]{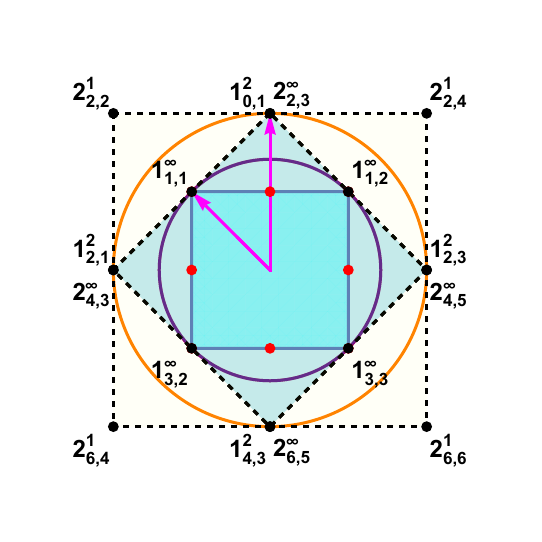}
\caption{$(2,\infty )$-frame, generated by $1_{0,1}^2$ and $1_{1,1}^\infty $. }
\label{f.4d_class_2i}
\end{subfigure}
\begin{subfigure}{.49\linewidth}
\centering
\includegraphics[scale=.75]{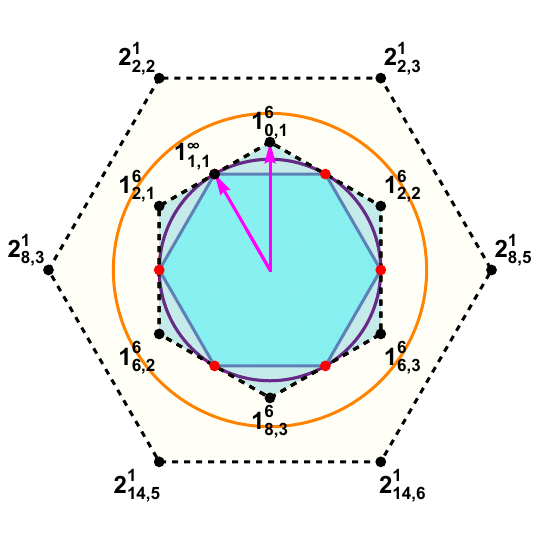}
\caption{$(6,\infty )$-frame, generated by $1_{0,1}^6$ and $1_{1,1}^\infty $. }
\label{f.4d_class_6i}
\end{subfigure}
\caption{Principal $\alpha$-vectors for 2d moduli spaces of 4d theories. The principal branes are labeled  by $p_{P_1,P_2}^n$, with respect to the duality frames, whose frame simplices are generated by the magenta arrows. $p$ refers to the spacetime dimension of the brane, $P_1$ and $P_2$ refer to the coordinates of the brane with respect to the frame, and $n$ labels the number of dimensions that decompactify from the KK-modes that come from the winding modes of the brane. The circles of radii $1/\sqrt{d-2}$ and $1$, are enclosed within the relevant convex hulls, thus satisfying the Sharpened DC and Brane DC. All of the $\vec\alpha_p/p$-vectors, labeled in red, lie on the boundary of the species polytope, labeled in cyan.}
\label{f.4d_class_str}
\end{figure}

The $(1,1)$-frame uniquely produces the following principal branes $p_{P_1,P_2}^n$,
\begin{align}
	(1,1):\qquad \begin{matrix}
		1_{0,1}^1,1_{1,0}^1,1_{1,2}^{\infty },1_{2,1}^{\infty },1_{2,3}^1,1_{3,2}^1,2_{2,2}^{\infty },2_{2,4}^1,2_{4,2}^1,2_{4,4}^{\infty }
	\end{matrix},\label{e.4d_class_11}
\end{align}
and they are depicted in Figure \ref{f.4d_class_11}. The $(1,2)$-frame uniquely produces the following principal branes $p_{P_1,P_2}^n$,
\begin{align}
	(1,2):\qquad \begin{matrix}
		1_{0,1}^1,1_{1,0}^2,1_{1,3}^5,1_{2,1}^5,1_{2,4}^2,1_{3,3}^1,2_{2,2}^5,2_{2,5}^2,2_{3,2}^5,2_{3,6}^5,2_{4,3}^2,2_{4,6}^5
	\end{matrix},\label{e.4d_class_12}
\end{align}
and they are depicted in Figure \ref{f.4d_class_12}. The $(1,3)$-frame uniquely produces the following principal branes $p_{P_1,P_2}^n$,
\begin{align}
	(1,3):\qquad \begin{matrix}
		1_{0,1}^1,1_{1,0}^3,1_{1,4}^3,1_{2,1}^3,1_{2,5}^3,1_{3,4}^1,2_{2,2}^3,2_{2,6}^3,2_{3,2}^2,2_{3,8}^2,2_{4,4}^3,2_{4,8}^3
	\end{matrix},\label{e.4d_class_13}
\end{align}
and they are depicted in Figure \ref{f.4d_class_13}. The $(1,4)$-frame produces nothing allowed. The $(1,5)$-frame reproduces the $(1,2)$-frame. The $(1,6)$-frame uniquely produces the following principal branes $p_{P_1,P_2}^n$,
\begin{align}
	(1,6):\qquad \begin{matrix}
		1_{0,1}^1,1_{1,0}^6,1_{1,6}^6,1_{2,2}^6,1_{2,8}^6,1_{3,7}^1,2_{2,3}^6,2_{2,9}^6,2_{3,2}^1,2_{3,14}^1,2_{4,7}^6,2_{4,13}^6
	\end{matrix},\label{e.4d_class_16}
\end{align}
and they are depicted in Figure \ref{f.4d_class_16}. The $(2,2)$-frame uniquely produces the following principal branes $p_{P_1,P_2}^n$,
\begin{align}
	(2,2):\qquad \begin{matrix}
		1_{0,1}^2,1_{1,0}^2,1_{1,3}^2,1_{3,1}^2,1_{3,4}^2,1_{4,3}^2,2_{2,2}^3,2_{2,3}^{\infty },2_{2,4}^3,2_{3,2}^{\infty },\\2_{3,5}^{\infty },2_{4,2}^3,2_{4,6}^3,2_{5,3}^{\infty },2_{5,6}^{\infty },2_{6,4}^3,2_{6,5}^{\infty },2_{6,6}^3
	\end{matrix},\label{e.4d_class_22}
\end{align}
and they are depicted in Figure \ref{f.4d_class_22}. The $(2,3)$-frame does not provide anything allowed. The $(2,4)$-frame at first provides four tentative tower polytopes. Three of these polytopes are inconsistent, as they produce $(1,\infty)$ faces, and, by analysis done below, the $(1,\infty)$-frame reproduces the $(1,1)$-frame. Thus, only one frame is allowed. The $2_{2,4}^\infty $ and $2_{6,8}^\infty $ strings are too short, and thus ignorable. The $2_{4,9}^8$ and $2_{4,3}^8$ strings are prohibited, since their respective products with  the $2_{4,10}^1$ and $2_{4,2}^1$ strings are too large, but these latter strings are required by the ESC. The $2_{2,5}^8$ and $2_{2,3}^5$ strings are prohibited, since the $2_{2,6}^2$ and $2_{2,2}^2$ strings are required by the $p_\text{max}=2$ Brane DC. The $2_{6,9}^8$ and $2_{6,7}^8$ strings are prohibited, as they have respective dot products with the $2_{6,10}^2$ and $2_{6,6}^2$ strings, which are required for the codimension-2 Brane DC. Altogether, what is left from the $(2,4)$-frame are the following principal branes $p_{P_1,P_2}^n$,
\begin{align}
	(2,4):\qquad \begin{matrix}
		1_{0,1}^2,1_{1,0}^4,1_{1,4}^4,1_{2,1}^{\infty },1_{2,5}^{\infty },1_{3,2}^4,1_{3,6}^4,1_{4,5}^2,2_{2,2}^2,2_{2,6}^2,2_{4,2}^1,2_{4,10}^1,2_{6,6}^2,2_{6,10}^2
	\end{matrix},\label{e.4d_class_24}
\end{align}
and they are depicted in Figure \ref{f.4d_class_24}. The $(2,5)$-frame produces the $(1,2)$-frame. The $(3,3)$-frame produces the $(1,3)$-frame. The $(3,4)$-frame uniquely produces the following principal branes $p_{P_1,P_2}^n$,
\begin{align}
	(3,4):\qquad \begin{matrix}
		1_{0,1}^3,1_{1,0}^4,1_{1,3}^{\infty },1_{2,5}^3,1_{3,1}^3,1_{4,3}^{\infty },1_{4,6}^4,1_{5,5}^3,2_{2,3}^4,2_{2,6}^1,2_{3,2}^3,2_{5,3}^4,2_{5,9}^4,\\2_{7,10}^3,2_{8,6}^1,2_{8,9}^4
	\end{matrix},\label{e.4d_class_34}
\end{align}
and they are depicted in Figure \ref{f.4d_class_34}. The $(1,\infty)$-frame reproduces the $(1,1)$-frame. The $(2,\infty)$-frame uniquely produces the following principal branes $p_{P_1,P_2}^n$,
\begin{align}
	(2,\infty ):\qquad \begin{matrix}
		1_{0,1}^2,1_{1,1}^{\infty },1_{1,2}^{\infty },1_{2,1}^2,1_{2,3}^2,1_{3,2}^{\infty },1_{3,3}^{\infty },1_{4,3}^2,2_{2,2}^1,2_{2,3}^{\infty },2_{2,4}^1,2_{4,3}^{\infty },\\2_{4,5}^{\infty },2_{6,4}^1,2_{6,5}^{\infty },2_{6,6}^1
	\end{matrix},\label{e.4d_class_2i}
\end{align}
and they are depicted in Figure \ref{f.4d_class_2i}. The $(3,\infty)$-frame reproduces the $(3,4)$-frame. The $(4,\infty)$-frame reproduces either produces theories with inconsistencies with previously-discuss frames, or it reproduces the $(2,4)$-frame. The $(5,\infty)$-frame produces nothing allowed. The $(6,\infty)$-frame produces a unique theory with points
\begin{align}
	(6,\infty ):\qquad \begin{matrix}
		1_{0,1}^6,1_{1,1}^{\infty },1_{2,1}^6,1_{2,2}^6,1_{6,2}^6,1_{6,3}^6,1_{7,3}^{\infty },1_{8,3}^6,2_{2,2}^1,2_{2,3}^1,2_{8,3}^1,2_{8,5}^1,2_{14,5}^1,2_{14,6}^1
	\end{matrix},\label{e.4d_class_6i}
\end{align}
and they are depicted in Figure \ref{f.4d_class_6i}. This reproduces the radion-dilaton profile of IIB string theory on a 6-torus (c.f., Figure 4(d) of \cite{Etheredge:2024amg}). Thus, there are only nine theories consistent with this analysis, listed in Table \ref{t.2d_4d_theories} and depicted in Figures \ref{f.4d_class_geo1}, \ref{f.4d_class_geo2}, and  \ref{f.4d_class_str}.

\begin{table}[H]
$$\begin{array}{c|c|c|c|c|c|c|c|c|c|c|}
(D_1-d,D_2-d)&p_{P_1,P_2}^n \\ \hline
(1,1)& \begin{matrix}
	1_{0,1}^1,1_{1,0}^1,1_{1,2}^{\infty },1_{2,1}^{\infty },1_{2,3}^1,1_{3,2}^1,2_{2,2}^{\infty },2_{2,4}^1,2_{4,2}^1,2_{4,4}^{\infty }
	\end{matrix}\\\hline
(1,2)& \begin{matrix}1
	_{0,1}^1,1_{1,0}^2,1_{1,3}^5,1_{2,1}^5,1_{2,4}^2,1_{3,3}^1,2_{2,2}^5,2_{2,5}^2,2_{3,2}^5,2_{3,6}^5,2_{4,3}^2,2_{4,6}^5
	\end{matrix}\\\hline
(1,3)& \begin{matrix}
	1_{0,1}^1,1_{1,0}^3,1_{1,4}^3,1_{2,1}^3,1_{2,5}^3,1_{3,4}^1,2_{2,2}^3,2_{2,6}^3,2_{3,2}^2,2_{3,8}^2,2_{4,4}^3,2_{4,8}^3
	\end{matrix}\\\hline
(1,6)& \begin{matrix}
	1_{0,1}^1,1_{1,0}^6,1_{1,6}^6,1_{2,2}^6,1_{2,8}^6,1_{3,7}^1,2_{2,3}^6,2_{2,9}^6,2_{3,2}^1,2_{3,14}^1,2_{4,7}^6,2_{4,13}^6
	\end{matrix}\\\hline
(2,2)& \begin{matrix}
	1_{0,1}^2,1_{1,0}^2,1_{1,3}^2,1_{3,1}^2,1_{3,4}^2,1_{4,3}^2,2_{2,2}^3,2_{2,3}^{\infty },2_{2,4}^3,2_{3,2}^{\infty },2_{3,5}^{\infty },2_{4,2}^3,2_{4,6}^3,\\2_{5,3}^{\infty },2_{5,6}^{\infty },2_{6,4}^3,2_{6,5}^{\infty },2_{6,6}^3
	\end{matrix}\\\hline
(2,4)& \begin{matrix}
	1_{0,1}^2,1_{1,0}^4,1_{1,4}^4,1_{2,1}^{\infty },1_{2,5}^{\infty },1_{3,2}^4,1_{3,6}^4,1_{4,5}^2,2_{2,2}^2,2_{2,6}^2,2_{4,2}^1,2_{4,10}^1,2_{6,6}^2,2_{6,10}^2
	\end{matrix}\\\hline
(3,4)& \begin{matrix}
	1_{0,1}^3,1_{1,0}^4,1_{1,3}^{\infty },1_{2,5}^3,1_{3,1}^3,1_{4,3}^{\infty },1_{4,6}^4,1_{5,5}^3,2_{2,3}^4,2_{2,6}^1,2_{3,2}^3,2_{5,3}^4,\\2_{5,9}^4,2_{7,10}^3,2_{8,6}^1,2_{8,9}^4
	\end{matrix}\\\hline
(2,\infty)& \begin{matrix}
	1_{0,1}^2,1_{1,1}^{\infty },1_{1,2}^{\infty },1_{2,1}^2,1_{2,3}^2,1_{3,2}^{\infty },1_{3,3}^{\infty },1_{4,3}^2,2_{2,2}^1,2_{2,3}^{\infty },2_{2,4}^1,2_{4,3}^{\infty },\\2_{4,5}^{\infty },2_{6,4}^1,2_{6,5}^{\infty },2_{6,6}^1
	\end{matrix}\\\hline
(6,\infty)& \qquad \begin{matrix}
	1_{0,1}^6,1_{1,1}^{\infty },1_{2,1}^6,1_{2,2}^6,1_{6,2}^6,1_{6,3}^6,1_{7,3}^{\infty },1_{8,3}^6,2_{2,2}^1,2_{2,3}^1,2_{8,3}^1,2_{8,5}^1,2_{14,5}^1,2_{14,6}^1
	\end{matrix}\\\hline
\end{array}$$ 
\caption{4d theories with 2d moduli spaces, labeled by their frame simplices and which principal branes $p_{P_1,P_2}^n$ they possess. $p$ refers to the spacetime dimension of the brane, $P_1$ and $P_2$ refer to the coordinates of the brane with respect to the frame, and $n$ labels the number of dimensions that decompactify from the KK-modes that come from the winding modes of the brane.  The branes are depicted in Figures \ref{f.4d_class_geo1}, \ref{f.4d_class_geo2} and  \ref{f.4d_class_str}.} \label{t.2d_4d_theories}
\end{table}

Notably, for all of these theories, the principal strings have $\vec \alpha_2/2$-vectors that lie on the boundary of the species polytope! It would be interesting to understand further why this is the case.

\subsubsection{3d}

The solutions to the  principal norm condition \eqref{e.principal_norm} are listed in Table \ref{t.2d_4d_norm} for all of the allowed frames of 2d moduli spaces of 3d theories.

\clearpage
\begin{table}[H]
$$\begin{array}{c|c|c|c|c|c|c|c|c|c|c|}
(D_1-d,D_2-d)& p_{P_1,P_2}^{n} \\ \hline
(1,1)&\begin{matrix}
	1_{0,1}^1,1_{1,0}^1,1_{1,3}^1,1_{3,1}^1,1_{3,4}^1,1_{4,3}^1
\end{matrix}\\\hline 
(1,2)&\begin{matrix}
	1_{0,1}^1,1_{1,0}^2,1_{1,4}^2,1_{2,1}^{\infty },1_{2,5}^{\infty },1_{3,2}^2,1_{3,6}^2,1_{4,5}^1
\end{matrix}\\\hline
(1,3)&\begin{matrix}
	1_{0,1}^1,1_{1,0}^3,1_{1,5}^3,1_{2,1}^5,1_{2,7}^5,1_{3,3}^3,1_{3,8}^3,1_{4,7}^1
\end{matrix}\\\hline
(1,4)&\begin{matrix}
	1_{0,1}^1,1_{1,0}^4,1_{1,6}^4,1_{2,1}^3,1_{2,9}^3,1_{3,4}^4,1_{3,10}^4,1_{4,9}^1
\end{matrix}\\\hline
(1,5)&\begin{matrix}
	1_{0,1}^1,1_{1,0}^5,1_{1,7}^5,1_{3,5}^5,1_{3,12}^5,1_{4,11}^1
\end{matrix}\\\hline
(1,6)&\begin{matrix}
	1_{0,1}^1,1_{1,0}^6,1_{1,8}^6,1_{1,10}^1,1_{2,1}^2,1_{2,2}^{24},1_{2,12}^{24},1_{2,13}^2,1_{3,4}^1,1_{3,6}^6,1_{3,14}^6,1_{3,16}^1,1_{4,13}^1
\end{matrix}\\\hline
(1,7)&\begin{matrix}
	1_{0,1}^1,1_{1,0}^7,1_{1,9}^7,1_{2,2}^7,1_{2,14}^7,1_{3,7}^7,1_{3,16}^7,1_{4,15}^1
\end{matrix}\\\hline
(2,2)&\begin{matrix}
	1_{0,1}^2,1_{1,0}^2,1_{1,3}^5,1_{2,4}^{\infty },1_{3,1}^5,1_{3,5}^5,1_{4,2}^{\infty },1_{5,3}^5,1_{5,6}^2,1_{6,5}^2
\end{matrix}\\\hline
(2,3)&\begin{matrix}
	1_{0,1}^2,1_{1,0}^3,1_{1,4}^3,1_{3,1}^2,1_{3,7}^2,1_{5,4}^3,1_{5,8}^3,1_{6,7}^2
\end{matrix}\\\hline
(2,4)&\begin{matrix}
	1_{0,1}^2,1_{1,0}^4,1_{5,10}^4,1_{6,9}^2
\end{matrix}\\\hline
(2,5)&\begin{matrix}
	1_{0,1}^2,1_{1,0}^5,1_{1,6}^2,1_{2,1}^{\infty },1_{3,2}^5,1_{3,10}^5,1_{4,11}^{\infty },1_{5,6}^2,1_{5,12}^5,1_{6,11}^2
\end{matrix}\\\hline
(2,6)&\begin{matrix}
	1_{0,1}^2,1_{0,4}^1,1_{1,0}^6,1_{1,6}^6,1_{2,1}^{18},1_{2,9}^{18},1_{3,1}^1,1_{3,13}^1,1_{4,5}^{18},\\1_{4,13}^{18},1_{5,8}^6,1_{5,14}^6,1_{6,10}^1,1_{6,13}^2,1_{6,16}^1
\end{matrix}\\\hline
(3,3)&\begin{matrix}
	1_{0,1}^3,1_{1,0}^3,1_{1,3}^{21},1_{3,1}^{21},1_{5,7}^{21},1_{7,5}^{21},1_{7,8}^3,1_{8,7}^3
\end{matrix}\\\hline
(3,4)&\begin{matrix}
	1_{0,1}^3,1_{1,0}^4,1_{1,4}^4,1_{4,1}^1,1_{4,2}^8,1_{4,8}^8,1_{4,9}^1,1_{7,6}^4,1_{7,10}^4,1_{8,9}^3
\end{matrix}\\\hline
(3,5)&\begin{matrix}
	1_{0,1}^3,1_{1,0}^5,1_{1,4}^{45},1_{1,6}^1,1_{3,1}^3,1_{3,8}^{45},1_{5,4}^{45},1_{5,11}^3,1_{7,6}^1,1_{7,8}^{45},1_{7,12}^5,1_{8,11}^3
\end{matrix}\\\hline
(4,4)&\begin{matrix}
	1_{0,1}^4,1_{1,0}^4,1_{1,3}^{\infty },1_{2,5}^4,1_{3,1}^{\infty },1_{3,7}^1,1_{5,2}^4,1_{5,8}^4,1_{7,3}^1,1_{7,9}^{\infty },1_{8,5}^4,1_{9,7}^{\infty },1_{9,10}^4,\\1_{10,9}^4,1_{11,11}^1
\end{matrix}\\\hline
(1,\infty)&\begin{matrix}
	1_{0,1}^1,1_{1,1}^{\infty },1_{1,3}^{\infty },1_{2,1}^1,1_{2,5}^1,1_{3,3}^{\infty },1_{3,5}^{\infty },1_{4,5}^1
\end{matrix}\\\hline
(2,\infty)&\begin{matrix}
	1_{0,1}^2,1_{1,1}^{\infty },1_{1,3}^1,1_{2,1}^2,1_{4,5}^2,1_{5,3}^1,1_{5,5}^{\infty },1_{6,5}^2
\end{matrix}\\\hline
(3,\infty)&\begin{matrix}
	1_{0,1}^3,1_{1,1}^{\infty },1_{1,2}^{\infty },1_{2,1}^3,1_{2,3}^3,1_{4,2}^{\infty },1_{4,4}^{\infty },1_{6,3}^3,1_{6,5}^3,1_{7,4}^{\infty },1_{7,5}^{\infty },1_{8,5}^3
\end{matrix}\\\hline
(4,\infty)&\begin{matrix}
	1_{0,1}^4,1_{1,1}^{\infty },1_{1,2}^4,1_{2,1}^4,1_{3,1}^1,1_{3,3}^{\infty },1_{5,2}^4,1_{5,4}^4,1_{7,3}^{\infty },1_{7,5}^1,1_{8,5}^4,1_{9,4}^4,1_{9,5}^{\infty },\\1_{10,5}^4,1_{11,5}^1
\end{matrix}\\\hline
(5,\infty)&\begin{matrix}
	1_{0,1}^5,1_{1,1}^{\infty },1_{1,2}^2,1_{2,1}^5,1_{6,2}^2,1_{6,4}^2,1_{10,5}^5,1_{11,4}^2,1_{11,5}^{\infty },1_{12,5}^5
\end{matrix}\\\hline
(6,\infty)&\begin{matrix}
	1_{0,1}^6,1_{1,1}^{\infty },1_{2,1}^6,1_{4,3}^2,1_{10,3}^2,1_{12,5}^6,1_{13,5}^{\infty },1_{14,5}^6
\end{matrix}\\\hline
(7,\infty)&\begin{matrix}
	1_{0,1}^7,1_{1,1}^{\infty },1_{1,2}^1,1_{2,1}^7,1_{2,2}^7,1_{7,2}^7,1_{8,2}^1,1_{8,4}^1,1_{9,4}^7,1_{14,4}^7,1_{14,5}^7,1_{15,4}^1,\\1_{15,5}^{\infty },1_{16,5}^7
\end{matrix}\\\hline
\end{array}$$ 
\caption{Principal norm candidates $p^n_{P_1,P_2}$, with respect to frames $(D_1-d,D_2-d)$, for 2d moduli spaces of 4d theories. $p$ refers to the spacetime dimension of the brane, $P_1$ and $P_2$ refer to the coordinates of the brane with respect to the frame, and $n$ labels the number of dimensions that decompactify from the KK-modes that come from the winding modes of the brane.} \label{t.2d_3d_norm}
\end{table}

\begin{figure}[H]
\centering
\begin{subfigure}{.49\linewidth}
\centering
\includegraphics[scale=.75]{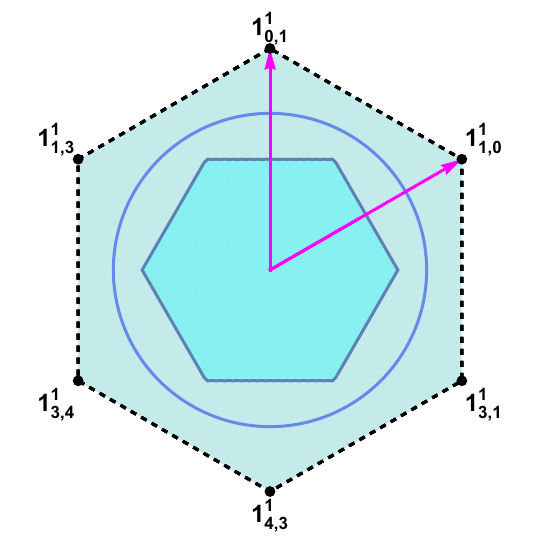}
\caption{$(1,1)$-frame, generated by $1_{0,1}^1$ and $1_{1,0}^1$. }
\label{f.3d_class_11}
\end{subfigure}
\begin{subfigure}{.49\linewidth}
\centering
\includegraphics[scale=.75]{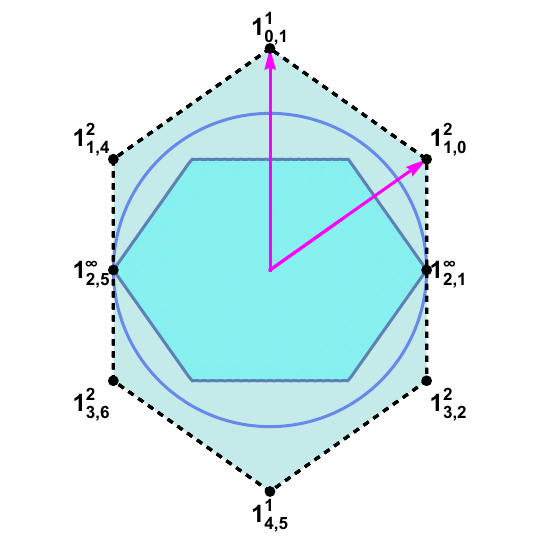}
\caption{$(1,2)$-frame, generated by $1_{0,1}^1$ and $1_{1,0}^2$. }
\label{f.3d_class_12}
\end{subfigure}
\begin{subfigure}{.49\linewidth}
\centering
\includegraphics[scale=.75]{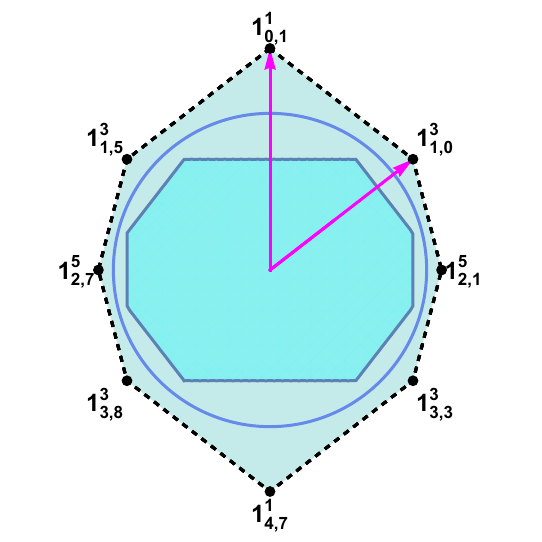}
\caption{$(1,3)$-frame, generated by $1_{0,1}^1$ and $1_{1,0}^3$. }
\label{f.3d_class_13}
\end{subfigure}
\begin{subfigure}{.49\linewidth}
\centering
\includegraphics[scale=.75]{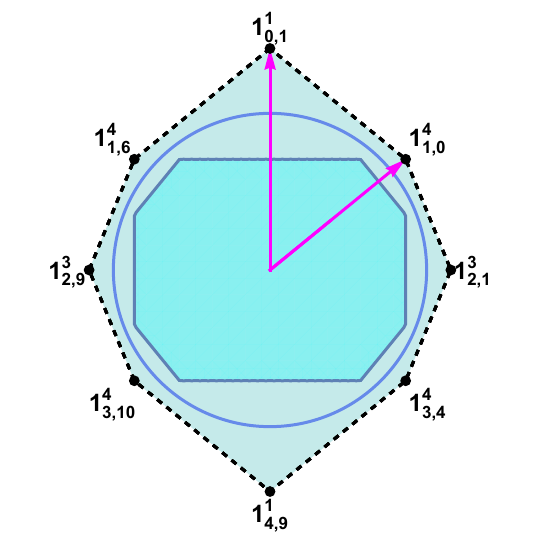}
\caption{$(1,4)$-frame, generated by $1_{0,1}^1$ and $1_{1,0}^4$. }
\label{f.3d_class_14}
\end{subfigure}
\caption{Principal $\alpha$-vectors for 2d frames in 3d, labeled by $p_{P_1,P_2}^n$. The magenta arrows point to the frame simplex vectors. $p$ refers to the spacetime dimension of the brane, $P_1$ and $P_2$ refer to the coordinates of the brane with respect to the frame, and $n$ labels the number of dimensions that decompactify from the KK-modes that come from the winding modes of the brane. The circle has radius 1 and is contained in the convex hull of particle towers, as required by the Sharpened DC. The species polytopes are cyan.}
\label{f.3d_class_geo1}
\end{figure}

\begin{figure}[H]
\centering
\begin{subfigure}{.49\linewidth}
\centering
\includegraphics[scale=.75]{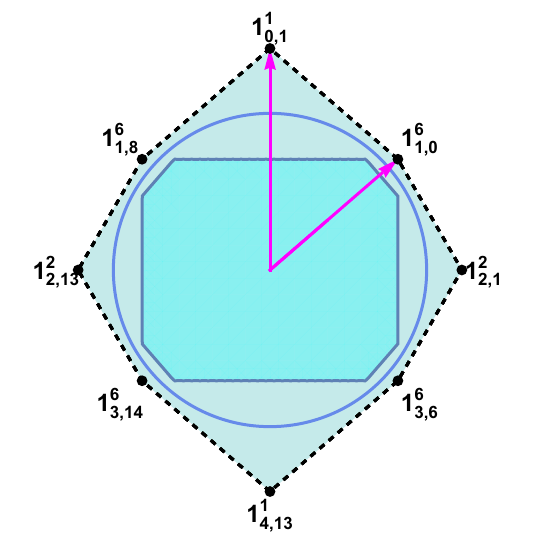}
\caption{$(1,6)$-frame, generated by $1_{0,1}^1$ and $1_{1,0}^6$. }
\label{f.3d_class_16}
\end{subfigure}
\begin{subfigure}{.49\linewidth}
\centering
\includegraphics[scale=.75]{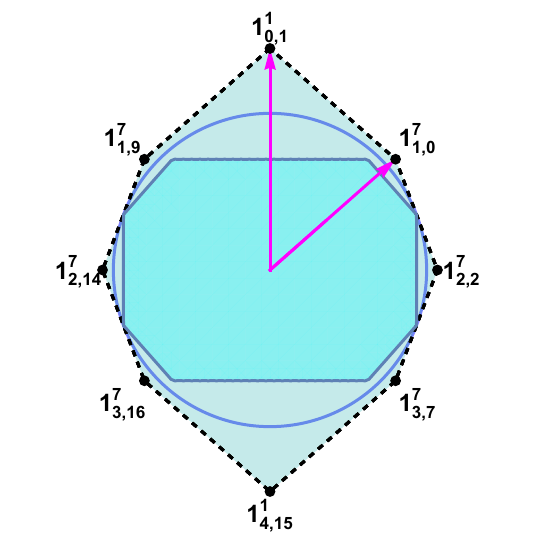}
\caption{$(1,7)$-frame, generated by $1_{0,1}^1$ and $1_{1,0}^7$. }
\label{f.3d_class_17}
\end{subfigure}
\begin{subfigure}{.49\linewidth}
\centering
\includegraphics[scale=.75]{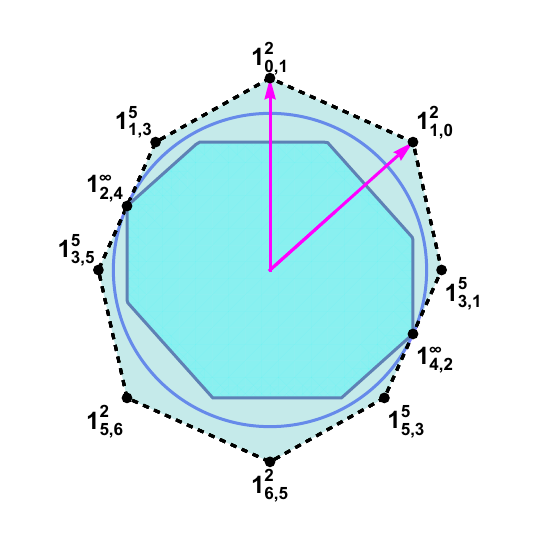}
\caption{$(2,2)$-frame, generated by $1_{0,1}^2$ and $1_{1,0}^2$. }
\label{f.3d_class_22}
\end{subfigure}
\begin{subfigure}{.49\linewidth}
\centering
\includegraphics[scale=.75]{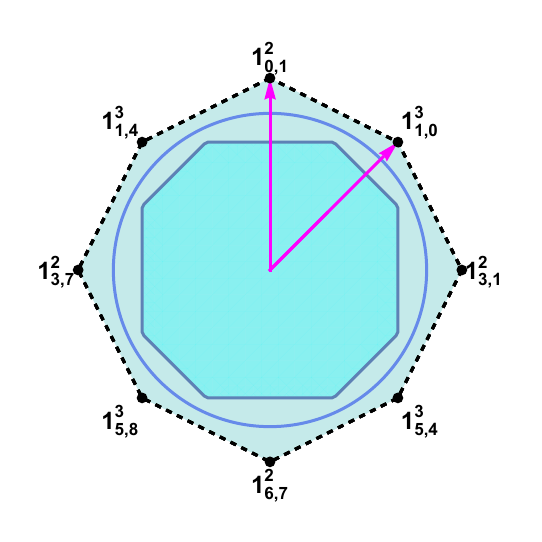}
\caption{$(2,3)$-frame, generated by $1_{0,1}^2$ and $1_{1,0}^3$. }
\label{f.3d_class_23}
\end{subfigure}
\caption{Principal $\alpha$-vectors for 2d frames in 3d, labeled by $p_{P_1,P_2}^n$. The magenta arrows point to the frame simplex vectors. $p$ refers to the spacetime dimension of the brane, $P_1$ and $P_2$ refer to the coordinates of the brane with respect to the frame, and $n$ labels the number of dimensions that decompactify from the KK-modes that come from the winding modes of the brane. The circle has radius 1 and is contained in the convex hull of particle towers, as required by the Sharpened DC. The species polytopes are cyan.}
\label{f.3d_class_geo2}
\end{figure}

\begin{figure}[H]
\centering
\begin{subfigure}{.49\linewidth}
\centering
\includegraphics[scale=.75]{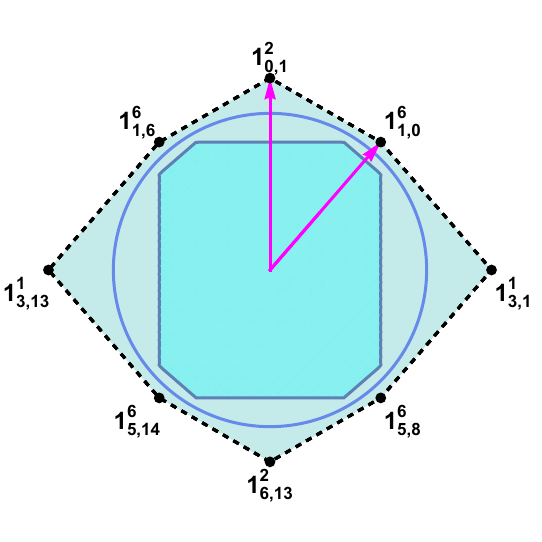}
\caption{$(2,6)$-frame, generated by $1_{0,1}^2$ and $1_{1,0}^6$. }
\label{f.3d_class_26}
\end{subfigure}
\begin{subfigure}{.49\linewidth}
\centering
\includegraphics[scale=.75]{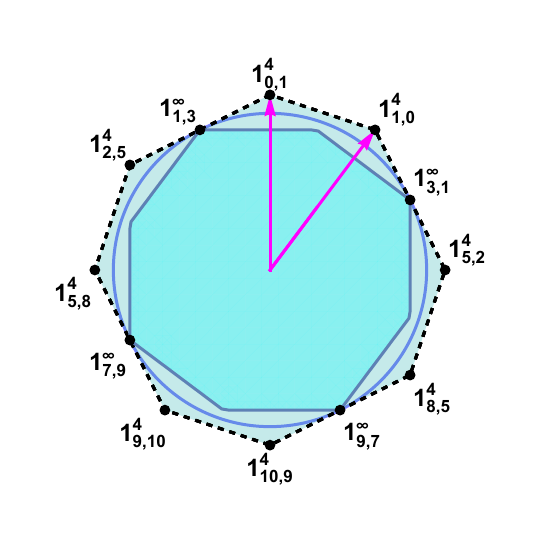}
\caption{$(4,4)$-frame, generated by $1_{0,1}^4$ and $1_{1,0}^4$. }
\label{f.3d_class_44}
\end{subfigure}
\begin{subfigure}{.49\linewidth}
\centering
\includegraphics[scale=.75]{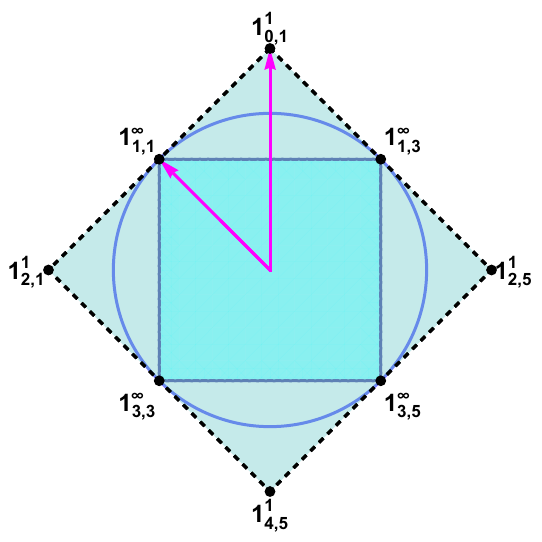}
\caption{$(1,\infty )$-frame, generated by $1_{0,1}^1$ and $1_{1,1}^\infty$. }
\label{f.3d_class_1i}
\end{subfigure}
\begin{subfigure}{.49\linewidth}
\centering
\includegraphics[scale=.75]{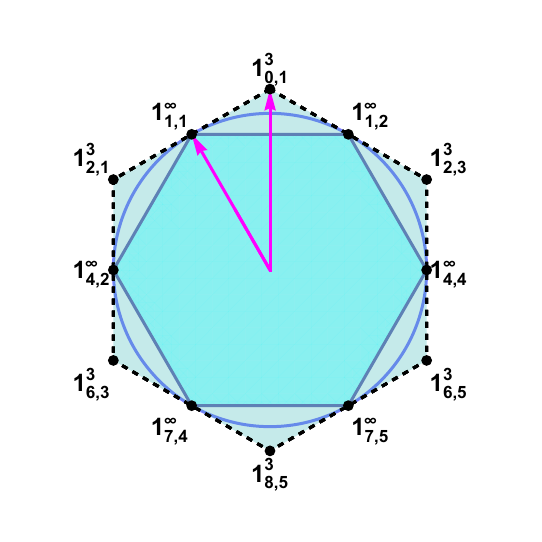}
\caption{$(3,\infty )$-frame, generated by $1_{0,1}^3$ and $1_{1,1}^\infty$. }
\label{f.3d_class_3i}
\end{subfigure}
\caption{Principal $\alpha$-vectors for 2d frames in 3d, labeled by $p_{P_1,P_2}^n$. The magenta arrows point to the frame simplex vectors. $p$ refers to the spacetime dimension of the brane, $P_1$ and $P_2$ refer to the coordinates of the brane with respect to the frame, and $n$ labels the number of dimensions that decompactify from the KK-modes that come from the winding modes of the brane. The circle has radius 1 and is contained in the convex hull of particle towers, as required by the Sharpened DC. The species polytopes are cyan.}
\label{f.3d_class_geo3}
\end{figure}

The $(1,1)$-frame produces a unique theory with principal branes $p_{P_1,P_2}^n$,
\begin{align}
	(1,1):\qquad \begin{matrix}
	1_{0,1}^1,1_{1,0}^1,1_{1,3}^1,1_{3,1}^1,1_{3,4}^1,1_{4,3}^1
	\end{matrix},\label{e.3d_class_11}
\end{align}
and they are depicted in Figure \ref{f.3d_class_11}. The $(1,2)$-frame produces a unique theory with branes
\begin{align}
	(1,2):\qquad \begin{matrix}1_{0,1}^1,1_{1,0}^2,1_{1,4}^2,1_{2,1}^{\infty },1_{2,5}^{\infty },1_{3,2}^2,1_{3,6}^2,1_{4,5}^1
	\end{matrix},\label{e.3d_class_12}
\end{align}
and they are depicted in Figure \ref{f.3d_class_12}. The $(1,3)$-frame produces a unique theory with branes
\begin{align}
	(1,3):\qquad \begin{matrix}1_{0,1}^1,1_{1,0}^3,1_{1,5}^3,1_{2,1}^5,1_{2,7}^5,1_{3,3}^3,1_{3,8}^3,1_{4,7}^1
	\end{matrix},\label{e.3d_class_13}
\end{align}
and they are depicted in Figure \ref{f.3d_class_13}. The $(1,4)$-frame produces a unique theory with branes
\begin{align}
	(1,4):\qquad \begin{matrix}1_{0,1}^1,1_{1,0}^4,1_{1,6}^4,1_{2,1}^3,1_{2,9}^3,1_{3,4}^4,1_{3,10}^4,1_{4,9}^1
	\end{matrix},\label{e.3d_class_14}
\end{align}
and they are depicted in Figure \ref{f.3d_class_14}. The $(1,5)$-frame is prohibited. The $(1,6)$-frame produces two theories, but one has a $(1,1)$-frame that is inconsistent with the previous $(1,1)$-frame analysis. Thus, the new theory has the branes
\begin{align}
	(1,6):\qquad \begin{matrix}1_{0,1}^2,1_{1,0}^3,1_{1,4}^3,1_{3,1}^2,1_{3,7}^2,1_{5,4}^3,1_{5,8}^3,1_{6,7}^2
	\end{matrix},\label{e.3d_class_16}
\end{align}
and they are depicted in Figure \ref{f.3d_class_16}. The $(1,7)$-frame produces a unique theory with branes
\begin{align}
	(1,7):\qquad \begin{matrix}1_{0,1}^1,1_{1,0}^7,1_{1,9}^7,1_{2,2}^7,1_{2,14}^7,1_{3,7}^7,1_{3,16}^7,1_{4,15}^1
	\end{matrix},\label{e.3d_class_17}
\end{align}
and they are depicted in Figure \ref{f.3d_class_17}. This theory reproduces the dilaton-volume modulus of IIB string theory on a circle (c.f., Figure 5 of \cite{Etheredge:2024amg}). The $(2,2)$-frame produces a unique theory with branes
\begin{align}
	(2,2):\qquad \begin{matrix}1_{0,1}^2,1_{1,0}^2,1_{1,3}^5,1_{2,4}^{\infty },1_{3,1}^5,1_{3,5}^5,1_{4,2}^{\infty },1_{5,3}^5,1_{5,6}^2,1_{6,5}^2
	\end{matrix},\label{e.3d_class_22}
\end{align}
and they are depicted in Figure \ref{f.3d_class_22}. The $(2,3)$-frame produces a unique theory with branes
\begin{align}
	(2,3):\qquad \begin{matrix}1_{0,1}^2,1_{1,0}^3,1_{1,4}^3,1_{3,1}^2,1_{3,7}^2,1_{5,4}^3,1_{5,8}^3,1_{6,7}^2
	\end{matrix},\label{e.3d_class_23}
\end{align}
and they are depicted in Figure \ref{f.3d_class_23}. The $(2,4)$-frame is prohibited. The $(2,5)$-frame reproduces the $(2,2)$-frame. The $(2,6)$-frame produces a unique theory with branes
\begin{align}
	(2,6):\qquad \begin{matrix}1_{0,1}^2,1_{1,0}^6,1_{1,6}^6,1_{3,1}^1,1_{3,13}^1,1_{5,8}^6,1_{5,14}^6,1_{6,13}^2
	\end{matrix},\label{e.3d_class_26}
\end{align}
and they are depicted in Figure \ref{f.3d_class_26}.  The $(3,3)$-frame is prohibited. The $(3,4)$-frame reproduces the $(1,4)$-frame. The $(3,5)$-frame reproduces the $(3,4)$-frame. Seven of the 8 polytopes with a $(4,4)$-frame have a $(1,\infty)$-frame, and these are inconsistent with the $(1,\infty)$-frame analysis (done below), and thus there is a unique $(4,4)$ theory, with the following branes
\begin{align}
	(4,4):\qquad \begin{matrix}1_{0,1}^4,1_{1,0}^4,1_{1,3}^{\infty },1_{2,5}^4,1_{3,1}^{\infty },1_{5,2}^4,1_{5,8}^4,1_{7,9}^{\infty },1_{8,5}^4,1_{9,7}^{\infty },1_{9,10}^4,1_{10,9}^4	\end{matrix},\label{e.3d_class_44}
\end{align}
and they are depicted in Figure \ref{f.3d_class_44}.  The $(1,\infty)$-frame produces a unique theory with branes
\begin{align}
	(1,\infty ):\qquad \begin{matrix}
		1_{0,1}^1,1_{1,1}^{\infty },1_{1,3}^{\infty },1_{2,1}^1,1_{2,5}^1,1_{3,3}^{\infty },1_{3,5}^{\infty },1_{4,5}^1
	\end{matrix},\label{e.3d_class_1i}
\end{align}
and they are depicted in Figure \ref{f.3d_class_1i}. The $(2,\infty)$-frame reproduces the $(1,2)$-frame. The $(3,\infty)$-frame produces a unique theory with branes
\begin{align}
	(3,\infty ):\qquad \begin{matrix}
		1_{0,1}^3,1_{1,1}^{\infty },1_{1,2}^{\infty },1_{2,1}^3,1_{2,3}^3,1_{4,2}^{\infty },1_{4,4}^{\infty },1_{6,3}^3,1_{6,5}^3,1_{7,4}^{\infty },1_{7,5}^{\infty },1_{8,5}^3
	\end{matrix},\label{e.3d_class_3i}
\end{align}
and they are depicted in Figure \ref{f.3d_class_3i}. The $(4,\infty)$-frame reproduces the $(4,4)$-frame. The $(5,\infty)$-frame reproduces the $(2,2)$-frame. The $(6,\infty)$-frame is prohibited. The $(7,\infty )$-frame reproduces the $(1,7)$-frame. Thus, there are only twelve theories consistent with this analysis, listed in Table \ref{t.2d_3d_theories} and depicted in Figures \ref{f.3d_class_geo1}, \ref{f.3d_class_geo2}, and \ref{f.3d_class_geo3}.

\begin{table}[H]
$$\begin{array}{c|c|c|c|c|c|c|c|c|c|c|}
(D_1-d,D_2-d)&p_{P_1,P_2}^n \\ \hline
(1,1)& \begin{matrix}
	1_{0,1}^1,1_{1,0}^1,1_{1,3}^1,1_{3,1}^1,1_{3,4}^1,1_{4,3}^1
	\end{matrix}\\\hline
(1,2)& \begin{matrix}
	1_{0,1}^1,1_{1,0}^2,1_{1,4}^2,1_{2,1}^{\infty },1_{2,5}^{\infty },1_{3,2}^2,1_{3,6}^2,1_{4,5}^1
	\end{matrix}\\\hline
(1,3)& \begin{matrix}
	1_{0,1}^1,1_{1,0}^3,1_{1,5}^3,1_{2,1}^5,1_{2,7}^5,1_{3,3}^3,1_{3,8}^3,1_{4,7}^1
	\end{matrix}\\\hline
(1,4)& \begin{matrix}
	1_{0,1}^1,1_{1,0}^4,1_{1,6}^4,1_{2,1}^3,1_{2,9}^3,1_{3,4}^4,1_{3,10}^4,1_{4,9}^1
	\end{matrix}\\\hline
(1,6)& \begin{matrix}1_{0,1},1_{1,0},1_{1,8},1_{2,1},1_{2,13},1_{3,6},1_{3,14},1_{4,13}
	\end{matrix}\\\hline
(1,7)& \begin{matrix}
	1_{0,1}^1,1_{1,0}^7,1_{1,9}^7,1_{2,2}^7,1_{2,14}^7,1_{3,7}^7,1_{3,16}^7,1_{4,15}^1
	\end{matrix}\\\hline
(2,2)& \begin{matrix}
	1_{0,1}^2,1_{1,0}^2,1_{1,3}^5,1_{2,4}^{\infty },1_{3,1}^5,1_{3,5}^5,1_{4,2}^{\infty },1_{5,3}^5,1_{5,6}^2,1_{6,5}^2
	\end{matrix}\\\hline
(2,3)& \begin{matrix}
	1_{0,1}^2,1_{1,0}^3,1_{1,4}^3,1_{3,1}^2,1_{3,7}^2,1_{5,4}^3,1_{5,8}^3,1_{6,7}^2
	\end{matrix}\\\hline
(2,6)& \begin{matrix}
	1_{0,1}^2,1_{1,0}^6,1_{1,6}^6,1_{3,1}^1,1_{3,13}^1,1_{5,8}^6,1_{5,14}^6,1_{6,13}^2
	\end{matrix}\\\hline
(4,4)& \begin{matrix}
	1_{0,1}^4,1_{1,0}^4,1_{1,3}^{\infty },1_{2,5}^4,1_{3,1}^{\infty },1_{5,2}^4,1_{5,8}^4,1_{7,9}^{\infty },1_{8,5}^4,1_{9,7}^{\infty },1_{9,10}^4,1_{10,9}^4	\end{matrix}\\\hline
(1,\infty)& \begin{matrix}
	1_{0,1}^1,1_{1,1}^{\infty },1_{1,3}^{\infty },1_{2,1}^1,1_{2,5}^1,1_{3,3}^{\infty },1_{3,5}^{\infty },1_{4,5}^1
	\end{matrix}\\\hline
(3,\infty)& \begin{matrix}
	1_{0,1}^3,1_{1,1}^{\infty },1_{1,2}^{\infty },1_{2,1}^3,1_{2,3}^3,1_{4,2}^{\infty },1_{4,4}^{\infty },1_{6,3}^3,1_{6,5}^3,1_{7,4}^{\infty },1_{7,5}^{\infty },1_{8,5}^3
	\end{matrix}\\\hline
\end{array}$$ 
\caption{3d theories with 2d moduli spaces, labeled by their frame simplices and which principal branes $p_{P_1,P_2}^n$ they possess. $p$ refers to the spacetime dimension of the brane, $P_1$ and $P_2$ refer to the coordinates of the brane with respect to the frame, and $n$ labels the number of dimensions that decompactify from the KK-modes that come from the winding modes of the brane. The branes are depicted in Figures \ref{f.3d_class_geo1}, \ref{f.3d_class_geo2} and  \ref{f.3d_class_geo3}.  } \label{t.2d_3d_theories}
\end{table}

\section{Application to the Dark Dimension Scenario\label{s.dark_dimension} }

The Dark Dimension Scenario \cite{Montero:2022prj} proposes that our universe is near a decompactification limit to 5d, with the compact extra ``Dark Dimension" having a length scale $l_\text{dark}$ related to the cosmological constant $\Lambda$ by $l_\text{dark}\sim \Lambda^{-1/4}$ in 4d Planck units.

Suppose that there exist two non-compact moduli in the 4d theory, where one modulus is a radion controlling $l_\text{dark}$, and suppose that we assume that the ESC holds, that moduli spaces are flat, and that there is no sliding. Then the principality conditions and classification in Section \ref{s.classification} require the existence of strings whose mass scale $m_s$ is bounded above by the Dark Dimension KK mass $m_\text{KK}$ by $m_s\lesssim m_\text{KK}^{1/3}$, in 4d Planck units. In the Dark Dimension scenario, the KK-mode scale is related to the cosmological constant $\Lambda$ by $m_\text{KK}\sim \Lambda^{1/4}$, and this implies that the string scale satisfies $m_s\lesssim \Lambda^{1/12}$. In the analysis of \cite{Heckman:2024trz}, the string scale is at or below the GUT scale.

\subsection{Motivating an extra modulus}

Consider the circle compactification of a $(D=4)$-dimensional theory to a $(d=3)$-dimensional theory. In the small radius limit, exotic branes are needed for the Sharpened DC and ESC to hold. This requires the 4d theory to have at least one modulus decompactifying $n$ dimensions, and I now determine how large $n$ must be. For the Sharpened DC to hold in the small radius limit, there must exist a particle tower that becomes light with radion value of
\begin{align}
	\alpha_\rho\leq -1.
\end{align}
By the radion lattice conditions \eqref{e.radion_lattice}, this is
\begin{align}
	\alpha_\rho=1-\frac 12P\leq -1,
\end{align}
thus $P\geq 4$. This requires the particle towers to be exotic. If the only types of exotic branes are those described in Section \ref{s.lattice.exotic}, then the exotic brane bound on $P$ is $P_\text{max}=D+d+n-d-2=n+2$. Having $P\geq 4$ requires $n$ to be at least $2$. But, this means that the 4d theory should have a decompactification limit to at least 6d. Thus, if the 4d theory has a decompactification limit to 5d and a modulus controlling that limit, there must also exist a second modulus that allows the 4d theory to be able to decompactify to at least 6d! For the case where $n=2$, the theory is depicted in Figure \ref{f.3d_class_11}.

This argument is not airtight, as it requires assuming that exotic branes originate through the mechanism described in Section \ref{s.lattice.exotic}. But, there might be other landscape lines of arguments for an extra modulus. For instance, the Dark Dimension scenario involves decompactification to 5d, but if the 5d theory comes from a 10d string theory or 11d M-theory, there could be a decompactification limit to these theories. For instance, 5d $\mathcal N=1$ theories from M-theory on CY threefolds can sometimes have a 0d vector multiplet moduli space but have complex structure moduli with a decompactification limit to M-theory.

\subsection{Dark strings}
I now argue that the existence of two or more moduli in the 4d theory requires the existence of strings, and I relate the tensions of these strings to the cosmological constant in the Dark Dimension Scenario. In particular, I show that these constraints require that there exist strings whose mass scale $m_s$ is bounded above by the Dark Dimension KK-mode mass $m_\text{KK}$ by $m_s\lesssim m_\text{KK}^{1/3}$.

Suppose that the size of the Dark Dimension is controlled by some canonically normalized radion $\rho$. Suppose that there exists some other canonically normalized non-compact modulus $\phi$ in the theory, and suppose that, at least asymptotically, $\rho$ and $\phi$ together combine to produce a two-dimensional flat moduli space.\footnote{The generalization of this argument to higher-dimensional moduli spaces does not significantly alter the following conclusions. }

Particle towers are not enough to satisfy the ESC under dimensional reduction, and strings are needed.

Consider first the convex hull of particle-tower $\alpha$-vectors. The ESC requires that this convex hull is a polygon, and the vertices that generate this polygon are $\alpha$-vectors of KK-modes and string oscillator modes \cite{Etheredge:2024tok}. One vertex of the tower polygon is given by the $\alpha$-vector of the Dark Dimension KK-modes, and its value is
\begin{align}
	\vec \alpha_\text{dark KK}=\sqrt{\frac32}\hat \rho.\label{e.alphadarkKK}
\end{align}
Consider a different vertex of the tower polytope $\vec \alpha_n$ such that $\vec \alpha_n$ is at one end of an edge of the tower polytope, and $\vec \alpha_\text{dark KK}$ at the other end of the same edge of the tower polytope. Without loss of generality, consider the case where $\vec \alpha_n$ has positive $\hat \phi$ component.

From \cite{Etheredge:2024tok}, the taxonomy rule \eqref{e.taxonomy} requires that the dot product between $\vec \alpha_n$ and $\vec \alpha_\text{dark KK}$ is given by
\begin{align}
	\vec \alpha_n\cdot \vec \alpha_\text{dark KK}=\frac 1{\sqrt 2}.\label{e.producttaxonomy}
\end{align}
We can assume that $\vec \alpha_n$ is a vertex of a KK-mode involving decompactification from 4d to $(4+n)$-dimensions for some value $n$.\footnote{As discussed earlier in this paper, and also in \cite{Etheredge:2024tok}, for the purposes of this discussion, it is consistent to formally consider vertices related to string oscillator modes as just KK-modes corresponding to decompactification of infinitely many dimensions. So, the case where $\vec \alpha_n$ is a string oscillator mode is obtained by formally setting $n=\infty$ \cite{Etheredge:2024tok}. } Its length must satisfy the tower taxonomy rule \eqref{e.taxonomy},
\begin{align}
	|\vec \alpha_n|^2=\frac{2+n}{2n}. \label{e.lengthtaxonomy}
\end{align}

\begin{figure}
\begin{center}
\includegraphics[width = .7\linewidth]{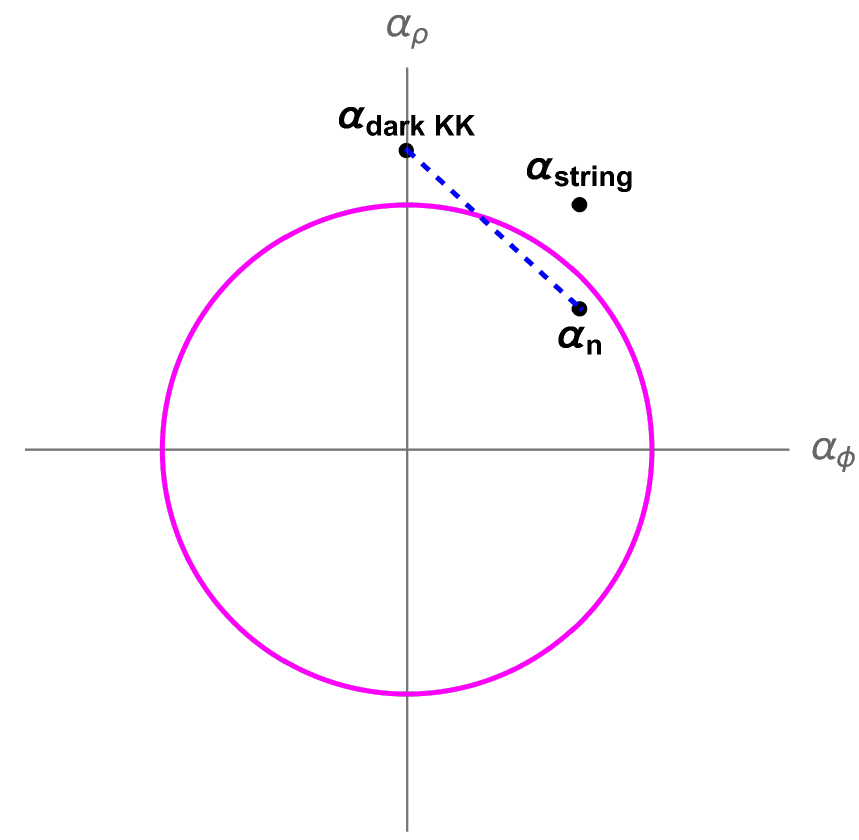}
\end{center}
\caption{The vector $\vec \alpha_\text{dark KK}$, the other vertex $\vec \alpha_n$ (here $n=3$ plotted), and the $\alpha$-vector of the required string. The magenta circle has radius 1, and so the line segment connecting the $\vec \alpha_\text{dark KK}$ and $\vec \alpha_n$ comes within a distance 1 of the origin, and thus the $p_\text{max}=2$ Brane DC requires the existence of a string.}
\label{f.darkstring}
\end{figure}

Let $\alpha_\text{min}$ be the closest distance between the origin and the line interval connecting $\vec \alpha_\text{dark KK}$ and $\vec \alpha_n$. The formulas \eqref{e.producttaxonomy} and \eqref{e.lengthtaxonomy} imply
\begin{align}
	\alpha_\text{min}^2= \frac{3+n}{2(1+n)}\leq 1.\label{e.alphamin}
\end{align}
The Brane DC \cite{Etheredge:2024amg} requires that, when $\alpha_\text{min} \leq 1$, there need to be additional strings whose $\alpha$-vectors are outside of the tower polytope. This is because in 4d the Brane DC requires that convex hull of $\alpha$-vectors of particle towers and strings needs to contain the ball of radius 1.\footnote{Strictly speaking, this argument requires such strings only when $\alpha_\text{min} <1$ (which happens when $n\geq 2$), and the argument that strings are required in the $\alpha_\text{min}=1$ case follows from the Sharpened DC saturation conditions discussed in \cite{Etheredge:2022opl, Etheredge:2024amg}, which is related to saturation of the Sharpened Distance Conjecture upon compactification to 3d.} And, indeed, $\alpha_\text{min}\leq 1$ for all $n$, thus requiring the existence of strings whose $\alpha$-vectors are outside of the tower polytope. See Figure \ref{f.darkstring}.

As discussed in \cite{Etheredge:2024amg}, or alternatively due to the lattice rules derived in earlier sections of this paper, there is a dot product relationship between the $\alpha$-vector of the string $\vec \alpha_\text{string}$ and the $\alpha$-vector $\vec \alpha_\text{dark KK}$ of the Dark Dimension KK mode,
\begin{align}
	\vec \alpha_\text{dark KK}\cdot \vec \alpha_\text{string}=1.\label{e.darkKKstringproduct}
\end{align}
Equations \eqref{e.darkKKstringproduct} and \eqref{e.alphadarkKK} imply that the radion component of $\vec \alpha_\text{string}$ is
\begin{align}
	\vec \alpha_\text{string}\cdot \hat \rho=\sqrt{\frac{2}{3}}.
\end{align}
Thus, the tension of the string scales with the radion $\rho$ and the Dark Dimension KK-mode mass $m_\text{dark KK}$ by
\begin{align}
	T_\text{string}\sim \exp\left(-\sqrt{\frac{2}{3}}\rho \right)\sim m_\text{dark KK}^{2/3}.
\end{align}

Meanwhile, from \cite{Montero:2022prj}, the Dark Dimension KK-mode has a mass that scales with the cosmological constant by
\begin{align}
	m_\text{dark KK}\sim \Lambda^{1/4},
\end{align}
and the mass scale of a string scales with the square root of the string tension,
\begin{align}
	m_\text{string}\sim \sqrt T_\text{string},
\end{align}
and so the mass scale of the string scales with the twelfth root of the cosmological constant,
\begin{align}
	m_\text{string}\sim \Lambda^{1/12}.\label{e.stringLambda}
\end{align}

The string that this argument produces can be interpreted as a string from the 5d theory that does not wrap the Dark Dimension. It is interesting that both the arguments in \cite{Heckman:2024trz} and also my ESC approach motivate the same bound \eqref{e.stringLambda} (which is related to the GUT scale in \cite{Heckman:2024trz}). The result here, though, follows from the ESC and from applying the lattice taxonomy rules and the principality conditions to the case where the Dark Dimension Scenario has an extra modulus (which I have motivated by requiring the ESC to hold upon reduction to 3d with exotic branes behaving in the way discussed in Section \ref{s.lattice.exotic}.)

The bound \eqref{e.stringLambda} is in fact an \emph{upper} bound on the string scale, since one can travel in also the $\phi$ direction in moduli space. However, there must be a string satisfying \eqref{e.stringLambda} whose $\alpha$-vector component in the $\phi$ direction is non-negative. Thus, traveling a Planckian distance in the $\phi$ direction in moduli space exponentially lowers the mass scale of the string. Thus,
\begin{align}
	m_\text{string}\lesssim \Lambda^{1/12}.\label{e.stringLambdalesssim}
\end{align}
It would be interesting to be able to place bounds in our universe on how far in the $\phi$ modulus we have traveled, as this could exponentially lower the string scale.

\section{Conclusions\label{s.conclusions} }

In flat slices of moduli space where $\alpha$-vectors are constant, I have argued that the ESC requires $\alpha$-vectors of particle-towers and branes to reside on lattices. This results in the existence of discretized, exponentially separated hierarchies of energy scales in asymptotic limits of moduli spaces. Given a duality frame with a set of leading towers, I have derived explicit formulas for the lattices in terms of the $\alpha$-vectors of the leading towers. Not all of these lattice points are necessarily populated by particle towers or branes, but I have identified principality conditions that determine whether a lattice site must be populated in order for the ESC and Sharpened DC to be preserved under dimensional reduction. In this paper, I have classified all duality frames, lattices, principal branes, and tower/species polytopes consistent with these assumptions for 0d, 1d, and 2d moduli spaces in theories with spacetime dimensions between 3 and 11, and I have argued that 11d is the maximum spacetime dimension. Remarkably, this classification reproduces the nontrivial particle-tower and brane content of various theories within the string landscape. If these assumptions apply to the Dark Dimension Scenario with an extra modulus, then I have argued for strings with tensions related to the cosmological constant by $T\lesssim \Lambda ^{1/6}$ in 4d Planck units.

In asymptotic limits of moduli spaces, this work implies the existence of hierarchies of energy scales that are exponentially separated, with the exponential separation governed by the lattice rules presented in this paper. With the analysis of principal branes, sometimes there are gaps---or ``deserts"---in the lattice of the logarithms of these energy scales. It would be interesting to investigate potential phenomenological implications of this, or identify these lattice hierarchy structures in particle phenomenology and cosmology. It would also be interesting if similar lattice rules govern coupling constants in phenomenology or the string landscape.

I have made strong assumptions in this paper, but it is likely that they can be relaxed. For instance, I have assumed that $\alpha$-vectors are constant (even after compactification and decompactification), and derived lattice rules and principality conditions, and classified the principal branes of 0d, 1d, and 2d moduli spaces consistent with these assumptions. This classification reproduces the $\frac 12$-BPS particle tower and brane content of maximal supergravity, where the assumptions are known to apply, but, surprisingly, in the examples I have investigated, it also governs the known particle tower and brane content of heterotic/Type I string theory and 5d $\mathcal N=1$ theories. In the examples I have checked, the known branes in regular limits of heterotic/Type I string theory reside on the lattices I have proposed, and I have also shown that in asymptotic limits of two examples of 5d $\mathcal N=1$ theories, the strings and particle towers have $\alpha$-vectors whose asymptotic behavior fits precisely within my classification. Also, in the limited cases where there is sliding that I have examined in this paper, I have pointed out that sliding always occurs between lattice sites. Does this hold more generally?

Relaxing these assumptions and understanding sliding would be useful in predicting new branes, especially in 10d heterotic and Type I string theory. If one hypothetically pretended that heterotic/Type I string theory did not have sliding under toroidal compactifications (which is known to not be the case \cite{Etheredge:2023odp}), then my analysis would require SO(32) heterotic string theory to have a full set of branes with D-odd-brane $\alpha$-vectors, (just like in IIB string theory) and $\mathrm{E_8\times E_8}$ heterotic string theory to have a full set of branes with D-even-brane $\alpha$-vectors (just like in IIA string theory). Furthermore, as discussed in Section \ref{s.principal.instantons}, sometimes principal branes imply the existence of lower-dimensional branes, and this might motivate the heterotic instanton proposed in \cite{Alvarez-Garcia:2024vnr}. With a better developed understanding of sliding, it would be interesting if the recently predicted non-BPS branes in heterotic string theory of \cite{Kaidi:2023tqo} are required by the ESC.

Even without sliding, it is not yet fully understood how the Sharpened DC and ESC are preserved under dimensional reduction. That is, there is not yet known a statement that is preserved under dimensional reduction and also automatically implies the Sharpened DC and ESC under dimensional reduction. While the principality conditions I have identified in this paper are necessary conditions, I have not shown whether they are sufficient conditions, (though it is possible that they are sufficient). It would be useful to have a precise statement that is preserved under reduction, even without sliding. That could potentially result in an extremely powerful Swampland conjecture that could heavily constrain string theory.

It would be useful to have results that depend on as little information as possible about other duality frames. The lattice formulas \eqref{e.lattice} and the principal norm condition \eqref{e.principal_norm} depend on only one duality frame, but the principal product condition \eqref{e.principal_product} could potentially be affected by branes with $\alpha$-vectors pointing in directions of other duality frames. It would be useful to reduce the set of objects that must be considered in testing the principal product condition \eqref{e.principal_product}. As discussed in Section \ref{s.principal}, it is possible that the species polytope could be the key to this. For instance, for a principal-norm candidate brane whose $\alpha_p/p$-vector is on the boundary of the species polytope, the principal product condition might need to be checked with only other branes whose $\alpha_q/q$-vectors are on the same facet of the species polytope.

The study of principal branes and the species polytope might be useful for obtaining a distance conjecture for instantons. For instance, the existence of principal branes with $\vec \alpha_p/p$-vectors on the boundary of a Planckian facet of the species polytope implies that these branes come from unwrapped branes of a higher-dimensional theory. But, these branes could also wrap the compactification manifold, and thus require the existence of lower dimensional branes that accompany higher-dimensional principal branes. In particular, this can be used to argue that instantons must in certain contexts accompany particle towers (and perhaps require the instanton of \cite{Alvarez-Garcia:2024vnr}). This is the subject of upcoming work \cite{Etheredge:Bitowers}.

 It is possible that these results could be combined with the Weak Gravity Conjecture (WGC). For example, the extremality bound of the WGC depends on $\alpha$-vectors \cite{Heidenreich:2020upe}. Also, the ESC requires $\alpha$-vectors to reside on lattices, and this is reminiscent of examples of the sublattice Weak Gravity Conjecture (sLWGC) \cite{Heidenreich:2015nta, Heidenreich:2016aqi, Etheredge:2025rkn}, where charge-to-tension ratios exist on lattices. There might exist nontrivial convex hull or lattice statements in the product space of $\alpha$-vector space and gauge-charge-to-tension-ratio space that unify the sLWGC with the statements in this paper.

To limit the scope of this paper, I have not classified moduli spaces of three dimensions or higher, but the extension to higher-dimensional spaces is likely straightforward. It is likely that there will be similar results, and finitely many allowed theories, especially since the maximum spacetime dimension of each moduli space decreases as one increases the dimension of moduli space under consideration. Explicitly doing this would significantly expand the classification program initiated in \cite{Etheredge:2024tok} and this paper.

I have shown that my classification reproduces theories in the 32, 16, and 8 supercharge string landscape. Do there exist examples within my classification that have not yet been discovered in the string landscape? This is more likely to occur in lower dimensions, where my classification produces a larger number of theories.

It would be interesting to further motivate the existence of an extra modulus $\phi$ in the Dark Dimension Scenario, and further justify my analysis. If there are ways of generating exotic branes beyond those discussed in Section \ref{s.lattice.exotic}, then my argument for an extra modulus is potentially invalidated. If there does exist an extra modulus $\phi$, and my analysis applies to the Dark Dimension Scenario, then I have shown that there must exist strings with tensions related to the cosmological constant by $T\lesssim \Lambda^{1/6}$ in 4d Planck units, and increasing the vev of $\phi$ decreases the tensions of these strings. For large vevs of $\phi$, these strings have exponentially lower tensions, and motivate the questions: What might be bounds on $\phi$ in our universe? What are lower-bounds on the Dark Dimension string tension?

The results of this paper might translate into CFT statements and refine the CFT Distance Conjecture \cite{Perlmutter:2020buo, Baume:2020dqd, Calderon-Infante:2024oed}. Translating statements about non-particle brane tensions into CFT statements might be difficult \cite{Bachas:2024nvh}, but my analysis governs masses of subleading particle towers, which are more easily translatable.  Since I have argued that particle towers have lattice-valued $\alpha$-vectors, perhaps via AdS/CFT the conformal-manifold gradients of the logarithms of the anomalous dimensions $\gamma$ of towers of currents are lattice-valued in asymptotic limits of conformal manifolds. Perhaps near higher spin points in conformal manifolds, there exist towers of higher-spin operators whose anomalous dimensions $\gamma$ approach zero, and perhaps they must satisfy rules where $(-\nabla \log \gamma)$-vectors lie on lattices. It would be interesting if there exist lattice rules that follow from replacing the $\alpha$-vectors for particle towers with $(-\nabla \log \gamma)$-vectors for anomalous dimensions. Could such formulas be derived from the CFT perspective?

It would be interesting to further compare the theories produced by my classification with more of the landscape. In this paper, I have reproduced structures that occur in maximal supergravity, heterotic string theory on tori, and vector multiplet moduli of 5d $\mathcal N=1$ theories. My comparison with the landscape of under 16 supercharges has been relatively limited (despite finding some non-trivial consistency), and future work aims to further analyze the landscape of 5d $\mathcal N=1$ theories and heterotic string theory on tori. It would also be interesting to see whether my analysis governs the particle towers and branes of string island examples and recently discovered theories in the landscape, and also non supersymmetric theories, (see, e.g., \cite{Dabholkar:1998kv, ParraDeFreitas:2022wnz, Fraiman:2022aik, Fraiman:2023cpa, Baykara:2025lhl}.) It would also be interesting to explore further whether the observations in \cite{Grieco:2025bjy} are examples of the lattice taxonomy rules proposed in this paper, and whether these lattice taxonomy rules explain the bottom-up observations of 4d $\mathcal N=1$ theories in \cite{Lanza:2021udy}.

\section*{Acknowledgements}
I am grateful for conversations with Rafael \'Alvarez-Garc\'\i a, Jos\'e Calder\'on-Infante, Bernardo Fraiman, Ben Heidenreich, H\'ector Parra de Freitas, Nick Pittman, Muthusamy Rajaguru, Sanjay Raman, Nicole Righi, Tom Rudelius, Ignacio Ruiz, Cumrun Vafa, and Irene Valenzuela. I also thank Gina Etheredge, Bernardo Fraiman, Ben Heidenreich, Nick Pittman, Yue Qiu, Nicole Righi, and Marin Sklan for feedback on the manuscript. I am grateful for the ``Landscape vs. the Swampland" workshop at the Erwin Schr\"odinger Institute, the 2025 Simons Summer Physics Workshop, the Harvard Swampland Initiative, and the Harvard CMSA Workshop on Symmetries and Gravity, where some of this work was inspired and completed. I received support from the NSF grant PHY-2112800.

\appendix

\section{5d $\mathcal N=1$ examples \label{s.5dN1}}

In this appendix, I study some examples in the vector multiplet moduli space of 5d $\mathcal N=1$ supergravity, and I compare the results with the lattices, principal branes, and polytopes found within the 5d classification sections in this paper.

I show in particular that the Rudelius Dirac-pair identity \cite{Rudelius:2023spc} is an example of both the dilaton and radion lattice conditions introduced in this paper that can be extended into the bulk of moduli space. I also show that the symmetric flop and GMSV conifold examples reproduce exactly the principal brane content in the classification of 1d moduli spaces of 5d theories in Section \ref{s.classification.1d.5d}. Thus this classification applies to some 8-supercharge examples and perhaps the assumptions in this paper can be relaxed with similar results still holding. Future work will explore this in more depth.

\subsection{Review}

Consider the vector-multiplet moduli of 5d $\mathcal N=1$ theories. From Section 2 of \cite{Alim:2021vhs}, the action is
\begin{align}
\begin{aligned}
	S&=\frac{1}{2\kappa_5^2}\int d^5x\sqrt{-g}\left(R-\frac 12 g_{ij}(\phi)\partial\phi^i\cdot \partial \phi^j\right)-\frac{1}{g_5^2}\int a_{IJ}(\phi)F^I\wedge \star F^J\\
	&+\frac{1}{6(2\pi)^2}\in C_{IJK}A^I\wedge F^J\wedge F^K,
\end{aligned}
\end{align}
where $I=0,\dots,n$ $i=1,\dots,n$, and $g_5^2=(2\pi)^{4/3}(2\kappa_5^2)^{1/3}$. The relevant quantities are determined by the prepotential $\mathcal F[Y]$ evaluated with the constraint
\begin{align}
	\mathcal F[Y(\phi)]=1,
\end{align}
and we parametrize $Y^I=Y^I(\phi)$ in terms of $n$ moduli $\phi^i$. Define $\mathcal F_{I}=\partial_I\mathcal F$ and $\mathcal F_{IJ}=\partial_I \partial_J\mathcal F$. Following \cite{Alim:2021vhs}, the gauge kinetic matrix is given by,
\begin{align}
	a_{IJ}=\mathcal F_I\mathcal F_J-\mathcal F_{IJ},\qquad g_{ij}=a_{IJ}\partial_iY^I\partial_j Y^J,
\end{align}
and the $\frac 12$-BPS bound for particles and strings is
\begin{align}
	m=\frac{g_5}{\sqrt 2\kappa_5}|q_IY^I(\phi)|\qquad T=\frac{\tilde g_5}{\sqrt 2\kappa_5}|\tilde q^I\mathcal F_I|.
\end{align}

\subsection{Dirac-pair identity as a lattice condition}
Rudelius observed in \cite{Rudelius:2023spc} that strings and particles that are Dirac pairs have the following dot product identity,
\begin{align}
	\vec \alpha_\text{str}\cdot \vec \alpha_\text{part}=\frac 23.\label{e.Rudelius}
\end{align}
I now show that this can be viewed as cases of both the dilaton and radion lattice rules.

Consider an emergent string limit, where a string with $\alpha$-vector $\vec \alpha_\text{str}$ becomes light. Then, by the dilaton lattice condition \eqref{e.dilaton_lattice}, the lattice conditions from the string imply that particles satisfy
\begin{align}
	\hat  \alpha_\text{osc}\cdot \vec \alpha_\text{part}=\frac 1{\sqrt{3}}+(1-P)\frac {\sqrt 3}2.\label{e.5d_osc_part}
\end{align}
Since the length of $\vec \alpha_\text{str}$ is $\frac 1{\sqrt 3}$, 
\begin{align}
	\hat  \alpha_\text{osc}=\frac {\sqrt{3}}2\vec \alpha_\text{str},
\end{align}
\eqref{e.5d_osc_part} implies
\begin{align}
	\vec \alpha_\text{str}\cdot \vec \alpha_\text{part}=\frac 23+1-P.
\end{align}
Thus, the Rudelius identity \eqref{e.Rudelius} is the $P=1$ case of the dilaton lattice condition.

Consider instead the case where $\vec \alpha_\text{part}$ is a KK-mode tower from decompactification to a $D$-dimensional theory and thus furnishes a radion. Then, by the radion lattice condition \eqref{e.radion_lattice}, strings satisfy
\begin{align}
	\hat \alpha_\text{part}\cdot \vec \alpha_\text{str}=\frac{2(D-2)-3P}{\sqrt{3(D-5)(D-2)}}.\label{e.5d_part_str}
\end{align}
Since the length of $\vec \alpha_\text{part}$ is $\sqrt{\frac{D-2}{3(D-5)}}$, \eqref{e.5d_part_str} implies that
\begin{align}
	\vec \alpha_\text{part}\cdot \vec \alpha_\text{str}=\frac{2(D-2)-3P}{3(D-5)}.
\end{align}
When $P=2$, the Rudelius identity \eqref{e.Rudelius} is satisfied by the dilaton lattice equations.

Thus, the Rudelius identity can be an example of both the dilaton and radion lattice conditions. This reasoning has been general and applies in the bulk of moduli space.

\subsection{Symmetric flop}

\begin{figure}
\centering
	\includegraphics{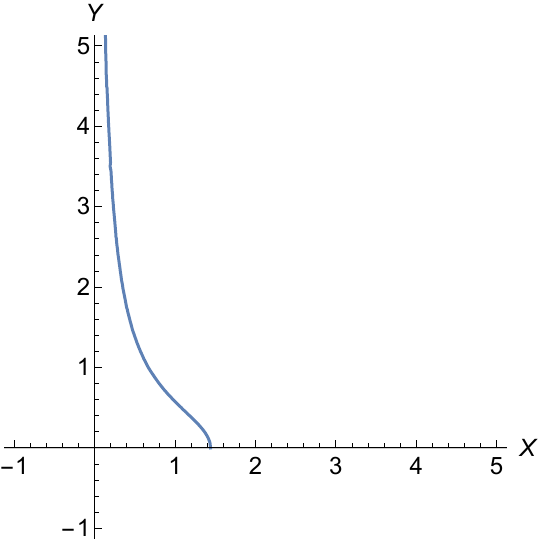}
	\caption{The $F=1$ constraint in $X$-$Y$-space for the 5d $\mathcal N=1$ symmetric flop theory. \label{f.5dN1symflop}}
\end{figure}

Consider first the $(h^{1,1},h^{2,1})=(2,86)$ CY explored in Section 7.1 of \cite{Alim:2021vhs}. The prepotential is
\begin{align}
	\mathcal F^\mathrm{(I)}=\frac{1}{3}X^3+2X^2Y,\qquad\mathcal K_\mathrm{I}=\{X,Y\geq 0\}.
\end{align}
The $\mathcal F=1$ constraint in $X$-$Y$-space is presented in Figure \ref{f.5dN1symflop}.

We can write $Y=\frac{6+X^3}{3X}$ to solve the $\mathcal F=1$ constraint, in which case
\begin{align}
	\mathcal F_I=\left(X^2+4XY,2X^2\right)=\left(\frac{6+X^3}{3X},2X^2\right).
\end{align}
In terms of the $X$ coordinate,
\begin{align}
	a_{IJ}&=\left(
\begin{array}{cc}
 X \left(X (X+4 Y)^2-2\right)-4 Y & 2 X \left(X^3+4 X^2 Y-2\right) \\
 2 X \left(X^3+4 X^2 Y-2\right) & 4 X^4 \\
\end{array}
\right)\nonumber\\
&=\left(
\begin{array}{cc}
 \frac{X^6+18}{9 X^2} & \frac{2 X^4}{3} \\
 \frac{2 X^4}{3} & 4 X^4 \\
\end{array}
\right),
\end{align}
and thus, in terms of the $X$ coordinate, the metric on moduli space is given by
\begin{align}
	g_{XX}=\frac 3{X^2}.
\end{align}
From Table 1 of \cite{Etheredge:2022opl}, the limit $X\rightarrow 0$ corresponds to an emergent string limit.

The BPS bounds for particles and strings are given by
\begin{align}
\begin{aligned}
	m&\propto |Q_1X+Q_2Y|=\left|Q_1 X-\frac{Q_2\left(X^3-3\right)}{6 X^2}\right|,\\
	T&\propto |Q^1 (X^2+4XY)+Q^22X^2|=\left|\frac{Q_1 \left(X^3+6\right)}{3 X}+2 Q_2 X^2\right|.
	\end{aligned}
\end{align}
The corresponding $\alpha$-vectors are
\begin{align}
	\vec \alpha_1=\frac{1}{\sqrt{3}}\left(\frac{9 Q_2}{X^3 (Q_2-6 Q_1)-3 Q_2}+1\right)\hat {\mathbf X},\qquad \vec \alpha_2=\frac{2 X^3 (Q_1+6 Q_2)-6 Q_1}{\sqrt{3} \left(X^3 (Q_1+6 Q_2)+6 Q_1\right)}\hat{\mathbf X}.
\end{align}
There are some special charges, namely integer multiples of
\begin{align}
	Q_I^s\in \{(1,0),(1,6)\},\qquad Q^I_s\in \{(0,1), (-6,1)\},
\end{align}
for which these $\alpha$-vectors are constant. For these special charges,
\begin{align}
	\vec \alpha_1\in\left\{\frac 1{\sqrt 3},-\frac 2{\sqrt 3}\right\}\hat {\mathbf X},\qquad \vec \alpha_2\in\left\{\frac 2{\sqrt 3},-\frac 1{\sqrt 3}\right\}\hat {\mathbf X}. \label{e.5DN1symfloplattice}
\end{align}
The charges form a pair of Dirac pairs and satisfy Rudelius's Dirac pair identity \eqref{e.Rudelius}.

\begin{figure}
\centering
\includegraphics[scale=1]{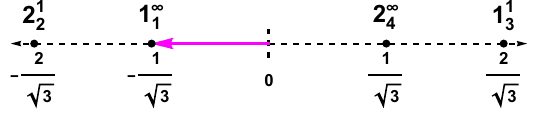}
\caption{Symmetric flop and GMSV examples. This matches exactly the classification in Figure \ref{f.5d_1dclass_1}.}
\label{f.5d_Flop}
\end{figure}

All of the other $\alpha$-vectors that are not these special charges asymptote to reside on \eqref{e.5DN1symfloplattice}! Thus, asymptotically, all of the other $\alpha$-vectors are also of the form \eqref{e.5DN1symfloplattice}. Thus, asymptotically, all of the $\alpha$-vectors in \eqref{e.5DN1symfloplattice} are on a lattice! Using the stringy frame generated by the frame simplex of just the point $\vec v_1=-\frac 1{\sqrt{3}}\unit X$, these $\alpha$-vectors have coordinates
\begin{align}
	1_1^{\infty },1_3^1,2_2^1,2_4^{\infty }.
\end{align}
These $\alpha$-vectors are depicted in Figure \ref{f.5d_Flop}. This matches exactly the classification in Section \ref{s.classification.1d.5d} of 1d moduli spaces in 5d, and the $\alpha$-vectors here correspond to the theory found and depicted in Figure \ref{f.5d_1dclass_1} (where Figure \ref{f.5d_1dclass_1} is depicted with respect to a KK-frame.) Thus, remarkably, the lattice and principality classification of Section \ref{s.classification.1d.5d} governs this 5d $\mathcal N=1$ theory, and shows that the classification program can potentially be extended to theories with only 8 supercharges.


As a final note, for the special charges, one can show that
\begin{align}
	\nabla^2 \log m=\nabla^2 \log T=0.
\end{align}
Laplacian eigenfunctions on moduli space will be the subject of future work.

\subsection{GMSV conifold}

Consider next the GMSV conifold.
\begin{align}
	\mathcal F=5/6 X^3 +2 X^2Y \qquad \text{with K\"ahler cone}\qquad \{X,Y\geq 0\}.
\end{align}

This example is similar to the symmetric flop example. There is an emergent string limit at $X\rightarrow 0$. Again, one can solve for $Y$ in terms of $X$. And, in the end, one finds that the $\alpha$-vectors of particle towers and strings are all, asymptotically, of the form
\begin{align}
	\vec \alpha_1\in\left\{-\frac 1{\sqrt 3},\frac 2{\sqrt 3}\right\}\hat {\mathbf X},\qquad \vec \alpha_2\in\left\{-\frac 2{\sqrt 3},\frac 1{\sqrt 3}\right\}\hat {\mathbf X} \label{e.5DN1GMSVlattice}
\end{align}
Using the stringy frame generated by the frame simplex of just the point $\vec v_1=-\frac 1{\sqrt{3}}\unit X$, these $\alpha$-vectors have coordinates
\begin{align}
	1_1^{\infty },1_3^1,2_2^1,2_4^{\infty }.
\end{align}
This matches exactly the symmetric flop example, and the $\alpha$-vectors are depicted in Figure \ref{f.5d_Flop}. Also, this matches exactly the same theory depicted in Figure \ref{f.5d_1dclass_1} (where Figure \ref{f.5d_1dclass_1} is depicted with respect to a KK-frame.)

Similarly to the symmetric flop, there are special particle charges $(Q_1,Q_2)=(Q_1,0)$ and $n(5,12)$, and special string charges $(Q_1,Q_2)=(0,Q_2)$ and $n(12,-5)$, with constant length $\alpha$-vectors everywhere in moduli space and that have tensions that are eigenfunctions of the moduli-space laplacian.

\bibliographystyle{JHEP}
\bibliography{ref}
\end{document}